\documentclass[11pt]{article}

\usepackage{amsmath,amsfonts,amssymb,bbm,epsfig,a4}
 \usepackage{tabularx}

\numberwithin{equation}{section} 

\renewcommand\slash[1]{\not \! #1}

\begin{document}

 \newcolumntype{L}[1]{>{\raggedright\arraybackslash}p{#1}}
 \newcolumntype{C}[1]{>{\centering\arraybackslash}p{#1}}
 \newcolumntype{R}[1]{>{\raggedleft\arraybackslash}p{#1}}

%
\def\be{\begin{equation}}
\def\ee{\end{equation}}
\def\bea{\begin{eqnarray}}
\def\eea{\end{eqnarray}}
 \newcommand{\ba}{\begin{eqnarray}}
 \newcommand{\ea}{\end{eqnarray}}
\def\rarr{\rightarrow}
\def\nn{\nonumber}
\def\fr{\frac}
\renewcommand\slash[1]{\not \! #1}
\newcommand\qs{\!\not \! q}
\def\del{\partial}
\def\gam{\gamma}
\newcommand\vphi{\varphi}
\def\tr{\mbox{tr}\,}
\newcommand\nin{\noindent}

\def\del{\partial}
\def\pbar{\bar{p}}
\begin{titlepage}
\begin{flushright}
\end{flushright}
\vspace{0.6cm}
\begin{center}
\boldmath
{\LARGE{\bf A Model for Soft High-Energy Scattering:}}\\[.2cm]
{\LARGE{\bf Tensor Pomeron and Vector Odderon}}\\
\unboldmath
\end{center}
\vspace{0.6cm}
\begin{center}
{\bf \Large
Carlo Ewerz\,$^{a,b,1}$, Markos Maniatis\,$^{c,2}$,\\
Otto Nachtmann\,$^{a,3}$}
\end{center}
\vspace{.2cm}
\begin{center}
$^a$
{\sl
Institut f\"ur Theoretische Physik, Universit\"at Heidelberg\\
Philosophenweg 16, D-69120 Heidelberg, Germany}
\\[.5cm]
$^b$
{\sl
ExtreMe Matter Institute EMMI, GSI Helmholtzzentrum f\"ur Schwerionenforschung\\
Planckstra{\ss}e 1, D-64291 Darmstadt, Germany}
\\[.5cm]
$^c$
{\sl
Departamento de Ciencias B\'a{}sicas, Universidad del B\'i{}o-B\'i{}o\\
Avda.\ Andr\'es Bello s/n, Casilla 447, Chill\'a{}n 3780000, Chile
}
\end{center}                                                                
\vfill
\begin{abstract}
\noindent
A model for soft high-energy scattering is developed. The model is 
formulated in terms of effective propagators and vertices for the 
exchange objects: the pomeron, the odderon, and the reggeons. 
The vertices are required to respect standard rules of QFT. The 
propagators are constructed taking into account the crossing 
properties of amplitudes in QFT and the power-law ans\"atze 
from the Regge model. We propose to describe the pomeron as an 
effective spin 2 exchange. This tensor pomeron gives, at high energies, 
the same results for the $pp$ and $p \bar{p}$ elastic amplitudes as the 
standard Donnachie-Landshoff pomeron. But with our tensor pomeron 
it is much more natural to write down effective vertices of all kinds which 
respect the rules of QFT. This is particularly clear for the coupling 
of the pomeron to particles carrying spin, for instance vector mesons. 
We describe the odderon as an effective vector exchange. We emphasise 
that with a tensor pomeron and a vector odderon the corresponding 
charge-conjugation relations are automatically fulfilled. 
We compare the model to some experimental data, in particular to data 
for the total cross sections, in order to determine the model parameters. 
The model should provide a starting point for 
a general framework for describing soft high-energy reactions. 
It should give to experimentalists an easily manageable tool for 
calculating amplitudes for such reactions and for obtaining 
predictions which can be compared in detail with data. 
\vfill
\end{abstract}
\vspace{5em}
\hrule width 5.cm
\vspace*{.5em}
{\small \noindent
$^1$ email: C.Ewerz@thphys.uni-heidelberg.de\\
$^2$ email: mmaniatis@ubiobio.cl\\
$^3$ email: O.Nachtmann@thphys.uni-heidelberg.de
}
\end{titlepage}

\tableofcontents

\section{Introduction}
\label{Introduction}

Today we have a well-established theory of strong interactions, 
quantum chromodynamics (QCD). 
Thus, in principle, all strong-interaction phenomena should be describable 
in terms of the fundamental QCD Lagrangian. In practice, this goal is in 
essence achieved for short distance phenomena due to asymptotic freedom 
\cite{Gross:1973id,Politzer:1973fx}. Pure long-distance phenomena can be 
treated by lattice methods introduced for QCD in \cite{Wilson:1974sk}. 
But soft high-energy scattering where the c.\,m.\ energy $\sqrt{s}$ of the 
collision becomes large but the momentum transfer $\sqrt{|t|}$ stays small 
is neither a pure short distance nor a pure long distance phenomenon. 
Thus, when an experimentalist asks a theorist to calculate the cross section 
for a given soft high-energy reaction the theorist will have to resort to 
models. And indeed, there is a time-honoured model, the Regge model, 
which describes many observed regularities of soft high-energy scattering; 
see for instance 
\cite{Regge:1959mz,Chew:1962eu,Gribov:2009zz,Collins:1977jy,Collins:1984tj,Donnachie:2002en,ref7,ref8}. 
Now we were indeed interested in calculating the amplitude 
for a specific process: photoproduction of a $\pi^+\pi^-$ pair on a proton
\be\label{1.1}
\gamma + p \longrightarrow \pi^+ + \pi^- + p \,.
\ee
For this calculation we wanted to include not only pomeron but also reggeon, 
photon, and odderon exchange. We were then confronted with a problem. 
Simple and handy rules for calculations of soft high-energy reactions were 
hard to find. We thought that such rules should be formulated in terms of 
effective propagators and vertices. For a given process these propagators 
and vertices should then be combinable according to the standard rules of 
quantum field theory (QFT). In the present paper we shall formulate such 
rules. We shall also present a formulation of the exchanges of the pomeron 
and the odderon as effective spin 2 and spin 1 exchanges, respectively. 
Let us recall that the pomeron has charge conjugation $C=+1$ and is 
ubiquitous in high-energy reactions. On the other hand, the odderon with 
$C=-1$, introduced theoretically in \cite{Lukaszuk:1973nt,Joynson:1975az}, 
is still elusive in experiments; see \cite{Ewerz:2003xi} for a review. 
As we shall see, considering the pomeron as an effective spin 2 exchange 
presents advantages compared to viewing it as the exchange of an effective 
$C=+1$ `photon'  
\cite{Landshoff:1971pw,Donnachie:1983hf,Donnachie:1985iz,Donnachie:1987gu}. 
With an effective spin 2 pomeron it is easy to respect all rules of QFT. 
Moreover, it turns out that in this framework applications of the vector 
meson-dominance (VMD) model to photon-induced reactions are 
straightforward and do not lead to gauge-invariance problems. 

Our paper is organised as follows. In section \ref{The reactions} we list the 
reactions which we want to consider and identify the exchanges and vertices 
needed for calculating the corresponding amplitudes. In section 
\ref{Propagators and vertices} the rules for these exchanges and vertices are 
given. In sections \ref{Vector mesons} to \ref{Pion-proton and rho-proton scattering} 
we explain how we arrive at these rules. Section \ref{Conclusions} contains 
our conclusions. In a companion paper \cite{searchfortheodderon} we will treat 
in detail reaction \eqref{1.1} which is suitable for an odderon search. 

\section{The Reactions}
\label{The reactions}

In this section we list the reactions to be studied in the framework of our 
model. These are high-energy reactions and decay processes. The former 
will be described in terms of Regge exchanges, and we use the conventional 
names for the Regge trajectories: ${\mathbbm P}$ (pomeron), 
${\mathbbm O}$ (odderon), $\rho_R$ ($\rho$ reggeon), 
$\omega_R$ ($\omega$ reggeon), $f_{2R}$ ($f_2$ reggeon), 
$a_{2R}$ ($a_2$ reggeon). The expressions which 
we use for the effective propagators and vertices of these objects are 
presented in section \ref{Propagators and vertices} below. In table \ref{table1} we list 
the quantum numbers $C$ of charge conjugation and $G$ of $G$ parity for 
those particles and exchange objects which are eigenstates of $C$ 
respectively $G$. We recall that the $G$ parity operator is defined as
\be\label{2.1}
G = \exp(i \pi I_2) \, C 
\ee
where $I_2$ is the second component of the isospin operator. $C$ invariance 
holds for strong and electromagnetic interactions, whereas $G$ invariance 
only holds for strong interactions. Here and in the following the particles 
$f_2$ and $a_2$ are understood as $f_2(1270)$ and the neutral $a_2(1320)$, 
respectively, in the PDG list \cite{Beringer:1900zz}. 
\begin{table}
\begin{center}
\begin{minipage}{.3\textwidth}
\renewcommand{\arraystretch}{1.1}
\begin{tabular}{c|r|r}
particles&$C$&$G$\\
\hline
$\pi^0$&$1$&$-1$\\
$\pi^\pm$&&$-1$\\
$\rho^0$&$-1$&$1$\\
$\rho^\pm$&&$1$\\
$\omega$&$-1$&$-1$\\
$f_2$&$1$&$1$\\
$a_2$&$1$&$-1$\\
\end{tabular}
\end{minipage}
\qquad\qquad
\begin{minipage}{.3\textwidth}
\renewcommand{\arraystretch}{1.1}
\begin{tabular}{c|r|r}
exchanges&$C$&$G$\\
\hline
${\mathbbm P}$&$1$&$1$\\
${\mathbbm O}$&$-1$&$-1$\\
$\rho_R$&$-1$&$1$\\
$\omega_R$&$-1$&$-1$\\
$f_{2R}$&$1$&$1$\\
$a_{2R}$&$1$&$-1$\\
\end{tabular}
\end{minipage}
\caption{\label{table1}The charge conjugation and 
$G$-parity quantum numbers of particles and exchange objects.}
\end{center}
\end{table}

For high energy reactions we use the Mandelstam variables, $s$ the c.\,m.\ 
energy squared, and $t$ the momentum transfer squared.

\begin{itemize}
\item Soft high energy reactions to be considered:
\begin{gather}\label{2.2}
p + p \longrightarrow p + p \,,
\\
\label{2.3}
p + \bar{p} \longrightarrow p + \bar{p} \,,
\\
\label{2.4}
p + n \longrightarrow p + n \,,
\\
\label{2.4a}
\bar{p} + n \longrightarrow \bar{p} + n \,,
\\
\label{2.5}
\pi + p \longrightarrow \pi + p \,,
\\
\label{2.6}
\rho + p \longrightarrow \rho + p \,.
\end{gather}
The diagrams for \eqref{2.2} are shown in figure \ref{model:fig1}. The diagrams 
for \eqref{2.3} to \eqref{2.4a} are analogous. The diagrams for \eqref{2.5} are 
shown in figure \ref{model:fig2}. Of course, single $\gamma$ exchange only 
contributes to $\pi^\pm p\to \pi^\pm p$ and is absent for 
$\pi^0 p\to \pi^0 p$. Note that $G$ parity invariance forbids 
the vertices $a_{2R}\pi\pi$, $\omega_R\pi\pi$ and ${\mathbbm O}\pi\pi$; 
see table \ref{table1}. Thus, the exchanges of $a_{2R},\omega_R$, and 
${\mathbbm O}$ cannot contribute to the reaction \eqref{2.5}. 
For $\rho p$ scattering \eqref{2.6} the diagrams are shown in figure 
\ref{model:fig3} where $\rho_R$ and $\gamma$ exchange only 
contribute to $\rho^\pm p\to \rho^\pm p$. This brings us to the decay 
reactions which we want to consider.
\begin{figure}[ht]
\begin{center}
\includegraphics[height=100pt]{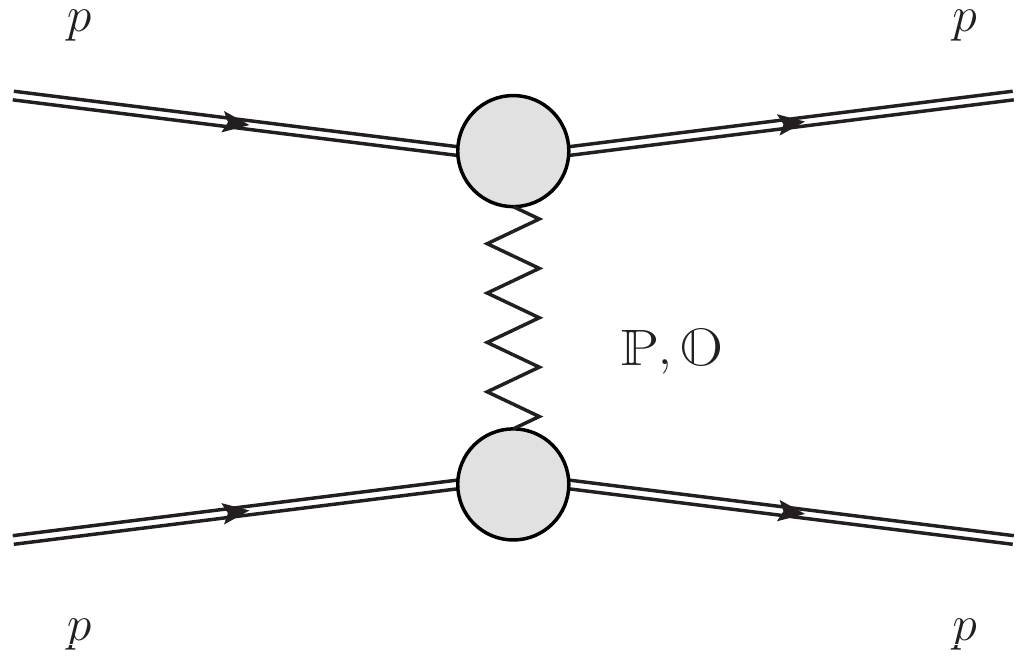}\qquad
\includegraphics[height=100pt]{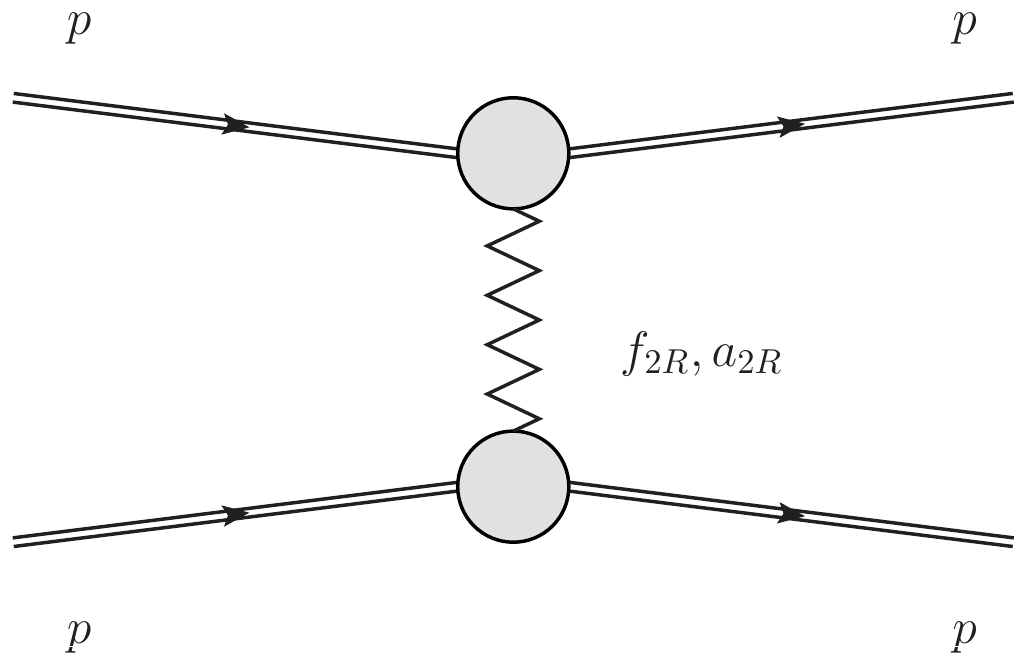}\\
\vspace{12pt}
\includegraphics[height=100pt]{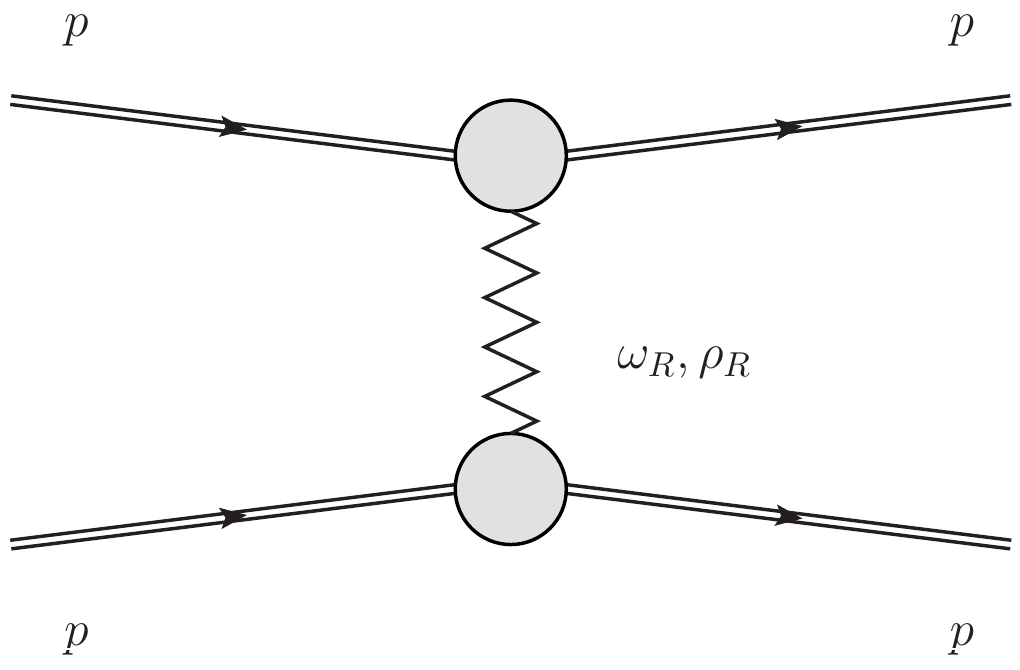}\qquad
\includegraphics[height=100pt]{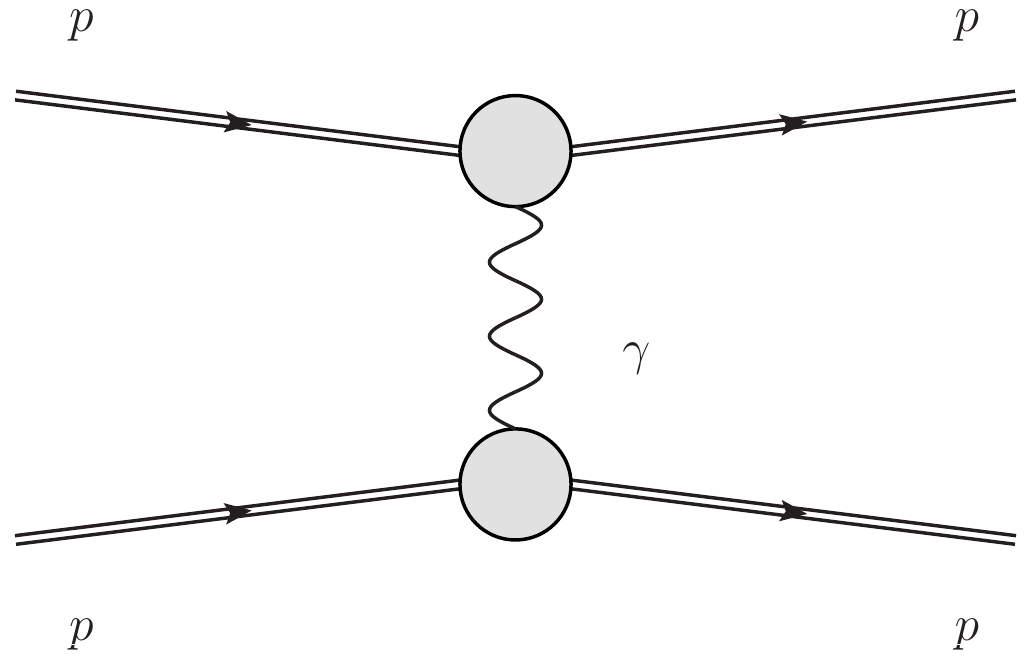}
\caption{Diagrams for $pp\to pp$ \eqref{2.2} at high energies.
\label{model:fig1}}
\end{center}
\end{figure}

\begin{figure}[ht]
\begin{center}
\includegraphics[height=100pt]{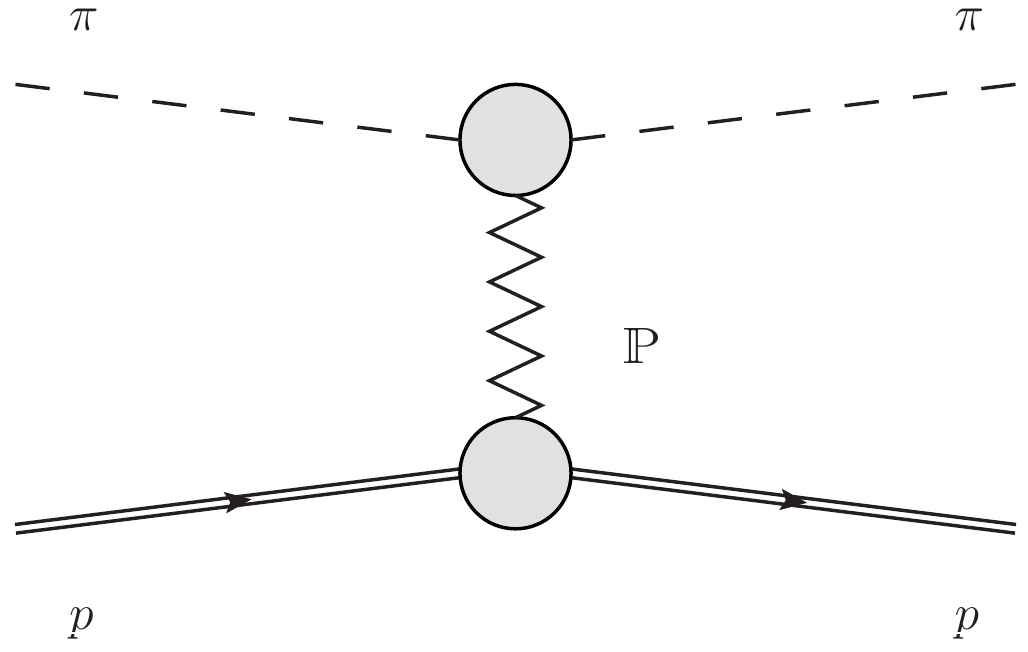}\qquad
\includegraphics[height=100pt]{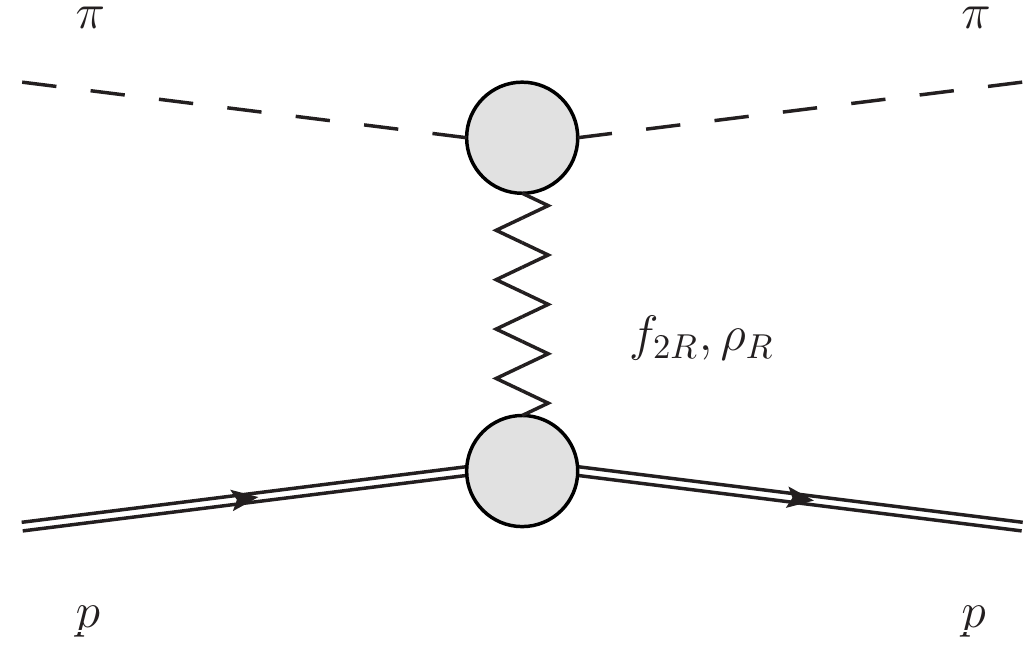}\\
\includegraphics[height=100pt]{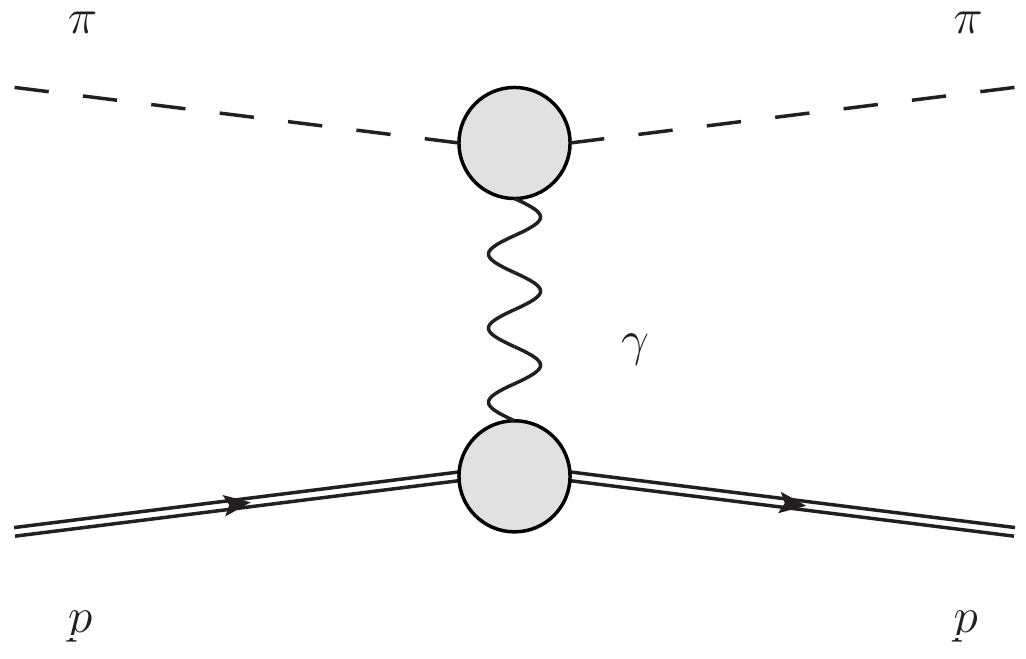}
\caption{Diagrams for $\pi p\to \pi p$ \eqref{2.5} at high energies. 
\label{model:fig2}}
\end{center}
\end{figure}

\begin{figure}[ht]
\begin{center}
\includegraphics[height=100pt]{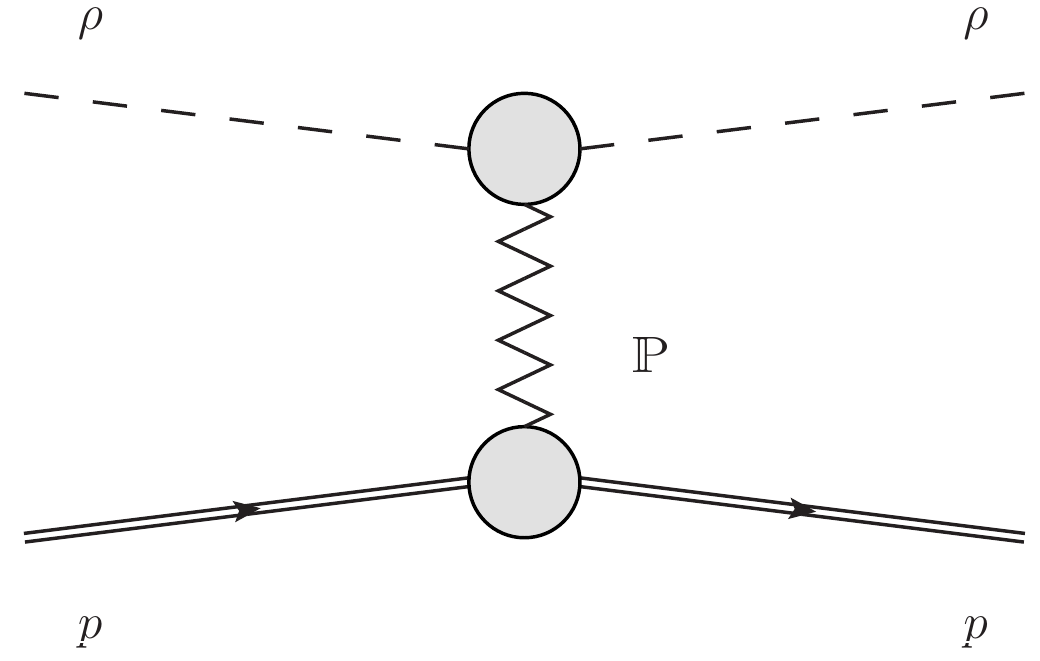}\qquad
\includegraphics[height=100pt]{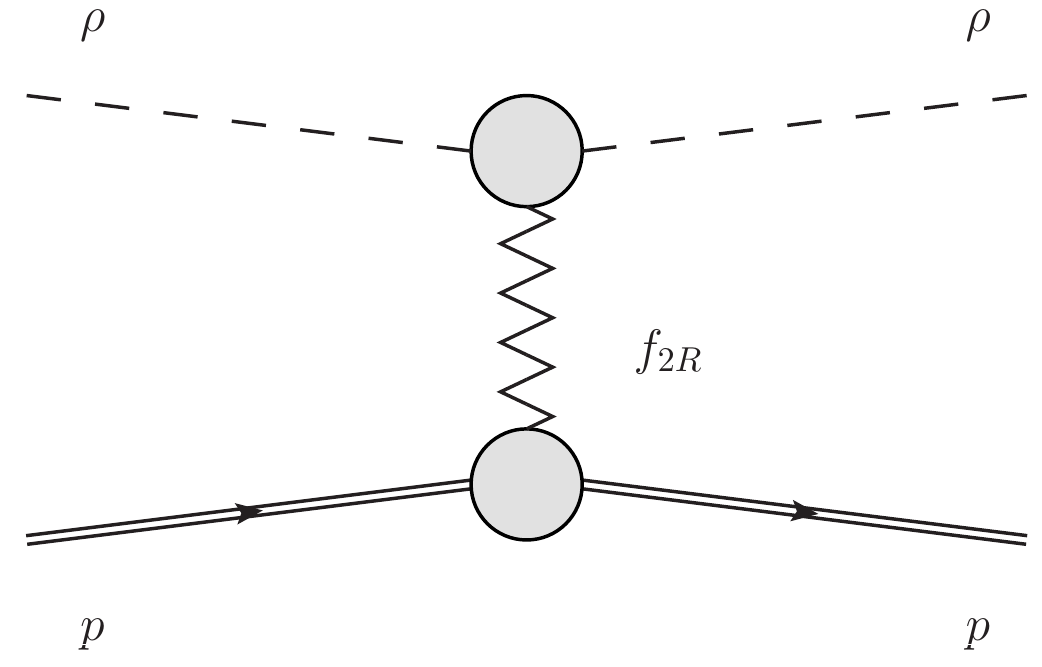}\\
\vspace{12pt}
\includegraphics[height=100pt]{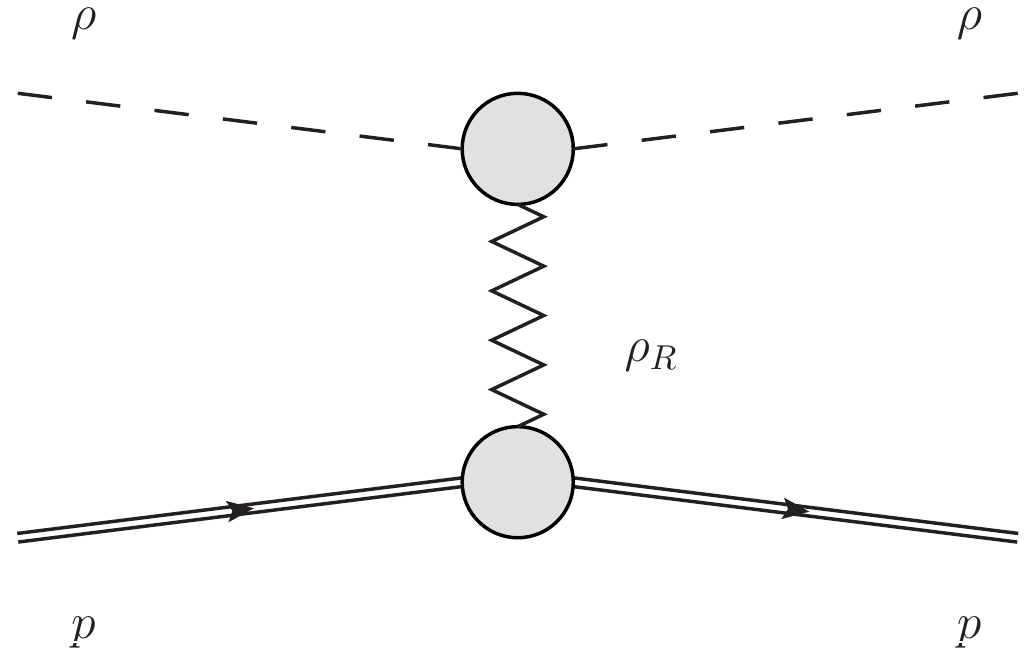}\qquad
\includegraphics[height=100pth]{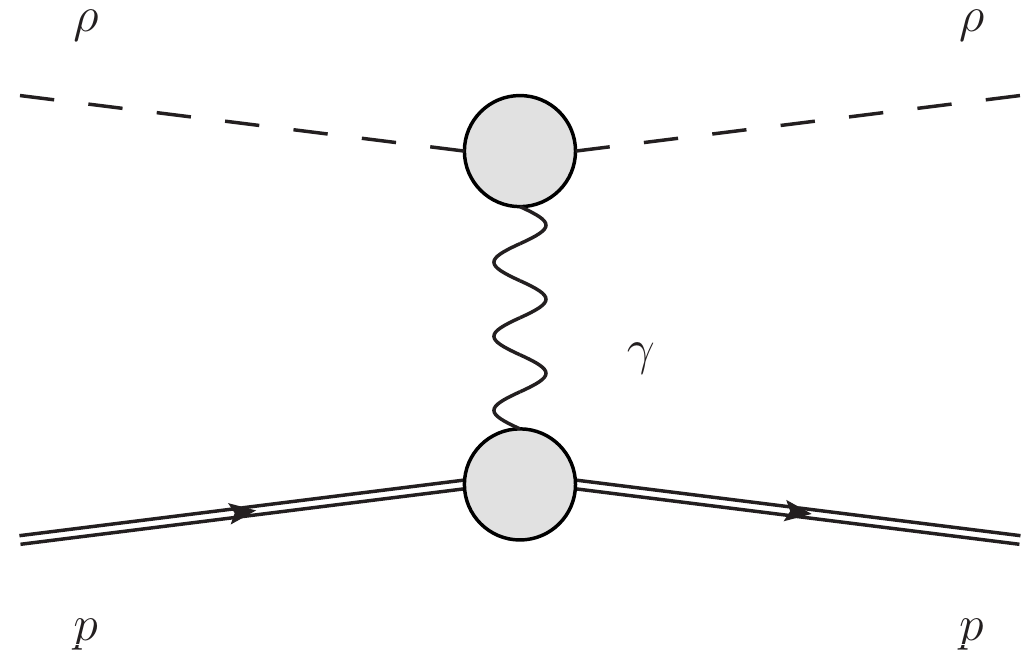}
\caption{Diagrams for $\rho p\to \rho p$ \eqref{2.6} at high energies.
\label{model:fig3}}
\end{center}
\end{figure}

\item Decay reactions to be considered:
\begin{align}\label{2.11}
\rho^0 &\longrightarrow \pi^+ + \pi^- \,,
\\
\label{2.12}
f_2 &\longrightarrow \pi^+ + \pi^-\,, \, \pi^0 + \pi^0 \,,
\\
\label{2.13}
f_2 &\longrightarrow \gamma + \gamma\,,
\\
\label{2.13a}
a_2 &\longrightarrow \gamma + \gamma\,.
\end{align}

\end{itemize}

In the following we will discuss the propagators and vertices needed 
for calculating the diagrams corresponding to the scattering reactions 
\eqref{2.2} to \eqref{2.6} and to the decay reactions \eqref{2.11} to 
\eqref{2.13a}. We shall encounter vertices with reggeons and particles, 
for instance the $f_{2R} \pi\pi$ and the $f_2 \pi\pi$ vertices. 
We formulate here the hypothesis that the coupling constants of 
corresponding reggeon-particle and particle-particle vertices are 
approximately equal. We will find support for this hypothesis in 
various cases from data. In other cases we shall use the hypothesis 
just to get estimates for coupling constants. 

The methods which we shall develop in the following for describing 
the reactions \eqref{2.2} to \eqref{2.13a} can easily be extended to 
the treatment of various other processes. For example, the inclusion 
of strange particles and corresponding trajectories would be 
straightforward. We leave this for future work. 

\section{Propagators and Vertices}
\label{Propagators and vertices}

In this section we give the analytic expressions for the propagators and 
vertices needed for calculating the amplitudes for the reactions listed in 
section \ref{The reactions}. We also give standard or default values 
for the parameters occurring. The justifications for these expressions 
and parameters will be given in later sections. The conventions for 
kinematics and Dirac $\gamma$ matrices follow \cite{Nachtmann:1990ta}. 

At this point it is appropriate to say something about the errors of the 
parameters listed in the following. Whenever errors are available, for 
instance for numbers taken from \cite{Beringer:1900zz}, these are quoted. 
Some numbers which we take from \cite{Donnachie:2002en} are given there 
without errors. We estimate these errors to be at the 5 to 10\% level in general. 
In some cases numbers which we derive have even larger errors, maybe up to 
30\%. Of course, we would like to know all parameters of our model as 
accurately as possible. The hope is that in the future, using our framework, 
one will be able to make global fits to data on soft reactions and determine 
the corresponding parameters together with realistic uncertainties. 

\subsection{Propagators and Effective Propagators}
\label{Propagators and effective propagators}

\noindent 
$\bullet$ photon $\gamma$
\newline
\vspace*{.2cm}
\hspace*{0.5cm}\includegraphics[width=100pt]{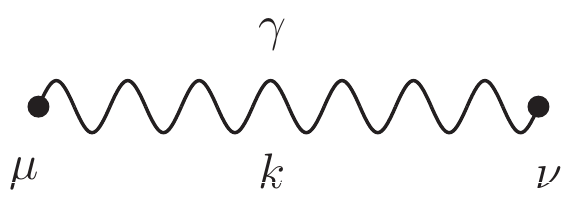} 
\begin{equation}
\label{3.1}
i \Delta_{\mu\nu}^{(\gamma)} (k) = \frac{- i g_{\mu \nu}}{k^2 + i \epsilon}\,.
\end{equation}
\newline

\noindent 
$\bullet$ vector mesons $V=\rho^0,\omega$ (see section \ref{Vector mesons})
\newline
\vspace*{.2cm}
\hspace*{0.5cm}\includegraphics[width=100pt]{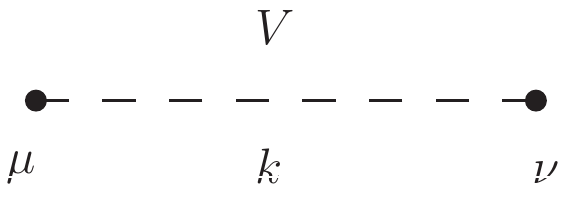} 
\begin{equation}
\label{3.2}
i \Delta_{\mu \nu}^{(V)}(k) = 
i \left( -g_{\mu \nu} + \frac{k_\mu k_\nu}{k^2 + i \epsilon}\right)
\Delta_T^{(V)} (k^2) 
- i \frac{k_\mu k_\nu}{k^2 + i \epsilon} \Delta_L^{(V)} (k^2) \,.
\end{equation}
Here $\Delta_T^{(V)} (k^2)$ and $\Delta_L^{(V)} (k^2)$ are the invariant 
functions in the transverse and longitudinal parts, respectively, of the 
propagators. 

For a careful discussion of the propagator matrix of the 
$\gamma$-$\rho$-$\omega$ system we refer to \cite{Melikhov:2003hs}. In Appendix B 
of \cite{Melikhov:2003hs} analytic formulae for the $\gamma$-$\rho$-$\omega$ 
propagator matrix are given. With the normalisation used there we get\footnote{In the 
symbols for vertices and propagators we do not always distinguish $\rho^0$ from 
$\rho^\pm$, the meaning should always be clear from the context.} 
\be\label{3.2a}
\begin{split}
\Delta_T^{(\rho)} (0) = -1/m_\rho^2 \,,
\\
\Delta_T^{(\omega)} (0) = -1/m_\omega^2 \,.
\end{split}
\ee
The longitudinal functions $\Delta_L^{(V)} (k^2)$ need 
not be specified since they will never enter our calculations. The 
transverse functions $\Delta_T^{(V)} (k^2)$ will be discussed in detail 
in section \ref{Vector mesons}. We quote here the values for the masses 
$m_V$ and the widths $\Gamma_V$ ($V=\rho^0,\omega$) from 
\cite{Beringer:1900zz}: 
\be
\begin{split}
\label{3.4a1}
m_{\rho^0} &= 775.49 \pm 0.34 \,\mbox{MeV} \,,
\:\:\:\:\:\:
\Gamma_{\rho^0} =146.2 \pm 0.7 \,\mbox{MeV} \,,
\\
m_\omega &= 782.65 \pm 0.12 \,\mbox{MeV} \,,
\:\:\:\:\:\:
\Gamma_\omega = 8.49 \pm 0.08 \,\mbox{MeV} \,.
\end{split}
\ee
We note that only for qualitative calculations not aiming at too high accuracy 
and neglecting $\rho$-$\omega$ interference one may use, for $k^2$ away 
from zero, the simple Breit-Wigner expressions for $\Delta^{(V)}_{T}(k^2)$, 
\be\label{3.4}
\Delta_T^{(V)} (k^2) = \frac{1}{k^2-m_V^2+ i m_V \Gamma_V}\,.
\ee
\newline

\noindent 
$\bullet$ tensor meson $f_2\equiv f_2 (1270)$ (see section \ref{Tensor mesons}) 
\newline
\vspace*{.2cm}
\hspace*{0.5cm}\includegraphics[width=100pt]{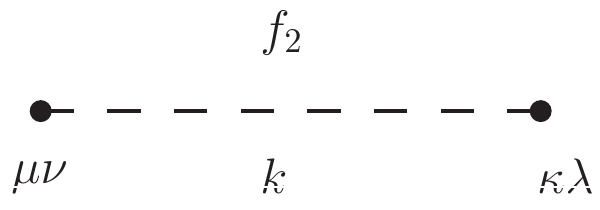}
\be
\label{3.5}
\begin{split}
i \Delta^{(f_2)}_{\mu\nu,\kappa\lambda} (k) = \, 
i \,\bigg\{ & \frac{1}{2} \left( - g_{\mu\kappa} + \frac{k_\mu k_\kappa}{k^2 + i \epsilon} \right) 
\left( - g_{\nu\lambda} + \frac{k_\nu k_\lambda}{k^2 + i \epsilon} \right)
\\
& + \frac{1}{2} \left( - g_{\mu\lambda} + \frac{k_\mu k_\lambda}{k^2 + i \epsilon} \right) 
\left( - g_{\nu\kappa} + \frac{k_\nu k_\kappa}{k^2 + i \epsilon} \right) 
\\
& - \frac{1}{3} \left( - g_{\mu\nu} + \frac{k_\mu k_\nu}{k^2 + i \epsilon} \right) 
\left( - g_{\kappa\lambda} + \frac{k_\kappa k_\lambda}{k^2 + i \epsilon} \right) 
\bigg\} \,\Delta^{(2)}(k^2)\,.
\end{split}
\ee
Here we only give the spin 2 part of the corresponding tensor-field propagator. 
The invariant function $\Delta^{(2)}(k^2)$ is the one where the $f_2$ meson appears; 
see \eqref{5.107} in section \ref{Tensor mesons}. We have the following relations: 
\be\label{3.5a}
\Delta^{(f_2)}_{\mu \nu, \kappa \lambda} (k) = 
\Delta^{(f_2)}_{\nu \mu, \kappa \lambda} (k) = 
\Delta^{(f_2)}_{\mu \nu, \lambda \kappa} (k) = 
\Delta^{(f_2)}_{\kappa \lambda,\mu \nu} (k) \,,
\ee
\be\label{3.5b}
\begin{split}
g^{\mu\nu} \Delta^{(f_2)}_{\mu \nu, \kappa \lambda} (k) &= 0 \,,
\\
g^{\kappa\lambda} \Delta^{(f_2)}_{\mu \nu, \kappa \lambda} (k) &= 0 \,.
\end{split}
\ee
From \cite{Beringer:1900zz} we have for the mass and width of the $f_2$ meson 
\begin{equation}
\label{3.6}
m_{f_2} = 1275.1 \pm 1.2 \,\mbox{MeV} \,, 
\:\:\:\:\:\:
\Gamma_{f_2} =185.1\,{}^{+2.9}_{-2.4}\,\mbox{MeV} \,.
\end{equation}
\newline

\noindent 
$\bullet$ pomeron ${\mathbbm P}$ (see section \ref{Pomeron exchange})
\newline
\vspace*{.2cm}
\hspace*{0.5cm}\includegraphics[width=125pt]{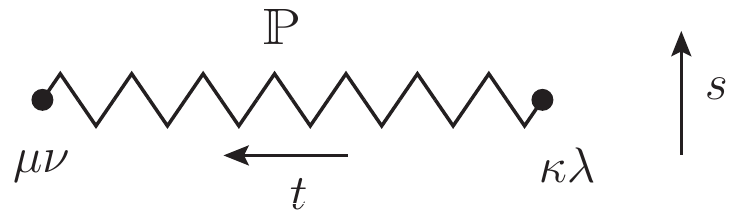} 
\begin{equation}
\label{3.7}
i\Delta^{(\mathbbm{P})}_{\mu\nu,\kappa\lambda} (s,t) 
= \frac{1}{4s} \left(g_{\mu\kappa} g_{\nu\lambda} + g_{\mu\lambda} g_{\nu\kappa} 
- \frac{1}{2} g_{\mu\nu} g_{\kappa\lambda} \right) 
\, (-i s \alpha'_\mathbbm{P})^{\alpha_\mathbbm{P}(t)-1} \,,
\end{equation}
\be\label{3.8}
\begin{split}
\alpha_\mathbbm{P}(t) &= 1 + \epsilon_\mathbbm{P} + \alpha'_\mathbbm{P} t \,,
\\
\epsilon_\mathbbm{P} &= 0.0808 \,,
\\
\alpha'_\mathbbm{P} &= 0.25 \,\mbox{GeV}^{-2} \,.
\end{split}
\ee
\newline

\noindent 
$\bullet$ reggeons ${\mathbbm R}_+=f_{2R},a_{2R}$ 
(see section \ref{Reggeon exchanges and total cross sections})
\newline
\vspace*{.2cm}
\hspace*{0.5cm}\includegraphics[width=125pt]{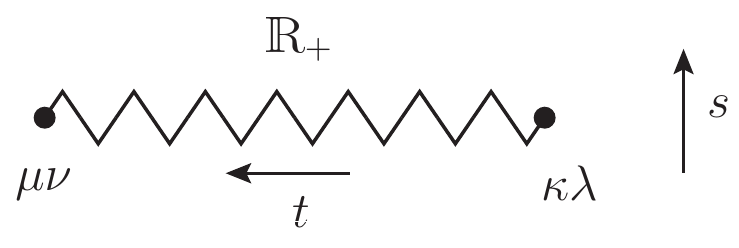} 
\begin{equation}\label{3.9}
i\Delta^{(\mathbbm{R}_+)}_{\mu\nu,\kappa\lambda} (s,t) 
= \frac{1}{4s} \left(g_{\mu\kappa} g_{\nu\lambda} + g_{\mu\lambda} g_{\nu\kappa} 
- \frac{1}{2} g_{\mu\nu} g_{\kappa\lambda} \right) 
(-i s \alpha'_{\mathbbm{R}_+})^{\alpha_{\mathbbm{R}_+}(t)-1} \,,
\end{equation}
\be\label{3.10}
\begin{split}
\alpha_{\mathbbm{R}_+}(t) &= \alpha_{\mathbbm{R}_+} (0)+ \alpha'_{\mathbbm{R}_+} t \,,
\\
\alpha_{\mathbbm{R}_+} (0) &= 0.5475 \,,
\\
\alpha'_{\mathbbm{R}_+} &= 0.9 \,\mbox{GeV}^{-2} \,.
\end{split}
\ee
\newline

\noindent 
$\bullet$ reggeons ${\mathbbm R}_-=\omega_R,\rho_R$ 
(see section \ref{Reggeon exchanges and total cross sections})
\newline
\vspace*{.2cm}
\hspace*{0.5cm}\includegraphics[width=125pt]{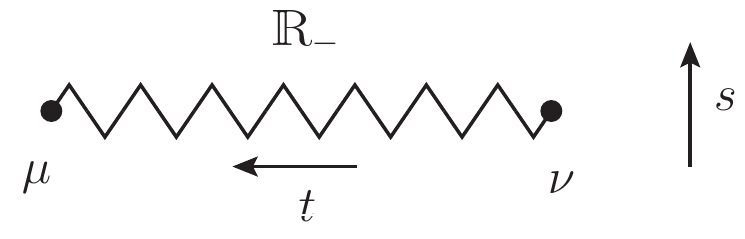} 
\begin{equation}\label{3.11}
i \Delta_{\mu\nu}^{(\mathbbm{R}_-)} (s,t) = i g_{\mu\nu} \,\frac{1}{M_-^2} \,
(-i s \alpha'_{\mathbbm{R}_-})^{\alpha_{\mathbbm{R}_-} (t)-1} \,,
\end{equation}
\be\label{3.12}
\begin{split}
\alpha_{\mathbbm{R}_-}(t) &= \alpha_{\mathbbm{R}_-} (0)+ \alpha'_{\mathbbm{R}_-} t \,,
\\
\alpha_{\mathbbm{R}_-} (0) &= 0.5475 \,,
\\
\alpha'_{\mathbbm{R}_-} &= 0.9 \,\mbox{GeV}^{-2} \,,
\\
M_- &= 1.41 \,\mbox{GeV} \,.
\end{split}
\ee
The numbers for the parameters in \eqref{3.8}, \eqref{3.10} and \eqref{3.12} 
are taken from \cite{Donnachie:2002en}, except for $M_-$ which is 
discussed in section \ref{Reggeon exchanges and total cross sections}. 
\newline

\noindent 
$\bullet$ odderon ${\mathbbm O}$ (see section \ref{Odderon exchange})
\newline
\vspace*{.2cm}
\hspace*{0.5cm}\includegraphics[width=125pt]{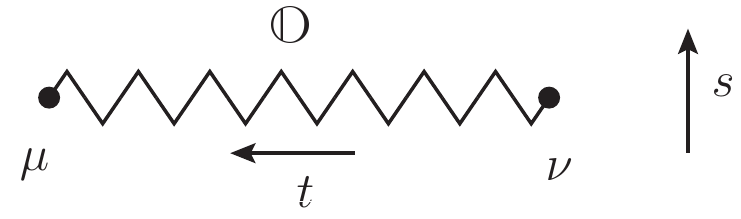} 
\begin{equation}\label{3.13}
i \Delta_{\mu\nu}^{(\mathbbm{O})} (s,t) = -i g_{\mu\nu} \,\frac{\eta_\mathbbm{O}}{M_0^2} \,
(-i s \alpha'_{\mathbbm{O}})^{\alpha_{\mathbbm{O}} (t)-1} \,,
\end{equation}
\be\label{3.14}
\begin{split}
\alpha_\mathbbm{O}(t) &= 1 + \epsilon_\mathbbm{O} + \alpha'_\mathbbm{O} t \,,
\\
M_0 &= 1\,\mbox{GeV} \,,
\\
\eta_\mathbbm{O} &= \pm 1 \,,
\\
\alpha'_\mathbbm{O} &= 0.25 \,\mbox{GeV}^{-2} \,.
\end{split}
\ee
The odderon parameters $\epsilon_\mathbbm{O}$, $\alpha'_\mathbbm{O}$, 
and $\eta_\mathbbm{O}$ are at present not known experimentally. 
In \eqref{3.13} we have put $M_0^{-2}$ for dimensional reasons. We note 
that another overall scale factor in the propagator instead of $M_0^{-2}$ 
can always be traded against scale factors in the odderon vertices; see 
section \ref{Odderon exchange}. We have, furthermore, assumed 
$\alpha'_\mathbbm{O} = \alpha'_\mathbbm{P}$ in \eqref{3.14} for 
lack of other information. 

\subsection{Vertices and Effective Vertices}
\label{Vertices and effective vertices}

In this section we list the vertices and effective vertices which we need 
for the discussion of the reactions of section \ref{The reactions}. 
For writing down these vertices we shall frequently use two rank-four 
tensor functions defined as follows
\begin{align}\label{3.15}
\Gamma_{\mu\nu\kappa\lambda}^{(0)} (k_1,k_2) =\,& 
[(k_1\cdot k_2) g_{\mu\nu} - k_{2\mu} k_{1\nu}] 
\left[k_{1\kappa}k_{2\lambda} + k_{2\kappa}k_{1\lambda} - 
\frac{1}{2} (k_1 \cdot k_2) g_{\kappa\lambda}\right] \,,
\\
\label{3.16}
\Gamma_{\mu\nu\kappa\lambda}^{(2)} (k_1,k_2) = \,
& (k_1\cdot k_2) (g_{\mu\kappa} g_{\nu\lambda} + g_{\mu\lambda} g_{\nu\kappa} )
+ g_{\mu\nu} (k_{1\kappa} k_{2\lambda} + k_{2\kappa} k_{1\lambda} ) 
\nn \\
& - k_{1\nu} k_{2 \lambda} g_{\mu\kappa} - k_{1\nu} k_{2 \kappa} g_{\mu\lambda} 
- k_{2\mu} k_{1 \lambda} g_{\nu\kappa} - k_{2\mu} k_{1 \kappa} g_{\nu\lambda} 
\\
& - [(k_1 \cdot k_2) g_{\mu\nu} - k_{2\mu} k_{1\nu} ] \,g_{\kappa\lambda} \,.
\nn
\end{align}
We have for $i=0,2$
\be\label{3.17}
\Gamma_{\mu\nu\kappa\lambda}^{(i)} (k_1,k_2) 
= \Gamma_{\mu\nu\lambda\kappa}^{(i)} (k_1,k_2) 
= \Gamma_{\nu\mu\kappa\lambda}^{(i)} (k_2,k_1) 
= \Gamma_{\mu\nu\kappa\lambda}^{(i)} (-k_1,-k_2) \,,
\ee
\be\label{3.18}
\begin{split}
& k_1^\mu \Gamma_{\mu\nu\kappa\lambda}^{(i)} (k_1,k_2) =0 \,,
\\
& k_2^\nu \Gamma_{\mu\nu\kappa\lambda}^{(i)} (k_1,k_2) =0 \,,
\end{split}
\ee
\be\label{3.19}
\Gamma_{\mu\nu\kappa\lambda}^{(i)} (k_1,k_2)\, g^{\kappa\lambda} =0 \,. 
\ee
Now here is our list of vertices. 
\newline

\noindent 
$\bullet$ $\gamma V$, where $V=\rho,\omega,\phi$ 
(see eq.\ (5.3) of \cite{Donnachie:2002en} and section \ref{Vector mesons}) 
\newline
\vspace*{.2cm}
\hspace*{0.5cm}\includegraphics[width=100pt]{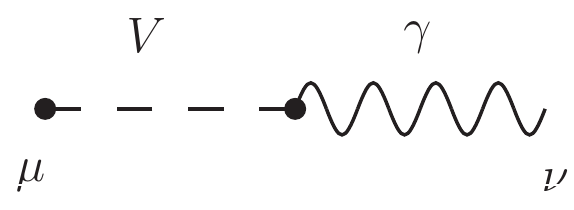} 
\begin{equation}\label{3.20}
 -i e \,\frac{m_V^2}{\gamma_V} \,g_{\mu\nu} \,,
\end{equation}
\be\label{3.20new}
e>0 \,,\quad \gamma_\rho >0\,,\quad \gamma_\omega>0 \,,\quad \gamma_\phi <0 \,,
\ee
\be\label{3.21}
\frac{4 \pi}{\gamma_\rho^2} = 0.496 \pm 0.023 \,,
\quad
\frac{4 \pi}{\gamma_\omega^2} = 0.042 \pm 0.0015 \,,
\quad
\frac{4 \pi}{\gamma_\phi^2} = 0.0716 \pm 0.0017 \,.
\ee
\newline

\noindent 
$\bullet$ $\gamma p p$ 
\newline
\vspace*{.2cm}
\hspace*{0.5cm}\includegraphics[height=85pt]{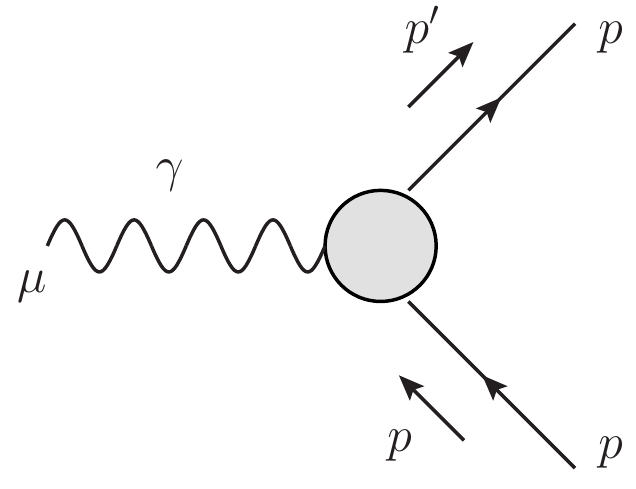} 
\begin{equation}\label{3.22} 
i \Gamma_\mu^{(\gamma p p) } (p',p) = 
-ie \left[ \gamma_\mu F_1(t) + \frac{i}{2 m_p}\, \sigma_{\mu\nu} (p'-p)^\nu F_2(t)\right] \,,
\end{equation}
\be\label{3.22A}
e>0 \,, \qquad
t=(p'-p)^2 \,,
\ee
\be\label{defsigmunu}
\sigma_{\mu\nu} = 
\frac{i}{2} \left(\gamma_\mu \gamma_\nu - \gamma_\nu \gamma_\mu\right)\,,
\ee
\begin{align}\label{3.23}
F_1(t) &= \left( 1 - \frac{t}{4m_p^2} \frac{\mu_p}{\mu_N}\right) 
\left( 1 - \frac{t}{4 m_p^2} \right)^{-1} G_D(t) \,,
\\
\label{3.24}
F_2(t) &= \left(\frac{\mu_p}{\mu_N} - 1 \right) \left( 1 - \frac{t}{4 m_p^2}\right)^{-1} G_D(t) \,,
\end{align}
\be\label{3.25}
\mu_N = \frac{e}{2 m_p} \,,
\qquad
\frac{\mu_p}{\mu_N} = 2.7928 \,,
\ee
\be\label{3.26}
G_D(t) = \left(1 - \frac{t}{m_D^2} \right)^{-2} \,,
\qquad
m_D^2 = 0.71 \,\mbox{GeV}^2 \,.
\ee
The $\gamma pp$ vertex is, of course, well known and we use here 
standard parametrisations for $F_{1,2}(t)$; see for instance chapter 2 
in \cite{Close:2007zzd}. $F_1$ and $F_2$ are the Dirac and Pauli 
form factors of the proton, respectively, and $G_D$ is the so-called 
dipole form factor. \newline

\noindent 
$\bullet$ $\gamma n n$ 
\newline
\vspace*{.2cm}
\hspace*{0.5cm}\includegraphics[height=85pt]{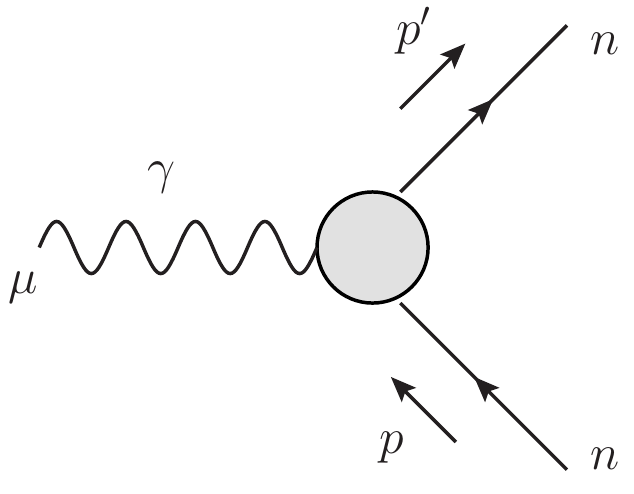} 
\hspace{0.1\linewidth}
\begin{equation}\label{3.22neutron} 
i \Gamma_\mu^{(\gamma n n) } (p',p) = 
-ie \left[ \gamma_\mu F^n_1(t) + \frac{i}{2 m_n} \sigma_{\mu\nu} (p'-p)^\nu F^n_2(t)\right] \,.
\end{equation}
The neutron form factors $F^n_{1,2}(t)$ are less well known than those of the 
proton. We have $F^n_1(0)=0$ as the neutron is electrically neutral. 
In the reactions that we want to consider the $\gamma n n$ vertex is not 
relevant and we will not discuss it further here. 
\newline

In the hadronic vertices listed in the following we have to take into 
account form factors since the hadrons are extended objects. For simplicity 
we use for the pomeron-nucleon coupling the electromagnetic Dirac form 
factor $F_1(t)$ of the proton \eqref{3.23}, as suggested in 
\cite{Donnachie:1983hf}; see also chapter 3.2 of 
\cite{Donnachie:2002en}. The same ansatz is made for the reggeon-nucleon 
and the odderon-nucleon couplings. In the corresponding couplings 
of pomeron, odderon and reggeons to mesons we always use the pion 
electromagnetic form factor in a simple parametrisation 
\be\label{3.29}
F_M(t) = F_\pi(t) = \frac{m_0^2}{m_0^2 -t} \,,
\ee
with $m_0^2=0.50$ GeV$^2$; see eq.\ (3.22) in chapter 3.2 of 
\cite{Donnachie:2002en} where it is also discussed why these 
assumptions cannot be true in general but should be a reasonable 
approximation for $0 \le -t\le 1$ GeV$^2$. Note that we do not use 
for $F_\pi(t)$ in \eqref{3.29} the canonical vector-meson-dominance 
(VMD) form which would 
correspond to replacing $m^2_0$ by $m^2_\rho \approx 0.60$ GeV$^2$. 
Indeed, for $t<0$ the measured pion form factor squared is {\em smaller} 
than given by the canonical VMD form (see for instance figure 4
of \cite{Melikhov:2003hs}) and this is taken into account by having 
$m^2_0<m^2_\rho$ in \eqref{3.29}. 
\newline

\noindent 
$\bullet$ $\rho^0\pi^+\pi^-$ 
(see section \ref{Vector mesons}, eq.\ \eqref{4.13b})
\newline
\vspace*{.2cm}
\hspace*{0.5cm}\includegraphics[height=85pt]{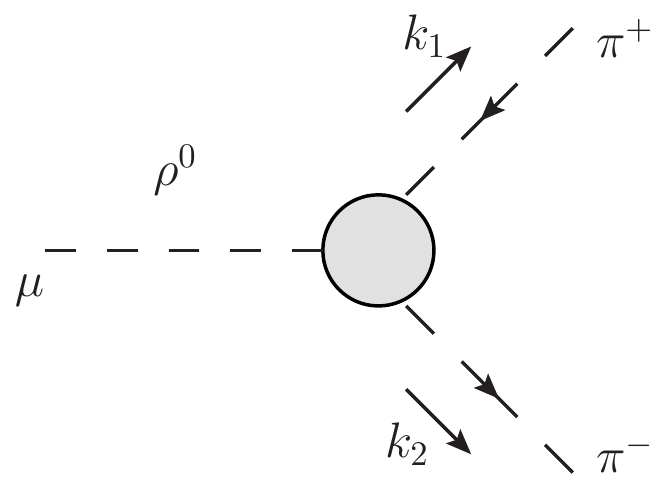} 
\begin{equation}\label{3.26a}
i \Gamma_\mu^{(\rho \pi\pi)} (k_1,k_2) = 
- \frac{1}{2} i g_{\rho \pi\pi} (k_1-k_2)_\mu \,,
\end{equation}
\be\label{3.27}
g_{\rho \pi\pi} = 11.51 \pm 0.07 \,.
\ee
\newline

\noindent 
$\bullet$ $f_2\pi\pi$ (see section \ref{f2topipi})
\newline
\vspace*{.2cm}
\hspace*{0.5cm}\includegraphics[height=85pt]{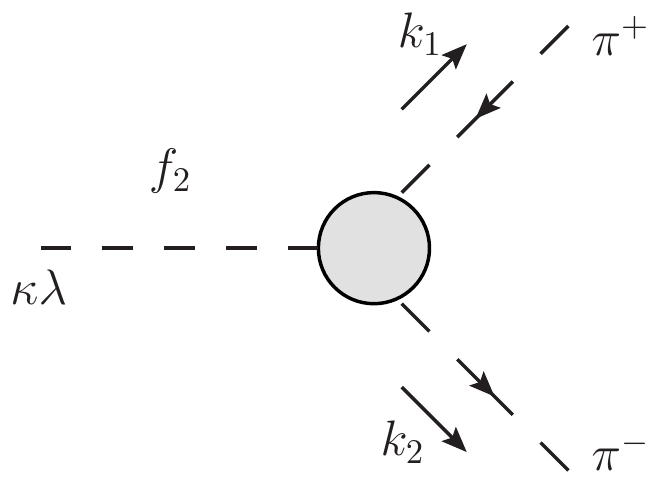} 
\hspace*{0.5cm}\includegraphics[height=85pt]{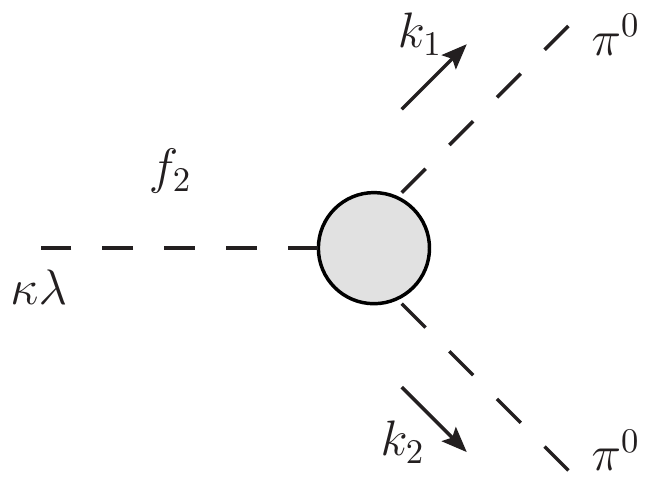} 
\newline
In the limit of strict isospin invariance both diagrams are described by 
the same expression 
\be\label{3.28}
i \Gamma_{\kappa \lambda}^{(f_2\pi\pi)} (k_1,k_2) 
= -i\, \frac{g_{f_2\pi\pi}}{2 M_0} 
\left[ (k_1-k_2)_\kappa (k_1-k_2)_\lambda - \frac{1}{4} g_{\kappa\lambda} (k_1-k_2)^2 \right] 
F^{(f_2\pi\pi)} (k^2) \,,
\ee
with $k=k_1+k_2$ and 
\be\label{3.28param}
g_{f_2\pi\pi} = 9.26 \pm 0.15 \,,\qquad
M_0=1\,\mbox{GeV} \,.
\ee
Here $F^{(f_2\pi\pi)}(k^2)$ is a form factor normalised to 
$F^{(f_2\pi\pi)}(m_{f_2}^2)=1$; see section \ref{f2topipi}. 

For production processes of $f_2$ decaying then to $\pi^+ \pi^-$, for example, 
we are also interested in the $\pi\pi$ invariant mass distribution. Thus, the 
couplings like $f_2 \pi\pi$ are not only needed for the "on-shell" $f_2$ 
but also away from the mass shell, for $k^2 \neq m_{f_2}^2$. Realistically, 
as done in \eqref{3.28}, 
we have to introduce form factors to describe the $k^2$ dependence of 
these couplings. These form factors will be normalised to $1$ at the 
on-shell point. 
\newline 

\noindent 
$\bullet$ $f_2\gamma\gamma$ (see sections \ref{f2togammagamma} and 
\ref{rho-proton scattering})
\newline
\vspace*{.2cm}
\hspace*{0.5cm}
\includegraphics[height=85pt]{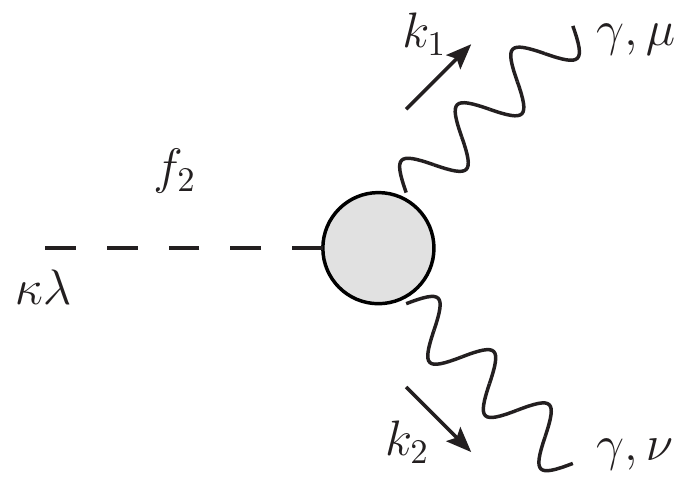} 
\begin{equation}\label{3.29new}
\begin{split}
i \Gamma_{\mu\nu\kappa\lambda}^{(f_2 \gamma \gamma)} (k_1,k_2) 
=\,& i F_M(k_1^2) F_M(k_2^2) F^{(f_2 \gamma \gamma)} (k^2) 
\\
& \times\left[
2 a_{f_2 \gamma\gamma} \Gamma_{\mu\nu\kappa\lambda}^{(0)}(k_1,k_2) 
- b_{f_2 \gamma\gamma} \Gamma_{\mu\nu\kappa\lambda}^{(2)}(k_1,k_2) 
\right] \,,
\end{split}
\end{equation}
with $k=k_1+k_2$ and 
\begin{equation}\label{3.29a}
a_{f_2 \gamma\gamma} = \frac{e^2}{4 \pi} \, 1.45 \,\mbox{GeV}^{-3} \,,
\qquad
b_{f_2 \gamma\gamma} = \frac{e^2}{4 \pi} \, 2.49 \,\mbox{GeV}^{-1} \,.
\end{equation}
Here $F^{(f_2 \gamma \gamma)} (k^2)$ is a form factor normalised to 
$F^{(f_2 \gamma \gamma)} (m_{f_2}^2)=1$. 
\newline \pagebreak

\noindent 
$\bullet$ $a_2\gamma\gamma$ (see sections \ref{a2togammagamma} and 
\ref{rho-proton scattering})
\newline
\vspace*{.2cm}
\hspace*{0.5cm}\includegraphics[height=85pt]{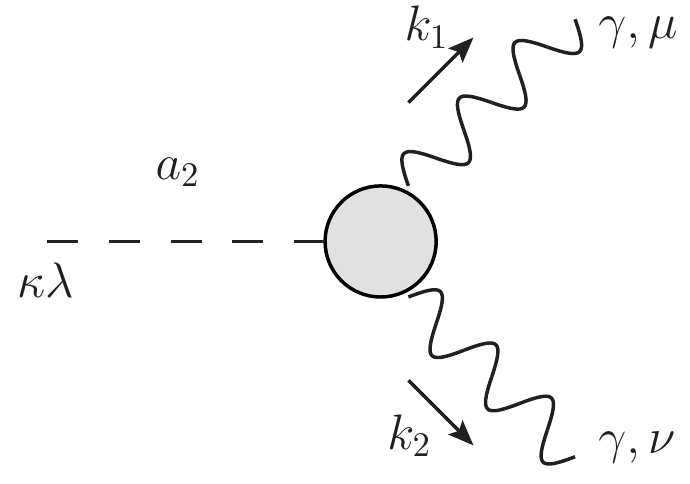} 
\begin{equation}\label{3.29b}
\begin{split}
i \Gamma_{\mu\nu\kappa\lambda}^{(a_2 \gamma \gamma)} (k_1,k_2) 
= \, & i F_M(k_1^2) F_M(k_2^2) F^{(a_2 \gamma \gamma)} (k^2)
\\
& \times \big[2 a_{a_2 \gamma\gamma} \Gamma_{\mu\nu\kappa\lambda}^{(0)}(k_1,k_2) 
- b_{a_2 \gamma\gamma} \Gamma_{\mu\nu\kappa\lambda}^{(2)}(k_1,k_2) \big]\,,
\end{split}
\end{equation}
with $k=k_1+k_2$ and 
\begin{equation}
\label{3.29bA}
\left| a_{a_2 \gamma\gamma}\right| = \frac{e^2}{4 \pi} \, 0.74 \,\mbox{GeV}^{-3} \,,
\qquad
\left| b_{a_2 \gamma\gamma} \right|= \frac{e^2}{4 \pi} \, 1.35 \,\mbox{GeV}^{-1} \,.
\end{equation}
The form factor $F^{(a_2 \gamma \gamma)} (k^2)$ is normalised to 
$F^{(a_2 \gamma \gamma)} (m_{a_2}^2)=1$. 
\newline

In the following we group diagrams together if the corresponding 
analytic expressions for the vertex functions are related by $C$ and/or 
isospin invariance. The vertex functions listed refer to each individual 
diagram of the corresponding group. Note that all $C$-invariance 
relations can be used in the correct and absolutely standard QFT 
way for our tensor pomeron and 
tensor reggeons $f_{2R}$, $a_{2R}$, as well as for the vector odderon 
and vector reggeons $\omega_R$, $\rho_R$. 
\newline

\noindent 
$\bullet$ ${\mathbbm P}NN$ (see section \ref{Pomeron exchange})
\newline
\vspace*{.2cm}
\hspace*{0.5cm}\includegraphics[height=85pt]{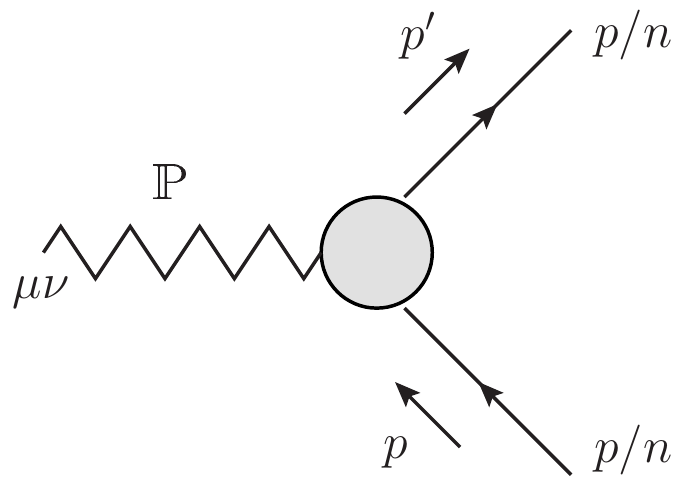} 
\hspace*{0.5cm}\includegraphics[height=85pt]{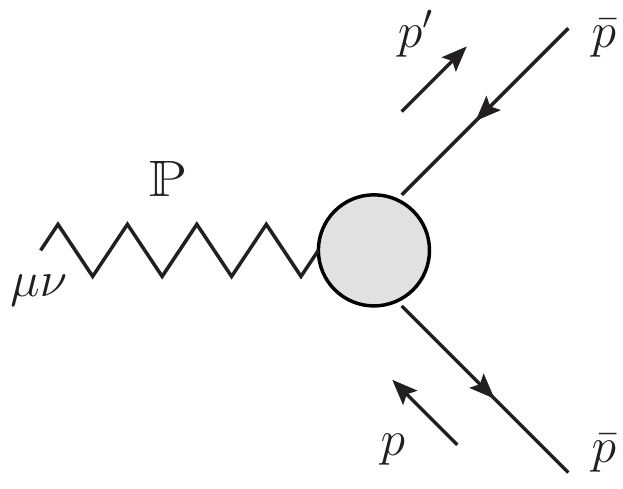} 
\newline
\be\label{3.30}
\begin{split}
i \Gamma_{\mu\nu}^{(\mathbbm{P}pp)}  (p',p) =& \,
i \Gamma_{\mu\nu}^{(\mathbbm{P}nn)} (p',p) = 
i \Gamma_{\mu\nu}^{(\mathbbm{P}\bar{p}\bar{p})} (p',p) 
\\
= &
-i \,3 \beta_{\mathbbm{P}NN} F_1[(p'-p)^2]
\\ & \times 
\left\{ \frac{1}{2} \left[ \gamma_\mu (p'+p)_\nu + \gamma_\nu (p'+p)_\mu \right] 
-\frac{1}{4} \, g_{\mu\nu} (\slash{p}' + \slash{p}) \right\} \,,
\end{split}
\ee
\be \label{3.30A}
\beta_{\mathbbm{P}NN} = 1.87 \,\mbox{GeV}^{-1} \,.
\ee
\newline \pagebreak

\noindent 
$\bullet$ ${\mathbbm P}\pi\pi$ (see section \ref{Pion-proton scattering})
\newline
\vspace*{.2cm}
\hspace*{0.5cm}\includegraphics[height=85pt]{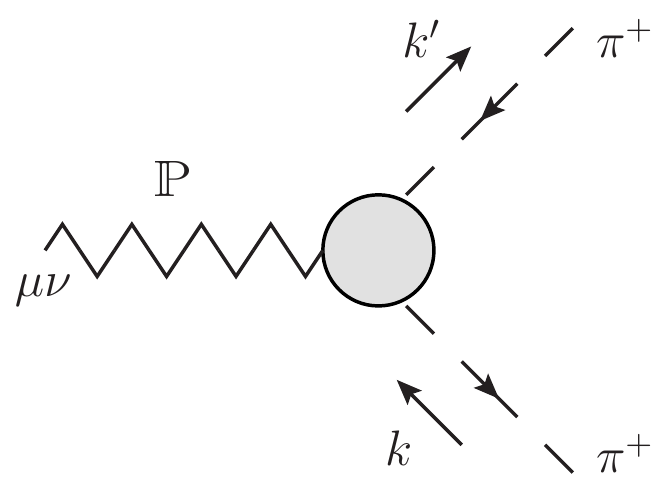} 
\hspace*{0.5cm}\includegraphics[height=85pt]{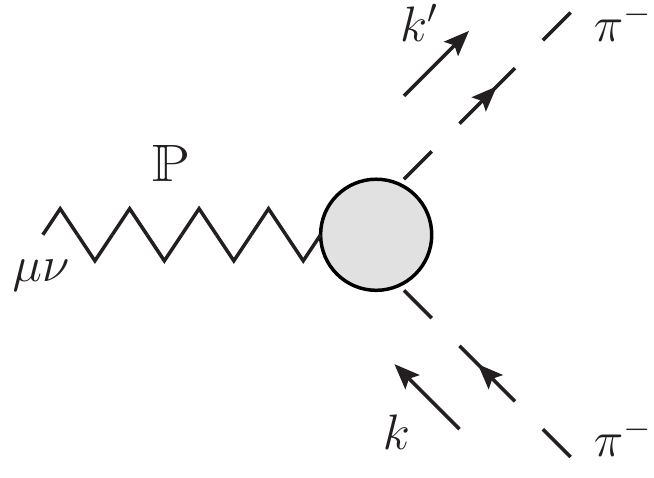} 
\hspace*{0.5cm}\includegraphics[height=85pt]{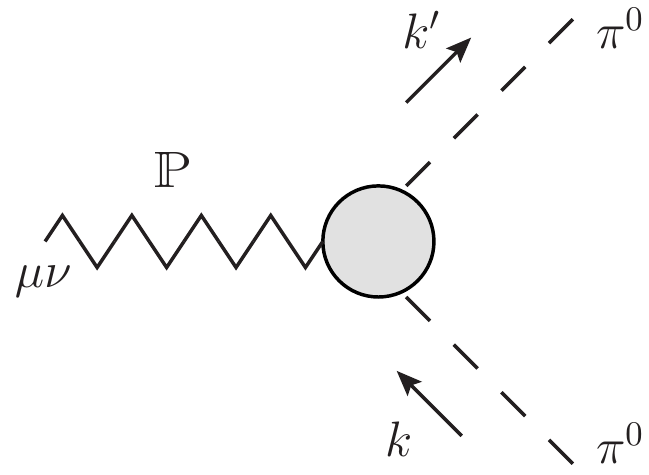} 
\newline
\be\label{3.31}
i \Gamma_{\mu\nu}^{(\mathbbm{P}\pi\pi)} (k',k) 
= -i \, 2 \beta_{\mathbbm{P}\pi\pi} F_M [(k'-k)^2] 
\left[ (k'+k)_\mu (k'+k)_\nu - \frac{1}{4} \, g_{\mu\nu} (k'+k)^2 \right] \,,
\ee
\be\label{3.31A}
\beta_{\mathbbm{P}\pi\pi} = 1.76 \,\mbox{GeV}^{-1} \,.
\ee
\newline

\noindent 
$\bullet$ ${\mathbbm P}\rho\rho$ (see section \ref{rho-proton scattering})
\newline
\vspace*{.2cm}
\hspace*{0.5cm}\includegraphics[height=85pt]{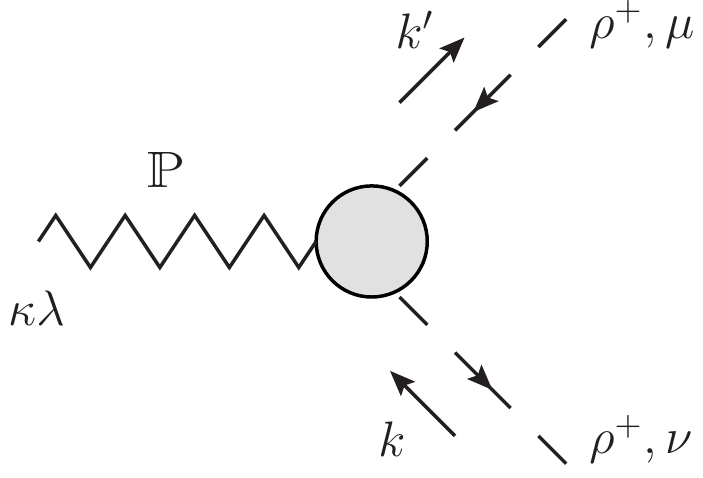} 
\hspace*{0.5cm}\includegraphics[height=85pt]{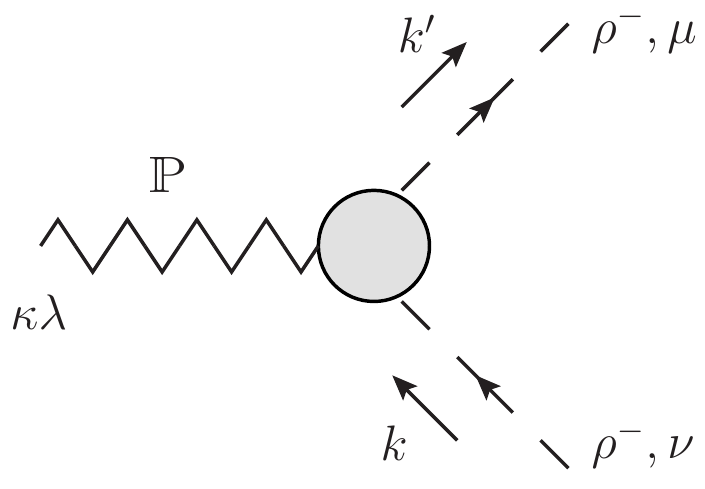} 
\hspace*{0.5cm}\includegraphics[height=85pt]{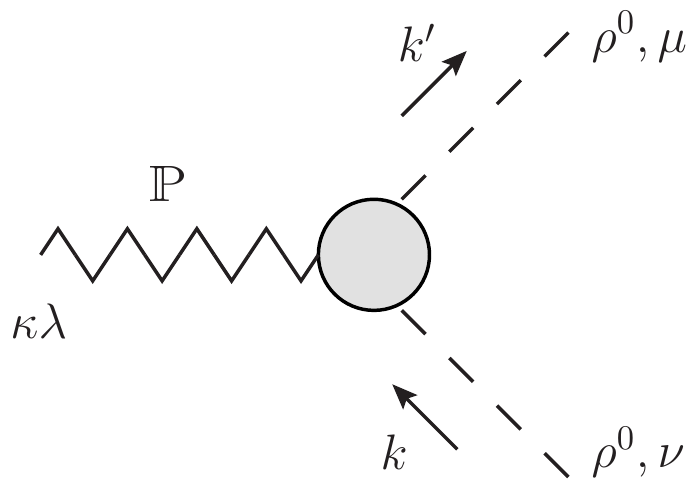} 
\newline
\be\label{3.32}
i \Gamma_{\mu\nu\kappa\lambda}^{(\mathbbm{P}\rho\rho)} (k',k) 
= iF_M[(k'-k)^2]
\left[2 a_{\mathbbm{P}\rho\rho} \Gamma_{\mu\nu\kappa\lambda}^{(0)}(k',-k) 
- b_{\mathbbm{P}\rho\rho}\Gamma_{\mu\nu\kappa\lambda}^{(2)}(k',-k) \right] \,.
\ee
In section \ref{rho-proton scattering} we give arguments that the 
following relation should hold: 
\be\label{3.32A}
2 m_\rho^2 a_{\mathbbm{P}\rho\rho} + b_{\mathbbm{P}\rho\rho} = 
4 \beta_{\mathbbm{P}\pi\pi} = 7.04 \,\mbox{GeV}^{-1} \,.
\ee
\newline

\noindent 
$\bullet$ $f_{2R}NN$ (see section \ref{Reggeon exchanges and total cross sections})
\newline
\vspace*{.2cm}
\hspace*{0.5cm}\includegraphics[height=85pt]{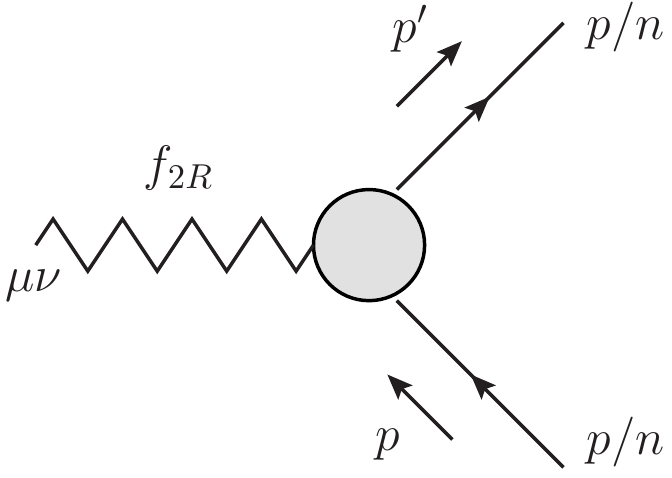} 
\hspace*{0.5cm}\includegraphics[height=85pt]{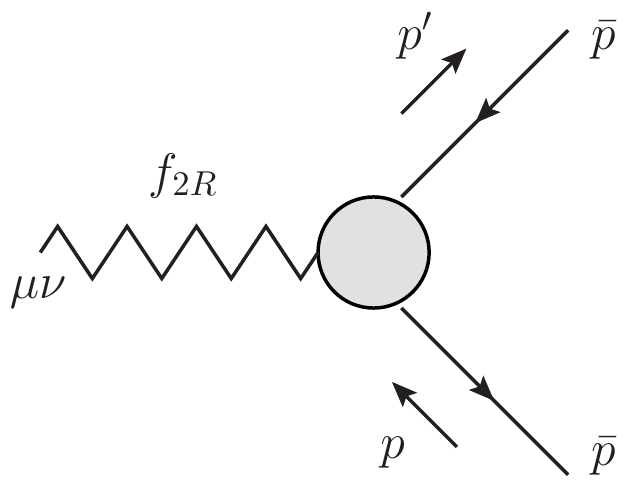} 
\newline
\be\label{3.33}
\begin{split}
i \Gamma_{\mu\nu}^{(f_{2R} pp)} (p',p) =&\,
i \Gamma_{\mu\nu}^{(f_{2R} nn)} (p',p) =
i \Gamma_{\mu\nu}^{(f_{2R} \bar{p}\bar{p})} (p',p) \\
=&- i g_{f_{2R}pp} \frac{1}{M_0}
F_1[ (p'-p)^2] \\
& \times \bigg\{ \frac{1}{2} [ \gamma_\mu (p'+p)_\nu + \gamma_\nu (p'+p)_\mu ]
- \frac{1}{4} g_{\mu\nu} (\slash{p}' + \slash{p}) \bigg\} \,,
\end{split}
\ee
\be\label{3.33A}
g_{f_{2R}pp} = 11.04\,, \qquad M_0 = 1~\text{GeV}\,.
\ee

\noindent 
$\bullet$ $a_{2R}NN$ (see section \ref{Reggeon exchanges and total cross sections})
\newline
\vspace*{.2cm}
\hspace*{0.5cm}\includegraphics[height=85pt]{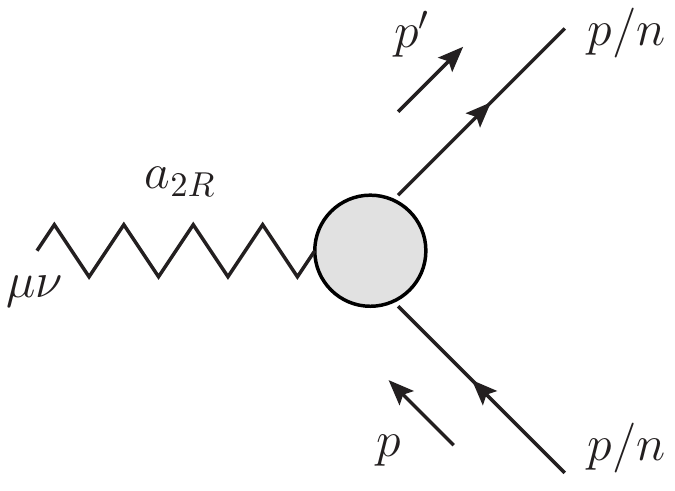} 
\hspace*{0.5cm}\includegraphics[height=85pt]{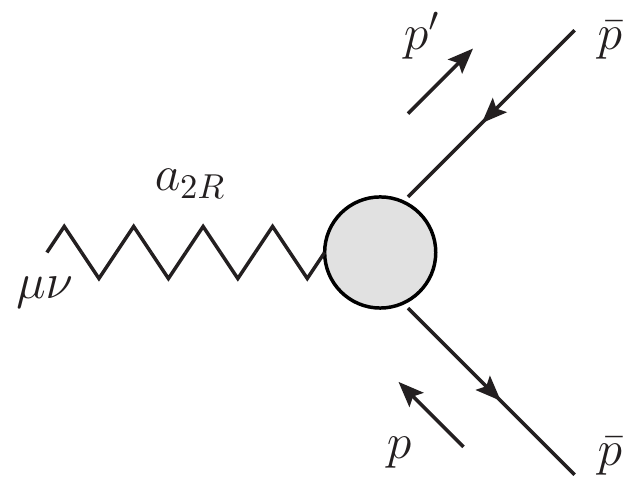} 
\newline
\be\label{3.34}
\begin{split}
i \Gamma_{\mu\nu}^{(a_{2R} pp)} (p',p) =&\,
- i \Gamma_{\mu\nu}^{(a_{2R} nn)} (p',p) =
i \Gamma_{\mu\nu}^{(a_{2R} \bar{p}\bar{p})} (p',p) \\
=&- i g_{a_{2R}pp} \frac{1}{M_0}
F_1[ (p'-p)^2]\\
&
\times \bigg\{ \frac{1}{2} [ \gamma_\mu (p'+p)_\nu + \gamma_\nu (p'+p)_\mu ]
- \frac{1}{4} g_{\mu\nu} (\slash{p}' + \slash{p}) \bigg\}\,,\\
\end{split}
\ee
\be\label{3.34A}
g_{a_{2R}pp} = 1.68\,,\qquad M_0 = 1~\text{GeV}\,.
\ee
\newline

\noindent 
$\bullet$ $f_{2R}\pi\pi$ (see section \ref{Pion-proton scattering})
\newline
\vspace*{.2cm}
\hspace*{0.5cm}\includegraphics[height=85pt]{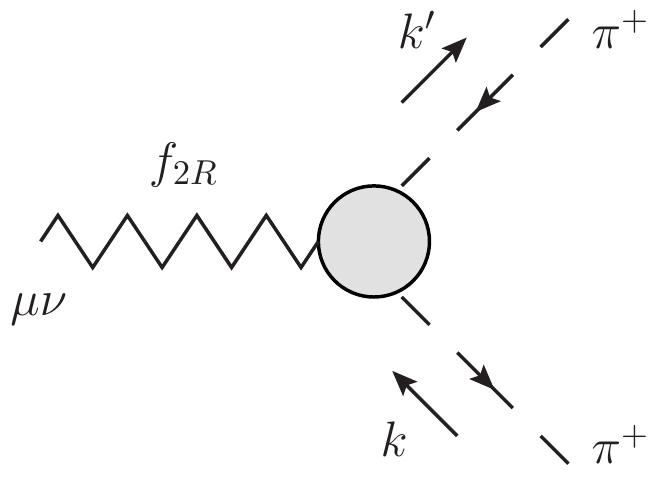} 
\hspace*{0.5cm}\includegraphics[height=85pt]{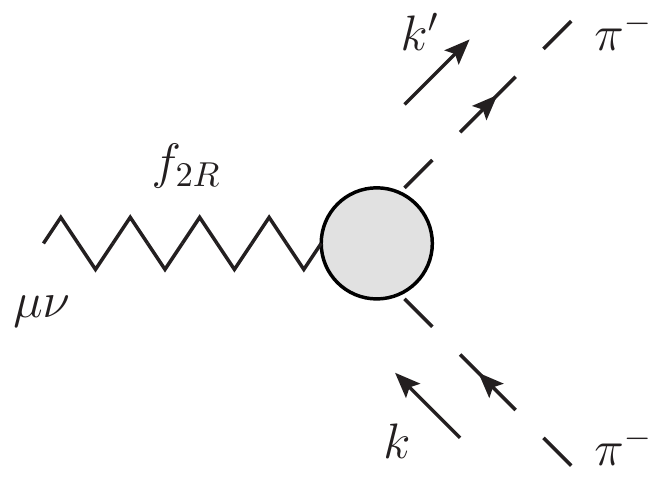} 
\hspace*{0.5cm}\includegraphics[height=85pt]{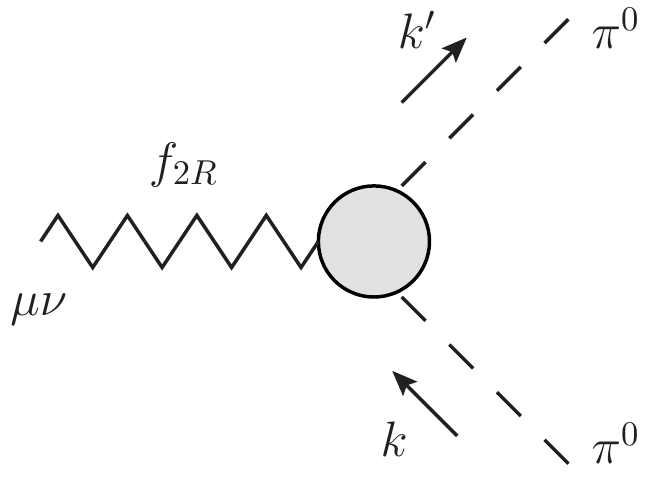} 
\newline
\be\label{3.36}
i \Gamma_{\mu\nu}^{(f_{2R} \pi\pi)} (k',k) =
- i \,\frac{g_{f_{2R}\pi\pi}}{2 M_0} F_M[(k'-k)^2]
\left[ (k'+k)_\mu (k'+k)_\nu - \frac{1}{4} g_{\mu\nu} (k'+k)^2 \right]\,,
\ee
\be\label{3.36A}
g_{f_{2R}\pi\pi} = 9.30\,,\qquad M_0 = 1~\text{GeV}\,.
\ee
\newline

\noindent 
$\bullet$ $f_{2R}\rho\rho$ (see section \ref{rho-proton scattering})
\newline
\vspace*{.2cm}
\hspace*{0.5cm}\includegraphics[height=85pt]{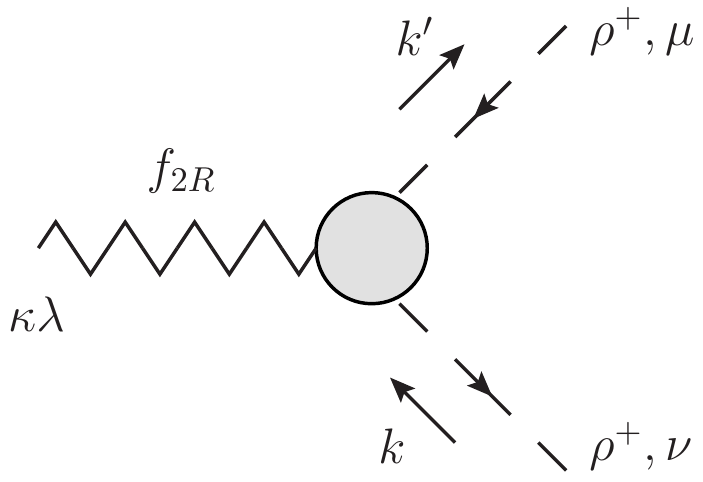} 
\hspace*{0.5cm}\includegraphics[height=85pt]{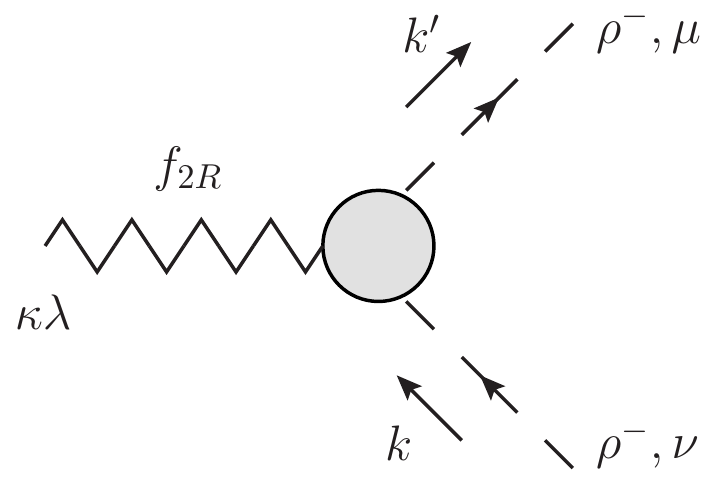} 
\hspace*{0.5cm}\includegraphics[height=85pt]{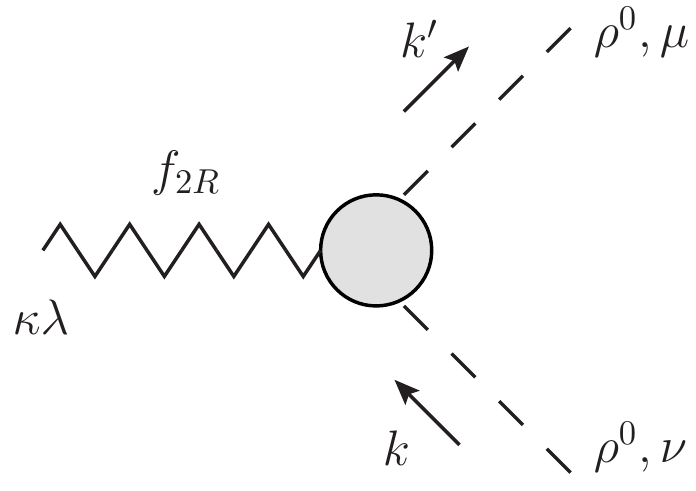} 
\newline
\be\label{3.37}
i \Gamma_{\mu\nu\kappa \lambda}^{(f_{2R} \rho\rho)} (k',k) =
i F_M[(k'-k)^2]
\left[ 2 a_{f_{2R}\rho\rho} \Gamma_{\mu\nu\kappa\lambda}^{(0)}(k',-k) -
b_{f_{2R}\rho\rho} \Gamma_{\mu\nu\kappa\lambda}^{(2)}(k',-k) \right]\,,
\ee
\be\label{3.37AA}
a_{f_{2R}\rho\rho} = 2.92~\text{GeV}^{-3}\,,\qquad
b_{f_{2R}\rho\rho} = 5.02~\text{GeV}^{-1}\,.
\ee

\noindent 
$\bullet$ $a_{2R} \omega \rho$ (see section \ref{rho-proton scattering})
\newline
\vspace*{.2cm}
\hspace*{0.5cm}\includegraphics[height=85pt]{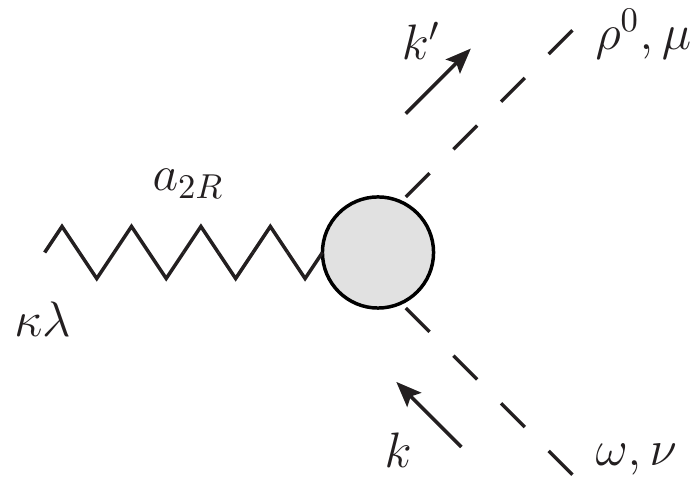} 
\newline
\begin{equation}\label{3.37a}
i \Gamma_{\mu\nu\kappa\lambda}^{(a_{2R} \omega\rho)} (k',k) =
i F_M[(k'-k)^2] 
\left[ 2 a_{a_{2R}\omega\rho} \Gamma_{\mu\nu\kappa\lambda}^{(0)}(k',-k) -
b_{a_{2R}\omega\rho} \Gamma_{\mu\nu\kappa\lambda}^{(2)}(k',-k) \right]\,,
\end{equation}
\begin{equation}\label{3.37ab}
\left| a_{a_{2R}\omega\rho} \right|= 2.56~\text{GeV}^{-3}\,,\qquad
\left| b_{a_{2R}\omega\rho} \right| = 4.68~\text{GeV}^{-1}\,.
\end{equation}
\newline

\noindent 
$\bullet$ $\omega_R NN$ (see section \ref{Reggeon exchanges and total cross sections})
\newline
\vspace*{.2cm}
\hspace*{0.5cm}\includegraphics[height=85pt]{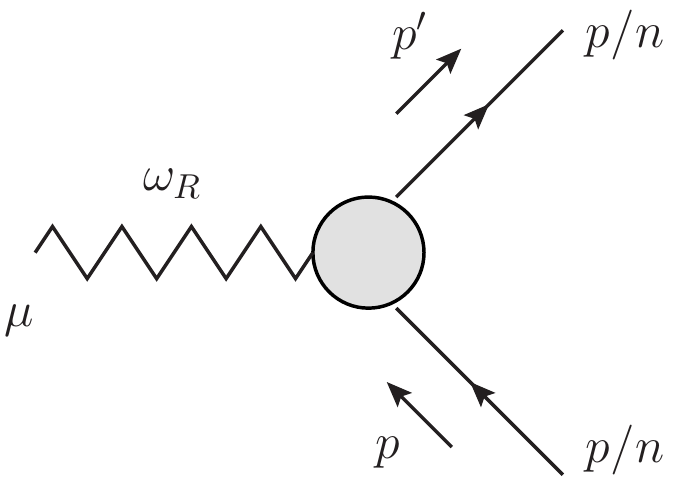} 
\hspace*{0.5cm}\includegraphics[height=85pt]{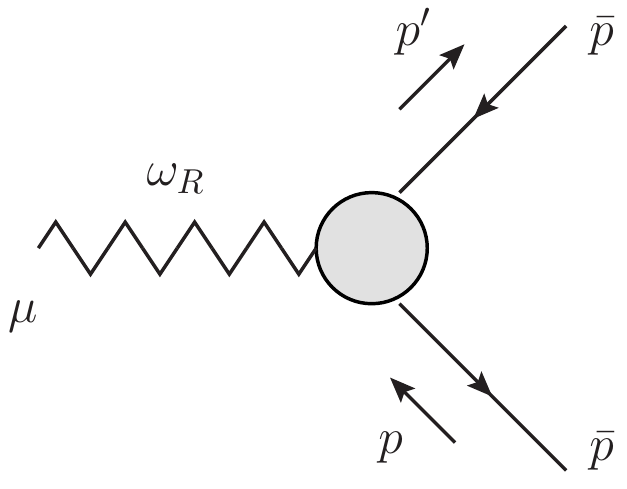} 
\newline
\be\label{3.38}
\begin{split}
i \Gamma_\mu^{(\omega_R pp)} (p',p) &=
 i \Gamma_\mu^{(\omega_R nn)} (p',p) =
- i \Gamma_\mu^{(\omega_R \bar{p}\bar{p})} (p',p) \\
&= - i g_{\omega_Rpp} F_1[ (p'-p)^2] \gamma_\mu\,,
\end{split}
\ee
\be\label{3.38A}
g_{\omega_Rpp} = 8.65\,.
\ee
\newline

\noindent 
$\bullet$ $\rho_R NN$ (see section \ref{Reggeon exchanges and total cross sections})
\newline
\vspace*{.2cm}
\hspace*{0.5cm}\includegraphics[height=85pt]{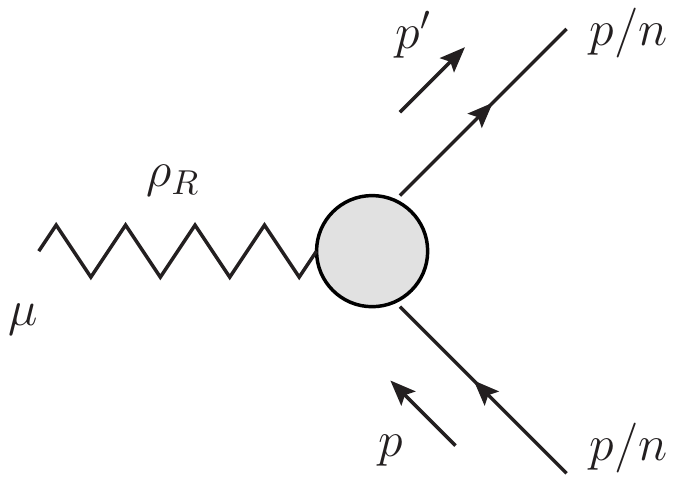} 
\hspace*{0.5cm}\includegraphics[height=85pt]{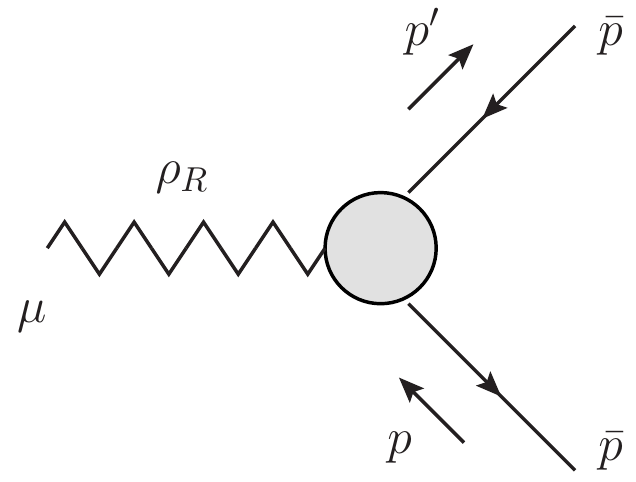} 
\newline
\be\label{3.40}
\begin{split}
i \Gamma_\mu^{(\rho_R pp)} (p',p) &=
- i \Gamma_\mu^{(\rho_R nn)} (p',p) =
- i \Gamma_\mu^{(\rho_R \bar{p}\bar{p})} (p',p) \\
&= - i g_{\rho_Rpp} F_1[ (p'-p)^2] \gamma_\mu\,,
\end{split}
\ee
\be\label{3.40A}
g_{\rho_Rpp} = 2.02\,.
\ee
\newline

\noindent 
$\bullet$ $\rho_R \pi\pi$ (see section \ref{Pion-proton scattering})
\newline
\vspace*{.2cm}
\hspace*{0.5cm}\includegraphics[height=85pt]{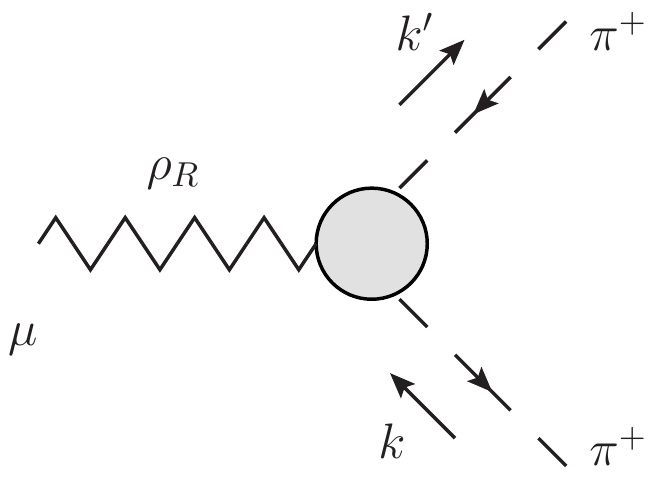} 
\hspace*{0.5cm}\includegraphics[height=85pt]{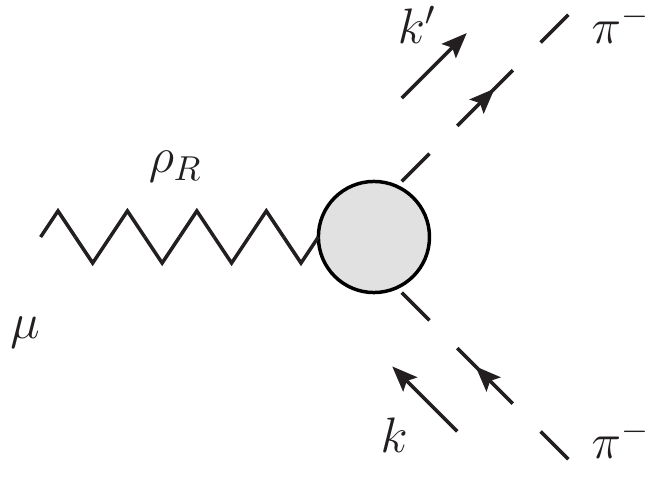} 
\newline
\be\label{3.42}
\begin{split}
i \Gamma_\mu^{(\rho_R \pi^+\pi^+)} (k',k) &=
- i \Gamma_\mu^{(\rho_R \pi^-\pi^-)} (k',k) \\
&= - \frac{i}{2} g_{\rho_R\pi\pi} F_M[ (k'-k)^2] (k'+k)_\mu\,,
\end{split}
\ee
\be\label{3.42A}
g_{\rho_R\pi\pi} = 15.63\,.
\ee
\newline

\noindent 
$\bullet$ $\omega_R\omega f_2$ (see section \ref{rho-proton scattering})
\newline
\vspace*{.2cm}
\hspace*{0.5cm}\includegraphics[height=85pt]{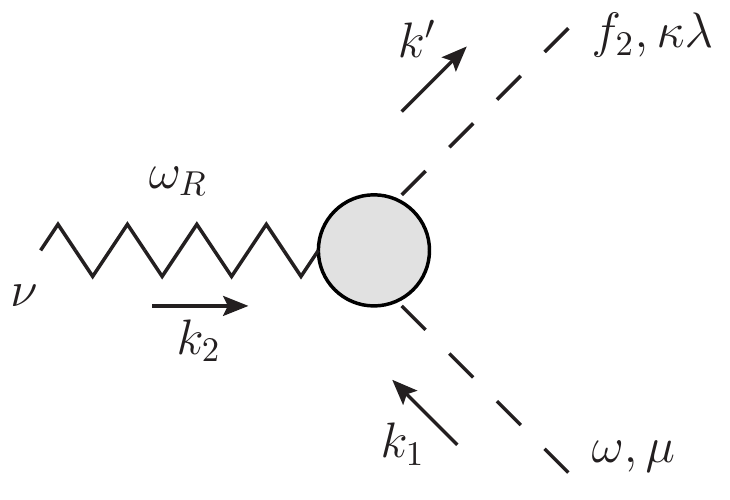} 
\begin{equation}\label{3.44}
i \Gamma_{\mu\nu\kappa\lambda}^{(\omega_R \omega f_2)} (k',k_1) =
i F_M(k_2^2) \left[ 2 a_{\omega_R\omega f_2} \Gamma_{\mu\nu\kappa\lambda}^{(0)}(k_1,k_2) -
b_{\omega_R\omega f_2} \Gamma_{\mu\nu\kappa\lambda}^{(2)}(k_1,k_2) \right] \,,
\end{equation}
with $k'=k_1 + k_2$\,.
\newline

\noindent 
$\bullet$ $\rho_R\rho^0 f_2$ (see section \ref{rho-proton scattering})
\newline
\vspace*{.2cm}
\hspace*{0.5cm}\includegraphics[height=85pt]{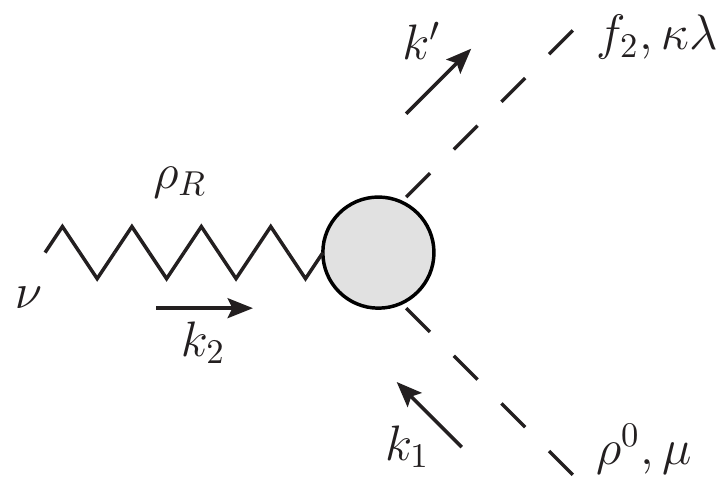} 
\begin{equation}\label{3.45}
i \Gamma_{\mu\nu\kappa\lambda}^{(\rho_R \rho^0 f_2)} (k',k_1) =
i F_M(k_2^2) \left[ 2 a_{\rho_R\rho^0 f_2} \Gamma_{\mu\nu\kappa\lambda}^{(0)}(k_1,k_2) -
b_{\rho_R\rho^0 f_2} \Gamma_{\mu\nu\kappa\lambda}^{(2)}(k_1,k_2) \right] \,,
\end{equation}
with $k'=k_1 + k_2$ and 
\be\label{3.45A}
a_{\rho_R\rho^0 f_2} = 2.92~\text{GeV}^{-3}\,, \qquad
b_{\rho_R\rho^0 f_2} = 5.02~\text{GeV}^{-1}\,.\\
\ee
\newline

\noindent 
$\bullet$ ${\mathbbm O} NN$ (see section \ref{Odderon exchange})
\newline
\vspace*{.2cm}
\hspace*{0.5cm}\includegraphics[height=85pt]{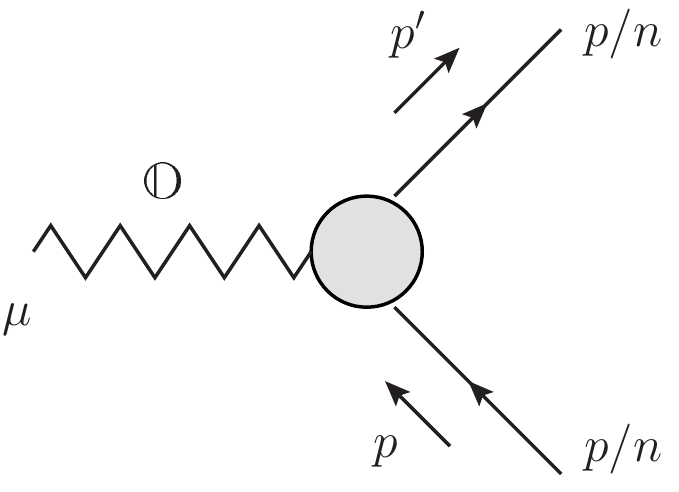} 
\hspace*{0.5cm}\includegraphics[height=85pt]{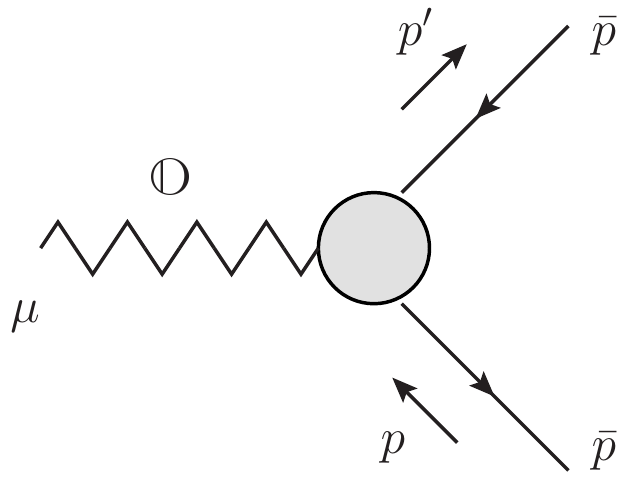} 
\be\label{3.46}
\begin{split}
i \Gamma_\mu^{({\mathbbm O} pp)} (p',p) &=
i \Gamma_\mu^{({\mathbbm O} nn)} (p',p) =
- i \Gamma_\mu^{({\mathbbm O} \bar{p}\bar{p})} (p',p) \\
&=- i 3 \beta_{{\mathbbm O}pp} M_0 F_1[ (p'-p)^2] \gamma_\mu\,,
\end{split}
\ee
\be\label{3.46A}
M_0 = 1~\text{GeV}\,.
\ee
\newline

\noindent 
$\bullet$ ${\mathbbm O}\gamma f_2$ (see section \ref{Odderon exchange})
\newline
\vspace*{.2cm}
\hspace*{0.5cm}\includegraphics[height=85pt]{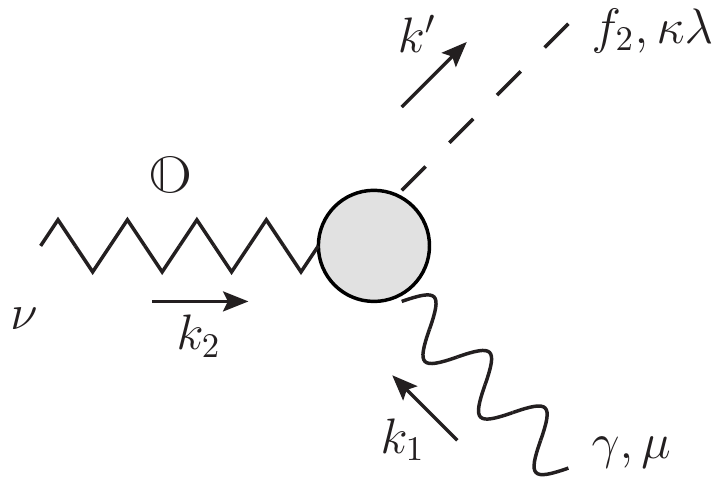} 
\begin{equation}\label{3.48}
i \Gamma_{\mu\nu\kappa\lambda}^{({\mathbbm O} \gamma f_2)} (k',k_1) =
i F_M(k_1^2) F_M(k_2^2) \left[ 2 a_{{\mathbbm O} \gamma f_2} \Gamma_{\mu\nu\kappa\lambda}^{(0)}(k_1,k_2) -
b_{{\mathbbm O}\gamma f_2} \Gamma_{\mu\nu\kappa\lambda}^{(2)}(k_1,k_2) \right]\,,
\end{equation}
with $k'=k_1 + k_2$ and 
\be\label{3.48A}
a_{{\mathbbm O}\gamma f_2} = e \hat{a}_{{\mathbbm O}\gamma f_2}\,, \qquad
b_{{\mathbbm O}\gamma f_2} = e \hat{b}_{{\mathbbm O}\gamma f_2}\,.
\ee

The vertices \eqref{3.44}, \eqref{3.45}, and \eqref{3.48} 
are written for `on-shell' $f_2$ mesons, that is for $k'^2=m_{f_2}^2$. 
For $k'^2 \neq m_{f_2}^2$ form factors as in \eqref{3.28} should be inserted 
in addition. 

The list of vertices given here can be extended. One obvious extension is to 
include particles containing strange quarks. Another extension would 
be to include vertices of three reggeons, the triple pomeron vertex etc.\ which 
would give rise to a reggeon field theory. 
This problem is very interesting but beyond the scope of the present paper. 

\section{Vector Mesons}
\label{Vector mesons}

Vector mesons were originally introduced in nuclear and high-energy physics 
on theoretical grounds in 
\cite{Johnson:1955zz,Duerr:1956zza,Duerr:1956zz,Nambu:1997vw,Frazer:1959gy,Frazer:1960zzb,Sakurai:1960ju} 
and the vector-meson-dominance (VMD) model in \cite{GellMann:1961tg}. 
The VMD model gives the $\gamma V$ vertices as listed in \eqref{3.20}-\eqref{3.21}. 
In our analysis we shall adopt these vertices which have 
been used extensively in data analyses. For reviews see for instance 
\cite{Donnachie:2002en,Bauer:1977iq,Schildknecht:2005xr,Close:2007zzd}. 
Thus, one may ask if anything new can be said about VMD. We hope to 
show that this is indeed the case. 

\boldmath
\subsection{The $\rho$ Propagator}
\unboldmath

In \cite{Melikhov:2003hs} a careful analysis of pion electromagnetic and weak 
form factors was presented. A dispersion theoretic 
approach was used based on general properties of propagators and vertex 
functions. Analytic expressions for the propagator matrix of the 
$\gamma$-$\rho$-$\omega$ system are given in Appendix B of \cite{Melikhov:2003hs}. 
The $(\rho,\rho)$ part of the inverse propagator matrix is given there, with 
$k^2=s$, as follows 
\be \label{4.100}
\left[\Delta^{-1}_T(s)\right]^{(\rho,\rho)} = -m_\rho^2 + s + B_{\rho \rho}(s) \,,
\ee
where 
\be \label{4.101}
\begin{split}
B_{\rho \rho}(s) =& \, g^2_{\rho\pi\pi} s \left[R(s,m_\pi^2) - R(m_\rho^2, m_\pi^2) 
+ \frac{1}{2} \, \left( R(s,m_K^2) - R(m_\rho^2,m_K^2) \right) \right]
\\
& + i g^2_{\rho \pi\pi} \left[ I(s,m_\pi^2) + \frac{1}{2}\,I(s,m_K^2) \right] 
\end{split}
\ee
with 
\begin{align}
\label{4.102}
I(s,m^2) &= \frac{1}{192 \pi}\,s \left(1-\frac{4m^2}{s} \right)^\frac{3}{2} 
\theta(s-4m^2) \,,
\\
\label{4.103}
R(s,m^2)&= \frac{s}{192 \pi^2}\, 
\mbox{V.\,P.}\int_{4m^2}^\infty 
\frac{ds'}{s' (s'-s)} \left(1-\frac{4m^2}{s'} \right)^\frac{3}{2} \,.
\end{align}
Here $\mbox{V.\,P.}$ stands for the principal value prescription. 
For the explicit expressions of $R(s,m^2)$ we refer to eq.\ (100) 
of \cite{Melikhov:2003hs}. 
The assumption going into \eqref{4.100} to \eqref{4.103} is that for 
calculating the self-energy part of the $\rho$ propagator it is sufficient 
to consider $\pi\pi$ and $K \bar{K}$ intermediate states with constant 
coupling parameters to the $\rho$. The corresponding $\rho\pi\pi$ 
coupling is taken over in the present paper in \eqref{3.26a}. 
Fits to the electromagnetic and weak form factors of the pion 
performed in \cite{Melikhov:2003hs} were very satisfactory. Fit III there 
gives $g_{\rho\pi\pi}$ as quoted in \eqref{3.27} and 
\be \label{4.104} 
m_{\rho^0} = 773.7 \pm 0.4 \,\mbox{MeV}
\ee
which is compatible with the PDG value \eqref{3.4a1}. 

If $\rho$-$\omega$ mixing is neglected we can set for the function 
$\Delta_T^{(\rho)}(s)$ in \eqref{3.2} 
\be \label{4.105}
\Delta_T^{(\rho)}(s) = \left[ \left(\Delta_T^{-1}(s) \right)^{(\rho,\rho)} \right]^{-1}
= \left[ -m_\rho^2 + s + B_{\rho\rho}(s) \right]^{-1} \,.
\ee
For the discussion of $\rho$-$\omega$ mixing and of $\Delta_T^{(\omega)}(s)$ 
we refer to \cite{Melikhov:2003hs}. We must also emphasise that our 
expression for the $\rho$ propagator function $\Delta_T^{(\rho)}(s)$ from 
\eqref{4.100} to \eqref{4.105} should be good for 
$|s| \lesssim 15\,\mbox{GeV}^2$ say. The expressions should not be used 
for much larger $|s|$ since then the assumption of constancy of the 
$\rho\pi\pi$ coupling will no longer hold. 

\begin{figure}[ht]
\begin{center}
\includegraphics[height=100pt]{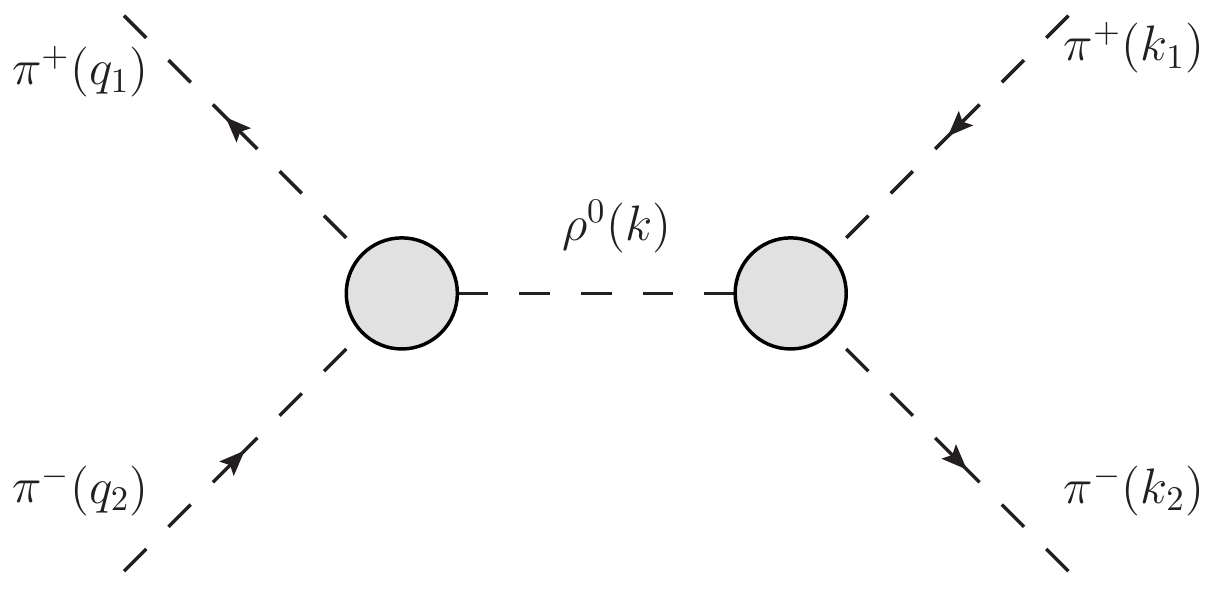}
\caption{Elastic scattering of $\pi^+\pi^-$ with $\rho$ exchange 
in the $s$ channel. 
\label{figpipirhopipi}}
\end{center}
\end{figure}

We explain now why we cannot use a simple Breit-Wigner ansatz for 
$\Delta_T^{(\rho)} (s)$ as given in \eqref{3.4}. For this we calculate 
the amplitude for the elastic pion-pion scattering 
\be \label{4.105a}
\pi^+ (q_1) + \pi^-(q_2) \longrightarrow \pi^+ (k_1) + \pi^- (k_2) 
\ee
through resonant $\rho$ exchange from the diagram of figure 
\ref{figpipirhopipi}. 
Using the $\rho\pi\pi$ coupling \eqref{3.26a} and the $\rho$ propagator 
\eqref{3.2} we get 
\be \label{4.106}
\begin{split}
\langle \pi^+(k_1), \pi^-(k_2) |\mathcal{T}|\pi^+(q_1), \pi^-(q_2)\rangle 
&= \frac{1}{4}\, g_{\rho\pi\pi}^2 (k_1-k_2)\cdot(q_1-q_2) \Delta_T^{(\rho)}(k^2) \,,
\\
k=k_1+k_2 &= q_1+q_2 \,.
\end{split}
\ee
With $\Delta_T^{(\rho)}(k^2)$ from \eqref{4.105} all partial-wave unitarity 
relations are satisfied. But this is not the case if we use $\Delta_T^{(\rho)}(k^2)$ 
from \eqref{3.4}. 

The careful reader of \cite{Melikhov:2003hs} will notice 
that the $\gamma V$ coupling used there looks very different from 
the VMD coupling \eqref{3.20}. We shall, therefore, make some remarks 
on the relation of these different forms for the $\gamma V$ vertex. 
The following section deals more with theoretical concepts related to 
VMD and can be skipped by a reader only interested in practical applications 
of our model to high-energy reactions. 

\boldmath
\subsection{Different Forms of the $\gamma V$ Vertex and Approximate VMD}
\unboldmath
\label{sec:siffforms}

We shall now show that from the dispersion theoretic approach of 
\cite{Melikhov:2003hs} we can shed new light on the time-honoured VMD model. 
For this we present the essence of the arguments of \cite{Melikhov:2003hs} in 
Lagrangian language. Consider an effective field theory containing -- to 
make the argument simple -- just $\pi^\pm$ and $\rho^0$ meson 
fields. The assumptions of the model presented in section 4.1 of 
\cite{Melikhov:2003hs} can now be phrased as follows. It is assumed that there 
are couplings of the photon to the $\pi$ mesons and to the $\rho^0$ meson. 
The Lagrangian reads then 
\be\label{4.1}
\begin{split}
{\cal L} (x) =& \left( \del_\mu \pi^-(x) \right) \del^\mu \pi^+(x) 
- m_\pi^2 \pi^-(x) \pi^+(x) 
\\
&+{\cal L}_{\rho^0} (x)
+{\cal L}'_{\gamma \pi\pi} (x) +{\cal L}'_{\gamma \rho^0} (x) 
+{\cal L}'_{\rho^0 \pi \pi} (x)\,,
\end{split}
\ee
where the $\rho^0$ kinetic term is 
\begin{equation}
\label{4.1a}
{\cal L}_{\rho^0} (x) = -\frac{1}{2} \left(\del_\nu \rho^0_\mu(x)\right) 
\left(\del^\nu \rho^{0\mu}(x)\right) + \frac{1}{2}\left(\del_\mu \rho^{0\mu}(x)\right)^2 
+ \frac{1}{2} m_\rho^2 \rho^0_\mu(x) \rho^{0\mu}(x)
\end{equation}
and the interaction parts are 
\begin{align}\label{4.2}
{\cal L}'_{\gamma \pi\pi} (x) &= -ie \big(\pi^-(x) \overleftrightarrow{\partial_\mu}
\pi^+(x) \big) A^\mu(x) + 
e^2 \pi^-(x) \pi^+(x) A^\mu(x) A_\mu(x)\,,
\\
\label{4.3}
{\cal L}'_{\gamma \rho^0} (x) &= - \frac{e}{2 \gamma_\rho}
\big( \partial_\mu \rho_\nu^0(x) - \partial_\nu \rho_\mu^0(x) \big)
\big( \partial^\mu A^\nu(x) - \partial^\nu A^\mu(x) \big) \,,
\\
\label{4.3a}
{\cal L}'_{\rho^0 \pi \pi} (x) &= - \frac{i}{2} g_{\rho\pi\pi} 
\big(\pi^-(x) \overleftrightarrow{\partial_\mu} \pi^+(x) \big) \rho^{0\mu}(x)
\,.
\end{align}
Here $A_\mu(x)$ is the photon vector potential. The $\gamma$-$\rho$ 
coupling constant $1/\gamma_\rho$ in \eqref{4.3} is related to $f_\rho$ 
used in \cite{Melikhov:2003hs} 
as follows\footnote{Then it is easy to see that from \eqref{4.3} we get 
exactly the $\gamma$-$\rho$-transition matrix element, without the 
dispersive part $B_{\gamma\rho}(s)$, in (53) of \cite{Melikhov:2003hs}.}
\be\label{4.4}
\frac{1}{\gamma_\rho} \equiv \frac{f_\rho}{m_\rho} \,.
\ee

\begin{figure}[htb]
\begin{center}
\includegraphics[height=100pt]{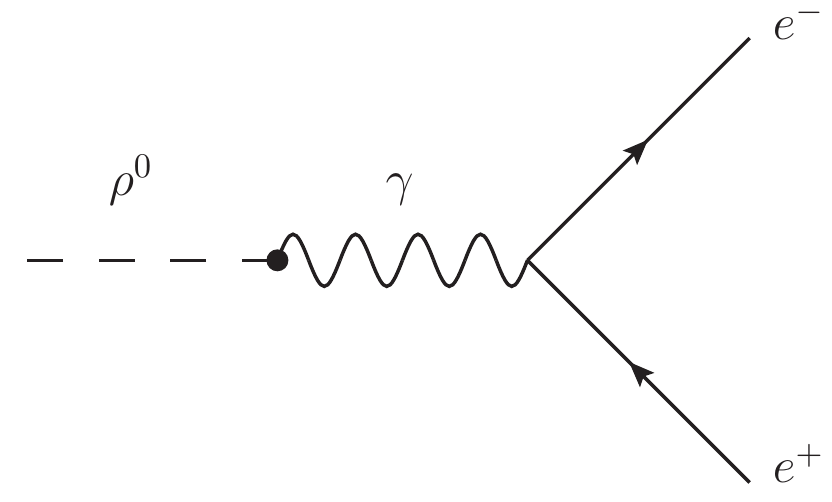}
\caption{Diagram for the decay $\rho^0 \to e^+ e^-$
\label{model:fig5}}
\end{center}
\end{figure}

Note that together with 
the standard kinetic term for the $\pi^\pm$ fields the coupling of 
the photon to $\pi^\pm$ and $\rho^0$ in \eqref{4.1} is perfectly 
gauge invariant. Clearly, the $\gamma\rho^0$ coupling in \eqref{4.3} 
looks very different from the standard VMD coupling \eqref{3.20}. 
But we can rewrite \eqref{4.3} as
\be\label{4.5}
{\cal L}'_{\gamma \rho^0} (x) = 
{\cal L}'^{(1)}_{\gamma \rho^0} (x)+
{\cal L}'^{(2)}_{\gamma \rho^0} (x)+
{\cal L}'^{(3)}_{\gamma \rho^0} (x) \,,
\ee
where 
\begin{align}\label{4.6}
{\cal L}'^{(1)}_{\gamma \rho^0} (x) &= 
-\frac{e}{\gamma_\rho} m_\rho^2 \rho_\mu^0(x) A^\mu(x)\,,
\\
\label{4.7}
{\cal L}'^{(2)}_{\gamma \rho^0} (x) &=
\frac{e}{\gamma_\rho} A^\mu(x)
\big[ (\square + m_\rho^2 ) \rho_\mu^0(x) - \partial_\mu \partial^\nu \rho_\nu^0(x) \big]\,,
\\
\label{4.8}
{\cal L}'^{(3)}_{\gamma \rho^0} (x) &=
-\frac{e}{\gamma_\rho} \partial^\mu
\big[ A^\nu(x) ( \partial_\mu \rho_\nu^0(x) - \partial_\nu \rho_\mu^0(x) ) \big]\,.
\end{align}
The term ${\cal L}'^{(1)}_{\gamma\rho^0}$ in \eqref{4.6} gives exactly 
the standard VMD coupling \eqref{3.20}. The term 
${\cal L}'^{(3)}_{\gamma\rho^0}$ in \eqref{4.8} is a total divergence 
and can be dropped. The term ${\cal L}'^{(2)}_{\gamma\rho^0}$ in 
\eqref{4.7} has the following properties. It vanishes on the $\rho^0$ 
mass shell. For instance, it does not contribute to the decay of 
an `on shell' $\rho^0$ to $e^+e^-$; see figure \ref{model:fig5}.

For processes with a virtual $\rho^0$ meson next to the coupling 
\eqref{4.7} the latter will eat up the $\rho^0$ propagator. As an 
example let us consider the reaction 
\be\label{4.9}
e^+(l_1) + e^-(l_2) \longrightarrow \pi^+(k_1) + \pi^-(k_2) 
\ee
and calculate the corresponding amplitude at tree level using 
the Lagrangian \eqref{4.1}. 

\begin{figure}[htb]
\begin{align}
(a)&\hspace*{1.3cm}
\includegraphics[height=100pt]{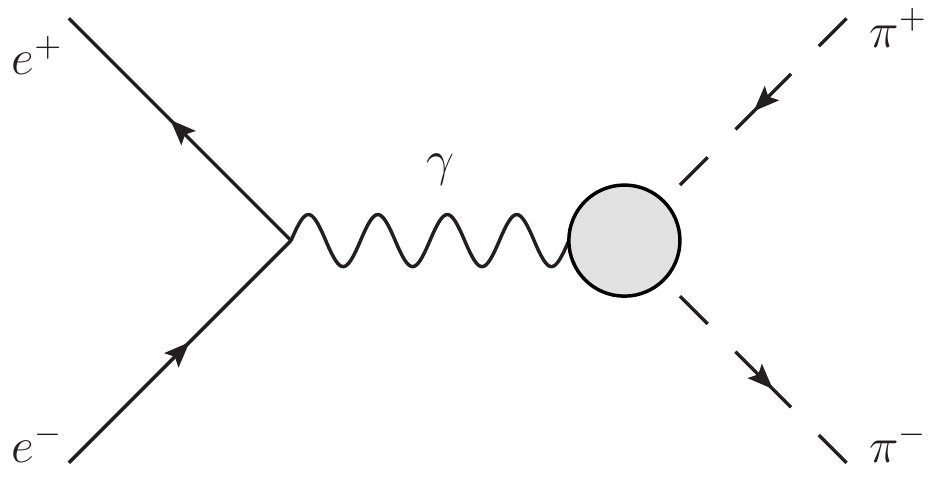} \nn \\
(b)&\hspace*{1.3cm}
\includegraphics[height=100pt]{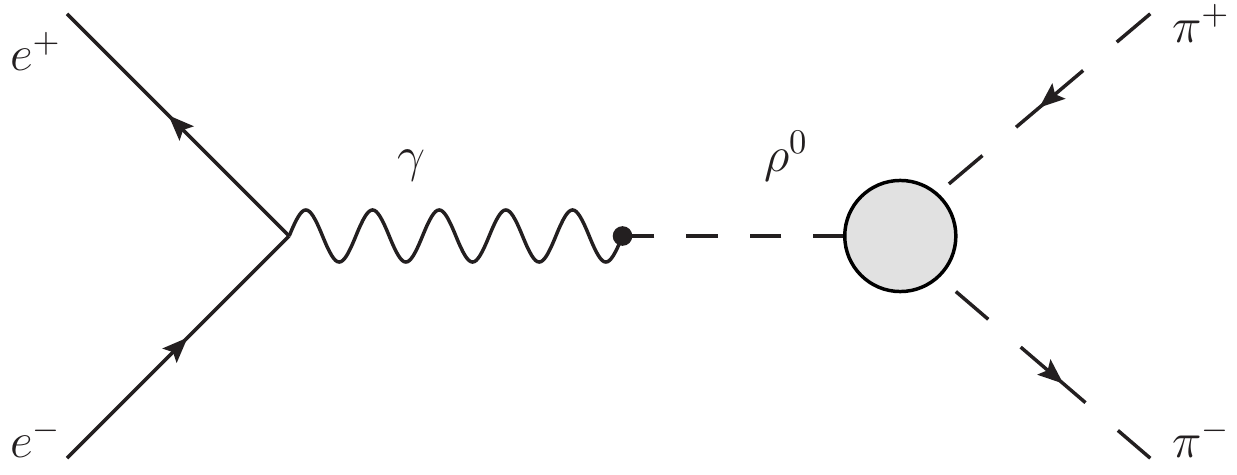} \nn\\
(c)&\hspace*{1.3cm}
\includegraphics[height=100pt]{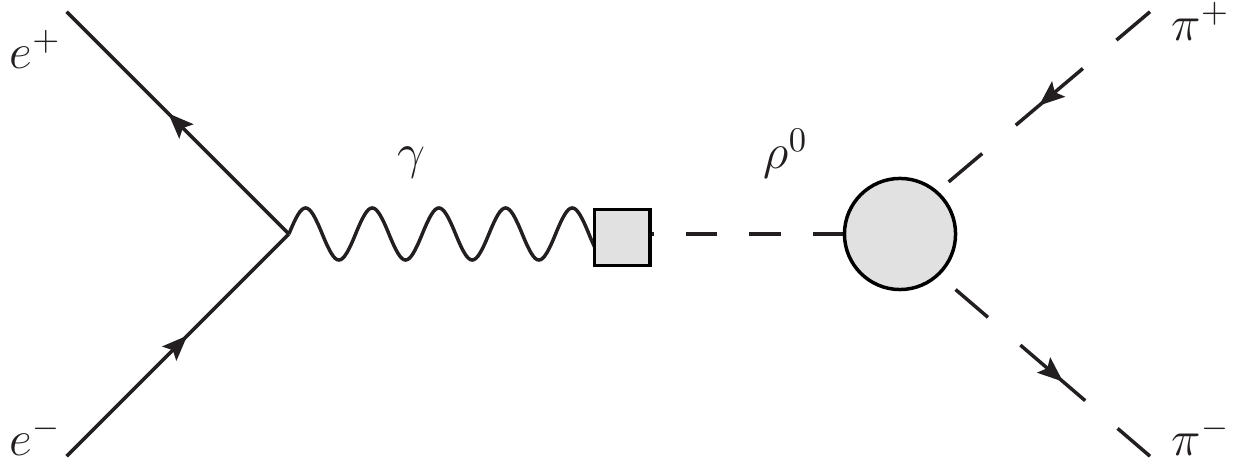} \nn
\end{align}
\caption{Tree level diagrams for $e^+e^-\to\pi^+\pi^-$ according 
to \eqref{4.1}. Here (a), (b), and (c) correspond to the direct $\gamma \pi\pi$ 
coupling term ${\cal L}'_{\gamma \pi\pi}$ \eqref{4.2}, the $\gamma \rho^0$ coupling term 
${\cal L}'^{(1)}_{\gamma \rho^0} $ \eqref{4.6}, and the $\gamma \rho^0$ 
coupling term ${\cal L}'^{(2)}_{\gamma \rho^0}$ \eqref{4.7}, respectively.
\label{model:fig6}}
\end{figure}

With \eqref{4.1} to \eqref{4.7} we have to consider three 
diagrams, see figure \ref{model:fig6}, the `direct' term due to 
${\cal L}'_{\gamma\pi\pi}$ and the terms with an intermediate 
virtual $\rho^0$ due to ${\cal L}'^{(1)}_{\gamma\rho^0}$ and 
${\cal L}'^{(2)}_{\gamma\rho^0}$. 

Using the $\rho^0$ propagator as obtained from \eqref{4.1a} 
we find for the amplitudes corresponding to figures \ref{model:fig6}(a) to (c) 
the following: 
\be
\label{4.10}
{\cal{M}}_\text{direct} = - i e^2 (k_1 - k_2)^\mu
\frac{1}{q^2} \,\bar{v}(l_1) \gamma_\mu u(l_2)\,,
\ee

\begin{align}
\label{4.11}
\begin{split}
{\cal{M}}_{\gamma \rho^0}^{(1)} &= -\frac{i}{2} g_{\rho\pi\pi} (k_1 - k_2)^\mu
i \Delta_{\mu\nu}^{(\rho)}(q) \left( -i \frac{e}{\gamma_\rho} m_\rho^2 \right)
\left( \frac{-i g^{\nu\lambda}}{q^2} \right)
\bar{v}(l_1) i e \gamma_\lambda u(l_2) 
\\
&= - i e^2 \frac{1}{2} g_{\rho\pi\pi}
\frac{ (- m_\rho^2)}{q^2 - m_\rho^2} \frac{1}{\gamma_\rho}
(k_1 - k_2)^\mu \frac{1}{q^2}
\bar{v}(l_1) \gamma_\mu u(l_2) \,,
\end{split}
\\
\label{4.12}
\begin{split}
{\cal{M}}_{\gamma\rho^0}^{(2)} &=
- \frac{i}{2} g_{\rho\pi\pi} (k_1 - k_2)^\mu
i \Delta_{\mu\kappa}^{(\rho)}(q)
\frac{i e}{\gamma_\rho}
\left[ (-q^2+m_\rho^2) \delta_\nu^\kappa + q^\kappa q_\nu \right]
\left(\frac{-i g^{\nu\lambda}}{q^2} \right)
\bar{v}(l_1) i e \gamma_\lambda u(l_2)
 \\  
&=
i e^2 \frac{1}{2} g_{\rho\pi\pi} \frac{1}{\gamma_\rho} (k_1 - k_2)^\mu
\frac{1}{q^2}
\bar{v}(l_1) \gamma_\mu u(l_2) \,.
\end{split}
\end{align}
In \eqref{4.10} to \eqref{4.12} we have set 
\be\label{4.13}
q = l_1+l_2 = k_1 + k_2\,.
\ee
Indeed, in ${\cal M}^{(2)}_{\gamma\rho^0}$ the $\rho^0$-pole is cancelled. 
This means that ${\cal L}'^{(2)}_{\gamma\rho}$ does {\em not} provide a 
term where the photon couples to a $\rho^0$ meson which then interacts 
with the pion, as one would expect from figure \ref{model:fig6}(c), but in 
fact a {\em direct} coupling of the photon to the pions {\em without} 
intermediate $\rho^0$. This opens the possibility of having 
a partial or even complete cancellation between ${\cal M}_\text{direct}$ 
\eqref{4.10} and ${\cal M}^{(2)}_{\gamma\rho^0}$ \eqref{4.12}. The VMD 
hypothesis assumes indeed that this cancellation is complete which requires
\be\label{4.13a}
\frac{1}{2} \, g_{\rho\pi\pi} \,\frac{1}{\gamma_\rho} = 1 \,.
\ee

This tree level calculation above is only meant to give an understanding 
how the various forms of the $\gamma \rho^0$ vertex are related. 
Including self-energy corrections in the $\rho^0$ propagator we find 
that the diagram of figure \ref{model:fig6}(c) due 
to $\mathcal{L}'^{(2)}_{\gamma\rho^0}$ gives a `direct' term plus 
a correction to the diagram of figure \ref{model:fig6}(b), that is, to 
the term due to $\mathcal{L}'^{(1)}_{\gamma\rho^0}$. 
From the rigorous dispersion theoretic 
analysis in \cite{Melikhov:2003hs} where VMD is {\em not} assumed we get, 
using the parameters from Table 4 (Fit III) there, 
the following:\footnote{These values for $g_{\rho\pi\pi}$ and $m_\rho$ 
have already been quoted in \eqref{3.27} and \eqref{4.104}, respectively.}
\be\label{4.13b}
g_{\rho\pi\pi} = 11.51 \pm 0.07 \,,
\quad
f_\rho = 155.4 \pm 1.7 \,\mbox{MeV}\,,
\quad
m_\rho = 773.7 \pm 0.4 \,\mbox{MeV}\,.
\ee
This leads, with \eqref{4.4}, to 
\be\label{4.13c}
\frac{1}{2} \, g_{\rho\pi\pi} \,\frac{1}{\gamma_\rho} 
=\frac{1}{2} \, g_{\rho\pi\pi} \,\frac{f_\rho}{m_\rho}  
= 1.16 \pm 0.01 \,,
\ee
\be\label{4.13d}
\frac{4 \pi}{\gamma_\rho^2} = 4 \pi \left( \frac{f_\rho}{m_\rho} \right)^2 
= 0.507 \pm 0.011 \,.
\ee
We see from \eqref{4.13c} that the VMD relation \eqref{4.13a} is, in this 
case, satisfied to an accuracy of around $15\%$. From \eqref{4.13d} we 
get a value for $4\pi/\gamma^2_\rho$ which is well compatible with 
the one used in \eqref{3.21} which is taken from eq.\ (5.3) of 
\cite{Donnachie:2002en}. 

To summarise: we have seen that the gauge invariant $\gamma\rho^0$ 
coupling ${\cal L}'_{\gamma\rho^0}$ \eqref{4.3} contains the standard 
VMD $\gamma\rho^0$ coupling term ${\cal L}'^{(1)}_{\gamma\rho^0}$ \eqref{4.6} 
plus a term ${\cal L}'^{(2)}_{\gamma\rho^0}$ \eqref{4.7} which, 
effectively, describes a {\em direct} coupling of the photons to pions 
{\em without} intermediate $\rho^0$. In strict VMD this latter term 
exactly cancels the contribution from the original direct coupling 
${\cal L}'_{\gamma\pi\pi}$.

\subsection{General VMD Hypothesis}
\label{subsec:genVMDhyp}

In view of the above discussion, we think that the general VMD 
hypothesis can be stated as follows. We have for the coupling of the 
photon to light hadrons, that is, hadrons made from $u,d$, and $s$ 
quarks, the following Lagrangian, comprising a direct term without 
intermediate vector mesons and a coupling with intermediate vector 
mesons $V=\rho^0,\omega,\phi$: 
\be\label{4.14}
{\cal L}'(x) = {\cal L}'_{\gamma,\mathrm{direct}}(x) 
+ \sum_{V=\rho^0,\omega,\phi} {\cal L}'_{\gamma V}(x) \,,
\ee
where for $V=\rho^0,\omega,\phi$ 
\be\label{4.15}
\begin{split}
{\cal L}'_{\gamma V}(x) &=-\frac{e}{2 \gamma_V} 
(\del_\mu V_\nu (x) - \del_\nu V_\mu(x)) 
(\del^\mu A^\nu(x) - \del^\nu A^\mu(x)) 
\\
&= {\cal L}'^{(1)}_{\gamma V}(x) +{\cal L}'^{(2)}_{\gamma V}(x) +{\cal L}'^{(3)}_{\gamma V}(x)\,.
\end{split}
\ee
Here the $\gamma_V$ are defined in \eqref{3.20}, \eqref{3.21} and 
${\cal L}'^{(i)}_{\gamma V}$ are defined as in \eqref{4.5} to \eqref{4.8} 
with $\rho^0$ replaced by $V$. The strict VMD hypothesis amounts to 
assuming that the coupling ${\cal L}'^{(2)}_{\gamma\rho^0}
+{\cal L}'^{(2)}_{\gamma\omega}+{\cal L}'^{(2)}_{\gamma\phi}$ -- which in 
fact is effectively a {\em direct} $\gamma$-hadron coupling -- precisely 
cancels the term ${\cal L}'_{\gamma,\mathrm{direct}}$ in \eqref{4.14}. 
Note that the remaining coupling, the standard VMD one \eqref{3.20}, 
which amounts to setting in \eqref{4.14} ${\cal L}'= 
{\cal L}'^{(1)}_{\gamma\rho^0}+{\cal L}'^{(1)}_{\gamma\omega}
+{\cal L}'^{(1)}_{\gamma\phi}$, will only be gauge invariant if the 
$\rho^0,\omega$, and $\phi$ fields are divergence free. It is, indeed, 
well known that with the standard VMD coupling \eqref{3.20} one 
has to be very careful in order to maintain gauge invariance in calculations.

Of course, it is also well known that the VMD model only gives approximate 
results. For soft reactions involving real photons these are typically 
correct at the $10\%$ to $20\%$ level. See, for instance, \eqref{4.13a} 
and \eqref{4.13c} and the extensive discussions of VMD in relation to 
experiments in \cite{Donnachie:2002en}, \cite{Schildknecht:2005xr} 
and \cite{Close:2007zzd}. Keeping this limited accuracy of VMD 
relations in mind we shall, in the present work, use the standard VMD 
couplings \eqref{3.20} for simplicity. But we shall always maintain 
strict gauge invariance since our vertices of section \ref{Propagators and vertices} 
incorporate the divergence condition $\partial^\mu V_\mu(x)=0$ for 
the vector meson fields. We note that with our description of the 
pomeron as an effective spin $2$ exchange there is no problem in 
maintaining this divergence condition also for the ${\mathbbm P}\rho\rho$ 
vertex; see \eqref{3.32}. Using VMD with the coupling \eqref{3.20} we 
get from this perfectly gauge invariant ${\mathbbm P}\gamma\rho$ 
and ${\mathbbm P}\gamma\gamma$ vertices. 
This is quite a non-trivial result. 

\section{Tensor Mesons}
\label{Tensor mesons}

In this section we discuss the vertices and the propagator for the tensor 
meson $f_2\equiv f_2(1270)$. We also make a few remarks 
on the $a_2\equiv a_2(1320)$ meson. 

\boldmath
\subsection{$f_2\to\pi\pi$}
\label{f2topipi}
\unboldmath

For the decays $f_2\to\pi^+\pi^-$ and $f_2\to\pi^0\pi^0$ there is 
only one invariant amplitude. It can be obtained from the following 
effective interaction Lagrangian $(M_0=1$ GeV): 
\begin{equation}\label{5.6}
\begin{split}
{\cal L}_{f_2\pi\pi}' (x) =& \,
\frac{1}{2 M_0} g_{f_2\pi\pi} \phi_{\kappa\lambda}(x)
\left( g^{\kappa\mu} g^{\lambda\nu} - \frac{1}{4} g^{\kappa\lambda}g^{\mu\nu} \right)\\
&
\times \sum_{a=1}^3 \left[
\frac{1}{2} \pi_a(x) \partial_\mu \partial_\nu \pi_a(x) + 
\frac{1}{2} (\partial_\mu \partial_\nu \pi_a(x) ) \pi_a(x)
- (\partial_\mu \pi_a(x) ) (\partial_\nu \pi_a(x) ) \right]\,.
\end{split}
\end{equation}
Here $\phi_{\kappa\lambda}$ is the $f_2$ field for which we require, 
since it corresponds to a neutral spin 2 particle, 
\be\label{5.6a}
\phi_{\kappa\lambda} = \phi_{\lambda\kappa} = \phi_{\kappa\lambda}^\dagger\,. 
\ee
The $\pi_a$ ($a=1,2,3$) are the pion fields and we have
\be\label{5.7}
\begin{split}
\pi^\pm (x) &= \frac{1}{\sqrt{2}} \, (\pi_1(x) \mp i \pi_2(x)) \,,
\\
\pi^0(x) &= \pi_3(x) \,.
\end{split}
\ee

The coupling \eqref{5.6} gives the $f_2\pi\pi$ vertex \eqref{3.28} 
where we have added a form factor taking into account that a 
variation of the coupling with the off-shellness of the $f_2$ must 
be expected. A convenient parametrisation of such a form factor is 
the exponential form 
\be\label{5.10a}
F^{(f_2\pi\pi)} (k^2) = \exp \left(-\frac{k^2-m_{f_2}^2}{\Lambda^2}\right)
\ee
with $\Lambda$ a parameter of the order 1 GeV. Here the normalisation 
condition
\be\label{5.10b} 
F^{(f_2\pi\pi)} (m_{f_2}^2) =1
\ee
is clearly satisfied.
 
From \eqref{3.28}  we find easily for an `on-shell' $f_2$ ($k^2=m_{f_2}^2$) 
\begin{equation}\label{5.10}
\Gamma(f_2 \to \pi^0\pi^0) =
\frac{1}{2}
\Gamma(f_2 \to \pi^+\pi^-) =
\frac{m_{f_2}}{960 \pi} |g_{f_2\pi\pi}|^2
\bigg( \frac{m_{f_2}}{M_0} \bigg)^2
\bigg( 1 - \frac{4 m_\pi^2}{m_{f_2}^2} \bigg)^{5/2} .
\end{equation}
With $\Gamma_{f_2}$ from \eqref{3.6} and 
\be\label{5.11}
\frac{\Gamma(f_2\to \pi\pi)}{\Gamma_{f_2}} = (84.8_{-1.2}^{+2.4} ) \%
\ee
from \cite{Beringer:1900zz} we get
\be\label{5.12}
\left| g_{f_2 \pi\pi} \right| = 9.26 \pm 0.15 \,.
\ee
With the arguments presented later in section \ref{Pion-proton scattering} 
we conclude that $g_{f_2\pi\pi}$ should be positive. Thus, we get finally 
as our best estimate, given in \eqref{3.28param},
\be\label{5.13}
g_{f_2 \pi\pi} = 9.26 \pm 0.15 \,.
\ee

\boldmath
\subsection{$f_2$ Propagator}
\label{f2 propagator}
\unboldmath

The propagator of a tensor field 
$\phi_{\mu\nu}(x) = \phi_{\nu\mu}(x) = \phi_{\mu\nu}^\dagger (x)$ 
is defined as 
\be\label{5.100} 
\Delta^{(T)}_{\mu\nu,\kappa\lambda} (k) = 
\frac{1}{i} \int d^4x\, e^{ikx} 
\langle 0 | \mathrm{T}^* (\phi_{\mu\nu} (x) \phi_{\kappa \lambda} (0)) |0\rangle 
\ee
where $\mathrm{T}^*$ is the covariantised T product. 
Let us first consider a free field corresponding to a tensor particle 
of mass $m_{f_2}$. To construct a basis for the polarisation tensors of 
such a particle of momentum $k$, $k^2=m_{f_2}^2$, $k^0>0$, 
we proceed as follows. 

We consider $\vec{k}\neq0$ and introduce orthonormal vectors 
$\vec{e}_1,\vec{e}_2,\vec{e}_3$ with
\be\label{5.2}
(\vec{e}_1 \times \vec{e}_2 ) \cdot \vec{e}_3 =1\,,
\qquad
\vec{e}_3 \parallel \vec{k} 
\ee
and define
\begin{equation}\label{5.3}
\epsilon^{(\pm)\mu} = \mp \frac{1}{\sqrt{2}}
\begin{pmatrix} 0 \\ \vec{e}_1 \pm i \vec{e}_2 \end{pmatrix}, \qquad
\epsilon^{(0)\mu} = \frac{1}{m_{f_2}}
\begin{pmatrix} |\vec{k}| \\ \vec{e}_3 k^0 \end{pmatrix}.
\end{equation}
A basis for the polarisation tensors of our tensor particle is given by
\be\label{5.1}
\begin{split}
\epsilon^{(m)\mu\nu}(k)=& \sum_{m_1,m_2} \langle 1,m_1;1,m_2|2,m\rangle
\epsilon^{(m_1)\mu}(k)
\epsilon^{(m_2)\nu}(k)\,,\\ 
&m=-2,...,2\,, \quad m_i=\pm1,0\,, \quad i=1,2\,.
\end{split}
\ee
Here $\langle  1,m_1;1,m_2|2,m\rangle$ are the usual Clebsch-Gordan 
coefficients. 
We have then
\begin{equation}\label{5.4}
\big(\epsilon_{\mu\nu}^{(m)}(k)\big)^* \epsilon^{(n)\mu\nu}(k) = \delta_{mn} \,.
\end{equation}
From \eqref{5.1} we get the spin sum as 
\begin{align}\label{5.5}
\sum_{m=-2}^2 & \epsilon_{\mu\nu}^{(m)}(k) 
\left(\epsilon_{\kappa\lambda}^{(m)}(k)\right)^* =
\frac{1}{2} \left( -g_{\mu\kappa}+\frac{k_\mu k_\kappa}{m_{f_2}^2} \right)
\left( -g_{\nu\lambda}+\frac{k_\nu k_\lambda}{m_{f_2}^2} \right)\\
&+\frac{1}{2} \left( -g_{\mu\lambda}+\frac{k_\mu k_\lambda}{m_{f_2}^2} \right)
\left( -g_{\nu\kappa}+\frac{k_\nu k_\kappa}{m_{f_2}^2} \right)
-\frac{1}{3} \left( -g_{\mu\nu}+\frac{k_\mu k_\nu}{m_{f_2}^2} \right)
\left( -g_{\kappa\lambda}+\frac{k_\kappa k_\lambda}{m_{f_2}^2} \right) \,.
\nn
\end{align}
The free propagator reads 
\begin{align}\label{5.101}
\Delta^{(T)}_{\mu\nu,\kappa\lambda} & (k)|_\textrm{free}= 
\frac{1}{k^2 - m_{f_2}^2 + i \epsilon} 
\left\{
\frac{1}{2} 
\left(-g_{\mu\kappa} + \frac{k_\mu k_\kappa}{m_{f_2}^2} \right) 
\left(-g_{\nu\lambda} + \frac{k_\nu k_\lambda}{m_{f_2}^2} \right) \right.
\\
& \left.
+\frac{1}{2} \left( -g_{\mu\lambda}+\frac{k_\mu k_\lambda}{m_{f_2}^2} \right)
\left( -g_{\nu\kappa}+\frac{k_\nu k_\kappa}{m_{f_2}^2} \right)
-\frac{1}{3} \left( -g_{\mu\nu}+\frac{k_\mu k_\nu}{m_{f_2}^2} \right)
\left( -g_{\kappa\lambda}+\frac{k_\kappa k_\lambda}{m_{f_2}^2} \right) 
\right\} \,.
\nn
\end{align}
This is well known; see for instance 
\cite{Sharp:1963zz,Weinberg:1964cn,Chang:1966zza,Bellucci:1994eb,Toublan:1995bk,Oh:2003aw}.

The full propagator \eqref{5.100} with $\phi_{\mu\nu}$ equal to 
the $f_2$ field should develop the resonance structure due to the $f_2$. 
But this structure should only appear in the true spin 2 part of the 
propagator \eqref{5.100}. Thus, we must now discuss the spin 
decomposition of the full propagator for $k^0>0$, $k^2 >0$. 
This is analogous to the decomposition of the vector propagator into 
transverse (spin 1) and longitudinal (spin 0) parts; 
see section \ref{Vector mesons}. 

The tensor-field propagator \eqref{5.100} contains, in general, spin 2 and 
spin 1 parts, and a $2 \times 2$ matrix as spin 0 part. 
The spin 1 and the spin 0 parts are due to the divergence of the field, 
$\del^\mu \phi_{\mu\nu}(x)$, and to $\del^\mu \del^\nu \phi_{\mu\nu}(x)$ 
and $g^{\mu\nu} \phi_{\mu\nu}(x)$, respectively. Thus, we can write 
down an expansion of the full propagator \eqref{5.100} in terms of 
tensors $P^{(i)}$, in essence projectors on the various spin components, 
times invariant functions $\Delta^{(i)}(k^2)$:
\be \label{5.102}
\Delta^{(T)}_{\mu\nu,\kappa\lambda}(k) = 
P^{(2)}_{\mu\nu,\kappa\lambda}(k) \Delta^{(2)}(k^2)
+ P^{(1)}_{\mu\nu,\kappa\lambda}(k) \Delta^{(1)}(k^2)
+ \sum_{a,b=1}^2 P^{(0,ab)}_{\mu\nu,\kappa\lambda}(k) \Delta^{(0)}_{ab} (k^2)
\,.
\ee
Here the spin 2 projector reads as in \eqref{5.5} but with $m_{f_2}^2$ 
replaced by $k^2 + i \epsilon$, 
\be\label{5.5ieps}
\begin{split}
P^{(2)}_{\mu\nu,\kappa\lambda}  (k) 
= & \frac{1}{2} \left( -g_{\mu\kappa}+\frac{k_\mu k_\kappa}{k^2+i \epsilon} \right)
\left( -g_{\nu\lambda}+\frac{k_\nu k_\lambda}{k^2+i \epsilon} \right)
\\
& 
+\frac{1}{2} \left( -g_{\mu\lambda}+\frac{k_\mu k_\lambda}{k^2+i \epsilon} \right)
\left( -g_{\nu\kappa}+\frac{k_\nu k_\kappa}{k^2+i \epsilon} \right)
\\
& -\frac{1}{3} \left( -g_{\mu\nu}+\frac{k_\mu k_\nu}{k^2+i \epsilon} \right)
\left( -g_{\kappa\lambda}+\frac{k_\kappa k_\lambda}{k^2+i \epsilon} \right) \,.
\end{split}
\ee
The other tensors $P^{(1)}$ and $P^{(0,ab)}$ are given explicitly in 
appendix \ref{appA}. 
For the tensor field corresponding to the $f_2$ meson, the $f_2$ pole should 
only appear in the invariant function $\Delta^{(2)}(k^2)$. In appendix 
\ref{appA} we present an analysis of $\Delta^{(2)}(k^2)$ along similar 
lines as done for $\Delta_T^{(\rho)}(k^2)$ in \cite{Melikhov:2003hs} and 
section \ref{Vector mesons}. The result is as follows where we set $k^2=s$: 
\be\label{5.104}
\left[ \Delta^{(2)} (s) \right]^{-1} = 
-m_{f_2}^2 + s + s \left[ R_{f_2}(s) - R_{f_2}(m_{f_2}^2) \right] 
+ i \, \mathrm{Im}\, B_{f_2}(s) \,,
\ee
with 
\begin{align}
\label{5.105}
\mathrm{Im}\,B_{f_2}(s) &= 
\frac{\Gamma_{f_2}}{\Gamma(f_2\to \pi\pi)} \frac{1}{320 \pi} 
\left|\frac{g_{f_2\pi\pi} F^{(f_2\pi\pi)}(s)}{M_0} \right|^2
s^2 \left(1-\frac{4m_\pi^2}{s}\right)^\frac{5}{2} \theta(s-4m_\pi^2) \,,
\\
\label{5.106}
R_{f_2}(s) &= \frac{s}{\pi} \,\mbox{V.\,P.}
\int_{4m_\pi^2}^\infty ds' \,
\frac{\mathrm{Im}\, B_{f_2}(s')}{s'^2 (s'-s)} \,.
\end{align}
Here V.\,P. means again the principal value prescription. 
A good representation of the $f_2$ propagator for, say, 
$0<k^2 < 10 \,\mbox{GeV}^2$ should then be given by 
\be\label{5.107}
i \Delta^{(f_2)}_{\mu\nu,\kappa\lambda} (k) = 
i P^{(2)}_{\mu\nu,\kappa\lambda} (k) \Delta^{(2)}(k^2) \,.
\ee
This is the propagator listed in \eqref{3.5}. Terms of the full tensor-field 
propagator \eqref{5.102} corresponding to spin 1 and spin 0 should not 
appear in \eqref{5.107} since these parts have nothing to do with the 
$f_2$ meson. 

\boldmath
\subsection{$f_2\to\gamma\gamma$}
\label{f2togammagamma}
\unboldmath

We discuss first the decay of an `on-shell' $f_2$ to two photons 
\be\label{5.14}
f_2(k,\epsilon) \longrightarrow \gamma(k_1, \epsilon_{\gamma,1})
+
\gamma(k_2, \epsilon_{\gamma,2})
\ee
where $k,k_1,k_2$ are the momenta, $\epsilon$ is the polarisation tensor 
of the $f_2$ and $\epsilon_{\gamma,1},\epsilon_{\gamma,2}$ are the 
polarisation vectors of the photons. Due to Bose symmetry of the photons 
and gauge invariance only two invariant amplitudes are allowed for 
\eqref{5.14}. The corresponding $f_2\gamma\gamma$ vertex is given 
in \eqref{3.29new} where for an `on-shell' $f_2$ and real photons all form 
factors equal 1. The coupling term generating this vertex is 
\begin{equation}\label{5.15}
\begin{split}
{\cal L}_{f_2\gamma\gamma}' = 
\left[
a_{f_2\gamma\gamma} (\partial_\kappa F_{\mu\nu})(\partial_\lambda F^{\mu\nu})
+ b_{f_2\gamma\gamma} F_{\mu\kappa} F_{\;\lambda}^\mu 
\right] 
\left(
g^{\kappa\kappa'} g^{\lambda\lambda'}
- \frac{1}{4}
g^{\kappa\lambda} g^{\kappa'\lambda'}
\right)
\phi_{\kappa'\lambda'}
\end{split}
\end{equation}
where $F_{\mu\nu}=\partial_\mu A_\nu-\partial_\nu A_\mu$ is the 
photon field-strength tensor. 

From \eqref{3.29new} we get for the ${\cal T}$-matrix elements of \eqref{5.14} 
in the $f_2$ rest system where $\vec{k}_2 = - \vec{k_1}$ 
\begin{equation}\label{5.15a}
\begin{split}
\langle \gamma(\vec{k}_1, \epsilon_{\gamma,1}^{(+)}),
\gamma(-\vec{k}_1, \epsilon_{\gamma,2}^{(+)}) |
{\cal T} | f_2(\epsilon^{(0)}) \rangle 
& =
\langle \gamma(\vec{k}_1, \epsilon_{\gamma,1}^{(-)}),
\gamma(-\vec{k}_1, \epsilon_{\gamma,2}^{(-)}) |
{\cal T} | f_2(\epsilon^{(0)}) \rangle  \\
& =
\frac{m_{f_2}^4}{\sqrt{6}} a_{f_2\gamma\gamma}\,,
\end{split}
\end{equation}
\begin{equation}
\label{5.16}
\begin{split}
\langle \gamma(\vec{k}_1, \epsilon_{\gamma,1}^{(+)}),
\gamma(-\vec{k}_1, \epsilon_{\gamma,2}^{(-)}) |
{\cal T} | f_2(\epsilon^{(2)}) \rangle 
& =
\langle \gamma(\vec{k}_1, \epsilon_{\gamma,1}^{(-)}),
\gamma(-\vec{k}_1, \epsilon_{\gamma,2}^{(+)}) |
{\cal T} | f_2(\epsilon^{(-2)}) \rangle  \\
& =
m_{f_2}^2 b_{f_2\gamma\gamma}\,.
\end{split}
\end{equation}
Here $\epsilon^{(m)}$ are the polarisation tensors of the $f_2$ at rest. 
These are constructed as in \eqref{5.2} to \eqref{5.1} with coordinate 
vectors $\vec{e}_1,\vec{e}_2,\vec{e}_3$ but choosing 
$\vec{e}_3=\vec{k}_1/|\vec{k}_1|$ as quantisation axis. 
The $\epsilon^{(\pm)}_{\gamma,i}$ ($i=1,2$) are the usual polarisation 
vectors of the photons. In the coordinate system considered we have
\begin{equation}\label{5.17}
\epsilon_{\gamma,1}^{(\pm)\mu} = \mp \frac{1}{\sqrt{2}}
\begin{pmatrix} 0 \\ \vec{e_1} \pm i\vec{e_2} \end{pmatrix}, \qquad
\epsilon_{\gamma,2}^{(\pm)\mu} = \mp \frac{1}{\sqrt{2}}
\begin{pmatrix} 0 \\ \vec{e_1} \mp i\vec{e_2} \end{pmatrix}.
\end{equation}
From \eqref{5.15a} and \eqref{5.16} we see that $a_{f_2\gamma\gamma}$ 
and $b_{f_2\gamma\gamma}$ parametrise the so-called helicity zero and 
helicity two $f_2\to\gamma\gamma$ amplitudes, respectively. In the basis 
$\epsilon^{(m)},\epsilon^{(n_i)}_{\gamma,i}$ ($i=1,2$) these are the only 
non-zero matrix elements. From this we get the decay rate
\begin{equation}\label{5.18}
\Gamma(f_2\to \gamma\gamma) =
\frac{m_{f_2}}{80 \pi}
\left[
\frac{1}{6} m_{f_2}^6 | a_{f_2\gamma\gamma}|^2
+ m_{f_2}^2 | b_{f_2\gamma\gamma}|^2
\right] \,.
\end{equation}
To obtain numbers for $a_{f_2\gamma\gamma}$ and $b_{f_2\gamma\gamma}$ 
we use the values
\be\label{5.19}
\begin{split}
&\Gamma(f_2 \to \gamma \gamma) = (3.14 \pm 0.20) \,\mbox{keV} \,,
\\
&\text{helicity zero contribution } \approx 13\% \:\mbox{of}\: 
\Gamma(f_2 \to \gamma \gamma) \,,
\end{split}
\ee
as quoted for the preferred solution A in \cite{Pennington:2008xd}. 
From this we obtain with the fine-structure constant $\alpha$ 
\begin{equation}\label{5.19a}
\frac{1}{6} | m_{f_2}^3 a_{f_2\gamma\gamma}|^2 +
|m_{f_2} b_{f_2\gamma\gamma}|^2 =
(11.62\pm0.74) \alpha^2
\end{equation}
and for the couplings
\begin{equation}\label{5.20}
|a_{f_2\gamma\gamma}| =\frac{e^2}{4\pi} 1.45~\text{GeV}^{-3}\,, \qquad
|b_{f_2\gamma\gamma}| =\frac{e^2}{4\pi} 2.49~\text{GeV}^{-1}
\end{equation}
with an estimated error of around $5\%$. From the arguments presented 
in section \ref{rho-proton scattering} below we conclude that we should have
\begin{equation}\label{5.21}
a_{f_2\gamma\gamma}>0\,, \qquad b_{f_2\gamma\gamma}>0\,.
\end{equation}
From \eqref{5.20} and \eqref{5.21} we get the numbers quoted in \eqref{3.29a}.

For off-shell $f_2$'s coupling to on- or off-shell photons we will again 
have to consider form factors. These are included in \eqref{3.29new}. 
Strictly speaking, this vertex should come with a single form factor 
$\tilde{F}^{(f_2\gamma\gamma)}(k^2,k_1^2,k_2^2)$. 
We assume that it can be approximated by a factorised expression 
\be\label{5.24a1}
\tilde{F}^{(f_2\gamma\gamma)}(k^2,k_1^2,k_2^2) \approx 
F_M(k_1^2) F_M(k_2^2) F^{(f_2\gamma\gamma)} (k^2) \,.
\ee
For lack of other information we will further set 
\be\label{5.24a}
F^{(f_2\gamma\gamma)} (k^2) = F^{(f_2\pi\pi)} (k^2) \,;
\ee
see \eqref{5.10a}. 

Finally we have compared our vertices for $f_2\pi\pi$ and $f_2\gamma\gamma$ 
in \eqref{3.28} and \eqref{3.29new}, respectively, to the ones used in 
\cite{Oh:2003aw}. For on-shell $f_2$'s we find a one to one correspondence. 
But we note that our $f_2\gamma\gamma$ vertex is strictly gauge invariant 
everywhere. That is, we get from \eqref{3.18} and \eqref{3.29new}
\be\label{5.24b}
\begin{split}
& k_1^\mu \Gamma_{\mu\nu\kappa\lambda}^{(f_2\gamma\gamma)} (k_1,k_2) =0 \,,
\\
& k_2^\nu \Gamma_{\mu\nu\kappa\lambda}^{(f_2\gamma\gamma)} (k_1,k_2) =0 \,.
\end{split}
\ee
This must be so since our coupling Lagrangian \eqref{5.15} is strictly 
gauge invariant. 

\boldmath
\subsection{$a_2\to\gamma\gamma$}
\label{a2togammagamma}
\unboldmath

The analysis of the vertices and the propagator of the $a_2 \equiv a_2(1320)$ 
meson could be done along the same lines as for the $f_2$. But we leave 
this for future work. Here we limit ourselves to making some remarks on the 
decay $a_2 \to \gamma\gamma$ which is the analogue of 
$f_2\to\gamma\gamma$ treated in section \ref{f2togammagamma}. 
The $a_2\gamma\gamma$ vertex function 
is as in \eqref{3.29new} with the constants $a_{f_2\gamma\gamma}$ and 
$b_{f_2\gamma\gamma}$ 
replaced by $a_{a_2\gamma\gamma}$ and 
$b_{a_2\gamma\gamma}$, respectively, which gives \eqref{3.29b}. 
The decay rate is
\begin{equation}\label{5.22}
\Gamma(a_2\to \gamma\gamma) =
\frac{m_{a_2}}{80 \pi}
\bigg\{
\frac{1}{6} m_{a_2}^6 | a_{a_2\gamma\gamma}|^2
+ m_{a_2}^2 | b_{a_2\gamma\gamma}|^2
\bigg\} \,.
\end{equation}
From \cite{Beringer:1900zz} we find
\begin{equation}\label{5.23}
m_{a_2} = 1318.3_{-0.6}^{+0.5}~\text{MeV}\,,\quad
\Gamma_{a_2} = 107 \pm 5~\text{MeV}\,,\quad
\frac{\Gamma(a_2\to \gamma\gamma)}{\Gamma_{a_2}} = (9.4\pm0.7 ) \times 10^{-6}\,.
\end{equation}
This gives 
\begin{equation}\label{5.24}
\frac{1}{6} | m_{a_2}^3 a_{a_2\gamma\gamma}|^2 +
|m_{a_2} b_{a_2\gamma\gamma}|^2 =
(3.60\pm 0.32) \alpha^2
\end{equation}
which is roughly a factor $3$ smaller than the corresponding quantity for 
the $f_2$; see \eqref{5.19a}. We discuss these coupling constants further 
in section \ref{rho-proton scattering} below. 

\boldmath
\section{Exchanges with Charge Conjugation \mbox{$C\!=\!+1$} and \mbox{$C\!=\!-1$} in Nucleon-Nucleon Scattering}
\label{Exchanges with charge conjugation}
\unboldmath

Let us consider $pp$ and $\bar pp$ elastic scattering
\begin{align}\label{6.1}
p(p_1) + p(p_2) &\longrightarrow p(p'_1) + p(p'_2)\,,
\\
\label{6.2}
\bar{p}(p_1) + p(p_2) &\longrightarrow \bar{p}(p'_1) + p(p'_2)\,.
\end{align}
The standard kinematic variables are 
\begin{equation}\label{6.3}
\begin{split}
&s = (p_1 + p_2)^2\,, \quad t = (p_1 - p_1')^2\,, \quad u = (p_1 - p_2')^2\,,\\
&\nu= \frac{1}{4}(s-u) = \frac{1}{4}(p_1+p_1') \cdot (p_2+p_2')\,.
\end{split}
\end{equation}
The ${\cal T}$-matrix elements for the reactions \eqref{6.1} and \eqref{6.2} 
can be expanded as follows 
\begin{align}\label{6.4}
\langle p(p_1'), p(p_2') | {\cal T} | p(p_1), p(p_2) \rangle &=
\bar{u}_{\alpha'}(p_1')
\bar{u}_{\beta'}(p_2')
\Gamma_{\alpha'\alpha,\beta'\beta}^{(pp)} (p_1', p_2', p_1, p_2)
u_\alpha(p_1)
u_\beta(p_2)\,,
\\
\label{6.5}
\langle \bar{p}(p_1'), p(p_2') | {\cal T} | \bar{p}(p_1), p(p_2) \rangle &=
\bar{v}_{\alpha}(p_1)
\bar{u}_{\beta'}(p_2')
\Gamma_{\alpha\alpha',\beta'\beta}^{(\bar{p}p)} (p_1', p_2', p_1, p_2)
v_{\alpha'}(p_1')
u_\beta(p_2)\,.
\end{align}
Here $\alpha,\beta,\alpha',\beta' \in \{1,\dots,4\}$ are Dirac indices. 
The substitution rule (crossing symmetry) requires
\begin{equation}\label{6.6}
\Gamma_{\alpha\alpha',\beta'\beta}^{(\bar{p}p)} (p_1', p_2', p_1, p_2)
= -\Gamma_{\alpha\alpha',\beta'\beta}^{(pp)} (-p_1, p_2', -p_1', p_2)\,.
\end{equation}
The $16\times 16$ matrices $\Gamma^{(pp)}$ and $\Gamma^{(\bar pp)}$ 
can be expanded in the basis 
$\gamma^\mu\otimes \gamma_\mu,~\gamma^\mu\gamma_5\otimes \gamma_\mu\gamma_5, \dots,~{\mathbbm 1}\otimes{\mathbbm 1}$. 
The standard assumption is that at high energies only the 
$\gamma^\mu\otimes \gamma_\mu$ structure is relevant; see 
\cite{Landshoff:1971pw,Donnachie:1983hf,Donnachie:1985iz,Donnachie:1987gu,Donnachie:2002en}. 
Following this assumption we can write 
\begin{align}\label{6.7}
\Gamma_{\alpha'\alpha,\beta'\beta}^{(pp)} (p_1', p_2', p_1, p_2)
&= T^{(pp)}(\nu,t) (\gamma^\mu)_{\alpha'\alpha} (\gamma_\mu)_{\beta'\beta} \,,
\\
\label{6.8}
\Gamma_{\alpha\alpha',\beta'\beta}^{(\bar{p}p)} (p_1', p_2', p_1, p_2)
&= T^{(\bar{p}p)}(\nu,t) (\gamma^\mu)_{\alpha\alpha'} (\gamma_\mu)_{\beta'\beta} \,. 
\end{align}
Inserting \eqref{6.7} and \eqref{6.8} in \eqref{6.4} and \eqref{6.5}, respectively, we get 
\begin{align}\label{6.8a}
\langle p(p_1'), p(p_2') | {\cal T} | p(p_1), p(p_2) \rangle &=
T^{(pp)}(\nu,t)\,
\bar{u}(p_1') \gamma^\mu u(p_1)
\bar{u}(p_2') \gamma_\mu u(p_2)\,,
\\
\label{6.8b}
\langle \bar{p}(p_1'), p(p_2') | {\cal T} | \bar{p}(p_1), p(p_2) \rangle &=
T^{(\bar{p}p)}(\nu,t)\,
\bar{v}(p_1) \gamma^\mu v(p_1')
\bar{u}(p_2') \gamma_\mu u(p_2)\,.
\end{align}

\begin{figure}[htb]
\begin{center}
\includegraphics[width=0.55\textwidth]{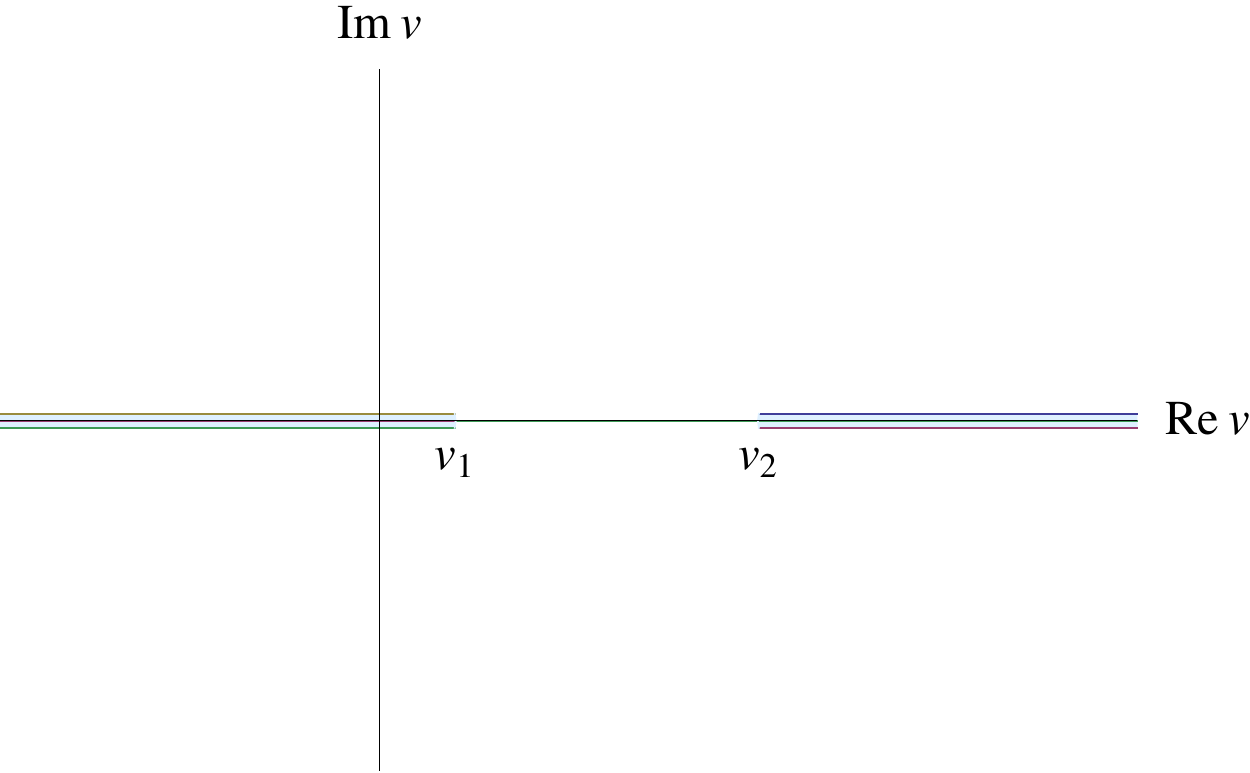}
\caption{The analytic structure of the amplitude $A(\nu,t)$ for fixed $t$ in the 
interval \eqref{6.10}. 
\label{model:fig7}}
\end{center}
\end{figure}

From QFT we know that, for fixed $t$, $T^{(pp)}(\nu,t)$ and $T^{(\bar pp)}(\nu,t)$ 
are boundary values of one analytic function in $\nu$, $A(\nu,t)$, with the 
cut structure shown in figure \ref{model:fig7}. 
The thresholds for the cuts are
\begin{equation}\label{6.9}
\nu_1 = m_p^2 -\frac{1}{2} m_\pi^2 - \frac{1}{4} t\,,\quad
\nu_2 = m_p^2 + \frac{1}{4} t\,.
\end{equation}
For 
\be\label{6.10}
0 \ge t > - m_\pi^2
\ee
we have
\be\label{6.12}
\nu_1 < \nu_2
\ee
and the amplitude $A(\nu,t)$ is then real in the interval
\be\label{6.13}
\nu_1 < \nu < \nu_2\,.
\ee
All this is easily seen from a study of the Landau singularities of the 
corresponding Feynman diagrams; see for instance \cite{Bjorken:1965zz} 
for an introduction to this topic. 
Then, for the $t$ interval \eqref{6.10}, we have 
\be\label{6.13a}
A^*(\nu^*,t) = A(\nu,t) \,.
\ee
The connection of $T^{(pp)},T^{(\bar pp)}$ to $A(\nu,t)$ is as follows. 
We have for real $\nu>\nu_2$
\begin{align}\label{6.14}
T^{(pp)} (\nu,t) &= \lim_{\epsilon \to 0} A(\nu+ i\epsilon,t)\,,
\\
\label{6.15}
T^{(\bar{p}p)} (\nu,t) &= -\lim_{\epsilon \to 0} A(-\nu- i\epsilon,t)\,.
\end{align}

So far all relations follow strictly from QFT. Now we shall invoke the 
Regge-pole model and assume that the amplitude $A(\nu,t)$ has 
a power behaviour in $\nu$ for $\nu\to\infty$. We shall treat $C=+1$ and 
$C=-1$ exchanges in turn. 

\subsection{Pomeron Exchange}
\label{Pomeron exchange}

There is a vast literature on the pomeron, for an overview see for 
instance \cite{Donnachie:2002en}. Let us here just mention some 
investigations of the pomeron in perturbative QCD 
\cite{Low:1975sv,Nussinov:1975mw,Gunion:1976iy,Kuraev:fs,Balitsky:ic,Lipatov:1996ts,Fadin:1998py,Ciafaloni:1998gs} 
and in non-perturbative QCD 
\cite{Landshoff:1986yj,Nachtmann:1991ua,Dosch:1992cu,Dosch:1994ym,Nachtmann:1996kt,Berger:1998gu,Shoshi:2002in,Meggiolaro:1996hf,Meggiolaro:2006mx,Giordano:2008ua}. 
In the present paper we shall adopt a modest phenomenological 
approach following the Regge-pole model of Donnachie and Landshoff 
\cite{Donnachie:1983hf,Donnachie:1985iz,Donnachie:1987gu,Donnachie:2002en}. 

Let us assume power behaviour for $A(\nu,t)$ for $\nu\to\infty$. 
The pomeron part of $A(\nu,t)$, corresponding to a $C=+1$ exchange, 
must be an odd function of $\nu$ for $\nu \to \infty$, see \eqref{6.14}, 
\eqref{6.15}, in order to give the same amplitudes for $pp$ and $\bar{p} p$ 
scattering. Therefore, our ansatz for the pomeron part of $A(\nu,t)$ reads 
\begin{equation}\label{6.16}
A^{\mathbbm P} (\nu,t) =
f(t)\; 2 \alpha_{\mathbbm P}'\; \nu\;
\left[ 4 \alpha_{\mathbbm P}'^2 (\nu_2 -\nu) (\nu-\nu_1) 
\right]^{\frac{1}{2} (\alpha_{\mathbbm P}(t) -2)}
\end{equation}
with $f(t)$ a real function. Here $\alpha_{{\mathbbm P}}(t)$ is as 
in \eqref{3.8} and we have inserted suitable factors 
$\alpha'_{{\mathbbm P}}=0.25$ GeV$^{-2}$ for dimensional reasons. 
The requirement that $A(\nu,t)$ is real in the interval \eqref{6.13} 
fixes the branches of the power functions to be taken in \eqref{6.16}. 
That is, we require that $A^{{\mathbbm P}}(\nu,t)$ has exactly the 
cut structure shown in figure \ref{model:fig7}. We obtain then for $\nu$ real, 
$\nu\to\infty$ from \eqref{6.14} and \eqref{6.15}
\begin{equation}\label{6.17}
T^{(pp) {\mathbbm P}}(\nu,t) =
T^{(\bar{p}p) {\mathbbm P}}(\nu,t) =
f(t)\, 2 \alpha_{\mathbbm P}'\, \nu\,
\left[ 4 \alpha_{\mathbbm P}'^2 \nu^2 e^{-i\pi} 
\right]^{\frac{1}{2}(\alpha_{\mathbbm P}(t)-2)}\,. 
\end{equation}
Setting now
\begin{equation}\label{6.18}
f(t) = [ 3 \beta_{{\mathbbm P}NN} F_1(t) ]^2
\end{equation}
with $\beta_{{\mathbbm P}NN}$ and $F_1(t)$ as in \eqref{3.30} and using 
in the high-energy limit 
\be\label{6.19}
\nu \approx \frac{s}{2} 
\ee
we get the celebrated Donnachie-Landshoff (DL) pomeron ansatz from 
\eqref{6.8a}, \eqref{6.8b} and \eqref{6.17} to \eqref{6.19}
\begin{align}\label{6.20}
\langle p(p_1'), p(p_2') | {\cal T} | p(p_1), p(p_2) \rangle\big|_{\mathbbm P} =&\,
i \,[ 3 \beta_{{\mathbbm P}NN} F_1(t) ]^2 
(- i s \alpha_{\mathbbm P}' )^{\alpha_{\mathbbm P}(t) -1} \nn \\
&\times \bar{u}(p_1') \gamma^\mu u(p_1)
\bar{u}(p_2') \gamma_\mu u(p_2)\,,
\\
\label{6.21}
\langle \bar{p}(p_1'), p(p_2') | {\cal T} | \bar{p}(p_1), p(p_2) \rangle\big|_{\mathbbm P} =&\,
i \,[ 3 \beta_{{\mathbbm P}NN} F_1(t) ]^2 
(- i s \alpha_{\mathbbm P}' )^{\alpha_{\mathbbm P}(t) -1} \nn\\
& \times \bar{v}(p_1) \gamma^\mu v(p_1')
\bar{u}(p_2') \gamma_\mu u(p_2)\,,
\end{align}
where the subscript ${\mathbbm P}$ indicates that we consider only 
the pomeron contribution. Here we have to set 
$-i = \exp(-i\pi/2)$ in the power functions. 

The $\gamma^\mu\otimes\gamma_\mu$ structure in \eqref{6.20} 
and \eqref{6.21} suggests to consider the pomeron as some effective 
vector exchange. And, indeed, the DL pomeron is frequently called 
a `$C=+1$ photon'. But this presents severe problems from the point 
of view of QFT. A QFT vector will couple to the proton and the antiproton 
with {\em opposite} signs, giving a relative minus sign between \eqref{6.20} 
and \eqref{6.21}. 
Of course, this is unacceptable since it would lead to 
the total cross sections for $pp$ and $\bar pp$ scattering having 
opposite signs. It is well known that a Regge-pole exchange as 
in \eqref{6.20} and \eqref{6.21} corresponds to the coherent sum of 
the exchanges of infinitely many spins. An explicit construction 
showing how \eqref{6.20} and \eqref{6.21} can be written in terms of 
a sum of quantum-field-theoretic exchanges with spin $2,4,6$ etc.\ 
can be found in section 6.2 of \cite{Nachtmann:1991ua}. But infinite 
sums are in general cumbersome to handle. We shall show in the 
following how we can write the pomeron exchange in \eqref{6.20} 
and \eqref{6.21} as an effective spin $2$ exchange satisfying the 
standard QFT crossing requirements. 
In appendix \ref{appB} we show how this effective spin 2 exchange 
can again be written as a coherent sum of exchanges of spin 2, 4, 6, etc. 
We also explain there why the pomeron should have nothing to do 
with an effective spin 0 exchange. 

Let us consider the reactions 
\eqref{6.1} and \eqref{6.2} in the c.\,m.\ system and let 
$\lambda_1,\lambda_2,\lambda'_1,\lambda'_2\in\{1/2,-1/2\}$ 
be the helicities of the nucleons. For high energies, $s\to\infty$, we have 
\begin{equation}\label{6.23}
\bar{u}_{\lambda'}(p') \gamma^\mu u_\lambda(p) =
\bar{v}_{\lambda}(p) \gamma^\mu v_{\lambda'}(p') 
\longrightarrow
(p'+p)^\mu \delta_{\lambda'\lambda}\,.
\end{equation}
Inserting this in \eqref{6.20} and \eqref{6.21} we get at high energies
\begin{equation}\label{6.24}
\begin{split}
\langle p(p_1',\lambda_1'), &\,p(p_2',\lambda_2') | {\cal T} | p(p_1,\lambda_1), p(p_2,\lambda_2) \rangle\big|_{\mathbbm P} 
\\
=&\,\langle \bar{p}(p_1',\lambda_1'), p(p_2',\lambda_2') | {\cal T} | \bar{p}(p_1,\lambda_1), p(p_2,\lambda_2) \rangle\big|_{\mathbbm P} 
\\
=&\, i \, 2 s\, [ 3 \beta_{{\mathbbm P}NN} F_1(t) ]^2 
(- i s \alpha_{\mathbbm P}' )^{\alpha_{\mathbbm P}(t) -1} 
\delta_{\lambda_1'\lambda_1}\delta_{\lambda_2'\lambda_2} \,.
\end{split}
\end{equation}

We shall now show that {\em exactly} the same expression is obtained 
in the high-energy limit by considering the pomeron as a symmetric 
traceless rank $2$ tensor object
\be\label{6.25}
{\mathbbm P}_{\mu\nu}(x) = {\mathbbm P}_{\nu\mu}(x)\,, \qquad
{\mathbbm P}_{\mu\nu}(x) g^{\mu\nu} =0 
\ee
and making suitable ans{\"a}tze for the ${\mathbbm P} pp$ vertex 
and the ${\mathbbm P}$ propagator. 
For the coupling Lagrangian of the tensor pomeron to proton and antiproton 
we make the ansatz
\be\label{6.26}
{\cal L}'_{{\mathbbm P}pp}(x) =
- 3\beta_{{\mathbbm P}NN} {\mathbbm P}_{\mu\nu}(x)
\frac{i}{2} \bar{\psi}_p(x)
\left[ 
\gamma^\mu \stackrel{\leftrightarrow}{\partial^\nu} 
+ \gamma^\nu \stackrel{\leftrightarrow}{\partial^\mu}
- \frac{1}{2} g^{\mu\nu} \gamma^\lambda
\stackrel{\leftrightarrow}{\partial_\lambda} 
\right] \psi_p(x)
\ee
where $\psi_p(x)$ is the proton field operator. With this tensor coupling 
we get in the standard way from QFT the ${\mathbbm  P} pp$ and 
${\mathbbm  P}\bar{p}\bar{p}$ vertices listed in \eqref{3.30}. 
Our ansatz for the effective pomeron propagator is given 
in \eqref{3.7}, \eqref{3.8}. 
We get now the amplitudes and, using also \eqref{6.23}, their expressions 
for large $s$ as follows: 
\begin{align}\label{6.27}
\langle p(p_1',\lambda_1'), &\,p(p_2',\lambda_2') | {\cal T} | p(p_1,\lambda_1), p(p_2,\lambda_2) \rangle\big|_{\mathbbm P} 
\nn \\
=&\, i [ 3 \beta_{{\mathbbm P}NN} F_1(t) ]^2 
(- i s \alpha_{\mathbbm P}' )^{\alpha_{\mathbbm P}(t) -1} \nn\\
&
\times \bar{u}(p_1',\lambda_1')
\left[
\frac{1}{2} \gamma^\mu (p_1'+ p_1)^\nu +
\frac{1}{2} \gamma^\nu (p_1'+ p_1)^\mu 
\right] u(p_1,\lambda_1)
\nn\\
&
\times \frac{1}{4s}
\left(
g_{\mu\kappa} g_{\nu\lambda} +
g_{\mu\lambda} g_{\nu\kappa} -
\frac{1}{2}
g_{\mu\nu} g_{\kappa\lambda} 
\right) \\
&
\times \bar{u}(p_2',\lambda_2')
\left[
\frac{1}{2} \gamma^\kappa (p_2'+ p_2)^\lambda +
\frac{1}{2} \gamma^\lambda (p_2'+ p_2)^\kappa 
\right] u(p_2,\lambda_2) 
\nn\\
& \!\!\!\!\!\!\!\!\xrightarrow[s \to \infty ]{}  i 2 s\,
[ 3 \beta_{{\mathbbm P}NN} F_1(t) ]^2 \;
(- i s \alpha_{\mathbbm P}' )^{\alpha_{\mathbbm P}(t) -1} \;
\delta_{\lambda_1'\lambda_1}
\delta_{\lambda_2'\lambda_2}\,,
\nn
\end{align}
\begin{align}
\label{6.28}
\langle \bar{p}(p_1',\lambda_1'),&\, p(p_2',\lambda_2') | 
{\cal T} | \bar{p}(p_1,\lambda_1), p(p_2,\lambda_2) \rangle\big|_{\mathbbm P} 
\nn\\
=&\, i [ 3 \beta_{{\mathbbm P}NN} F_1(t) ]^2 
(- i s \alpha_{\mathbbm P}' )^{\alpha_{\mathbbm P}(t) -1} \nn\\
&
\times \bar{v}(p_1,\lambda_1)
\left[
\frac{1}{2} \gamma^\mu (p_1'+ p_1)^\nu +
\frac{1}{2} \gamma^\nu (p_1'+ p_1)^\mu 
\right] v(p_1',\lambda_1') 
\nn\\
&
\times \frac{1}{4s}
\left(
g_{\mu\kappa} g_{\nu\lambda} +
g_{\mu\lambda} g_{\nu\kappa} -
\frac{1}{2}
g_{\mu\nu} g_{\kappa\lambda} 
\right) \\
&
\times \bar{u}(p_2',\lambda_2')
\left[
\frac{1}{2} \gamma^\kappa (p_2'+ p_2)^\lambda +
\frac{1}{2} \gamma^\lambda (p_2'+ p_2)^\kappa 
\right] u(p_2,\lambda_2) 
\nn\\
& \!\!\!\!\!\!\!\!\xrightarrow[s \to \infty ]{} 
i 2 s\,
[ 3 \beta_{{\mathbbm P}NN} F_1(t) ]^2 \;
(- i s \alpha_{\mathbbm P}' )^{\alpha_{\mathbbm P}(t) -1} \;
\delta_{\lambda_1'\lambda_1}
\delta_{\lambda_2'\lambda_2}\,.
\nn
\end{align}
Comparing \eqref{6.27} and \eqref{6.28} with \eqref{6.24} we find in the 
high energy limit complete agreement of the amplitudes calculated 
from our tensor-pomeron exchange with the standard DL amplitudes. 

To summarise: in QFT a second rank tensor -- like for gravity -- gives 
the same sign for the coupling of particles and of antiparticles. Thus, 
our tensor pomeron has automatically the {\em same} coupling to 
$p$ and $\bar p$ as required by phenomenology. The resulting expressions 
for the amplitudes of elastic $pp$ and $\bar pp$ scattering are for 
$s\to\infty$ {\em exactly} as for the DL pomeron if the effective 
tensor-pomeron propagator and the couplings to the nucleons are 
chosen appropriately. This, together with the requirement of standard 
QFT structures, gives the justification {\sl a posteriori} of the 
ans{\"a}tze \eqref{3.7} and \eqref{3.30}. 
Thus, one may ask if it pays to consider the pomeron as a tensor object. 
We shall show that once we consider the coupling of vector mesons 
to the pomeron, viewing the pomeron as a tensor object, 
presents enormous advantages, see section \ref{7.2} below. 

\subsection{Odderon Exchange}
\label{Odderon exchange}

The odderon -- so far only seen clearly in theoretical papers -- is a $C=-1$ 
exchange object. The corresponding amplitude $A^{\mathbbm O}(\nu,t)$ 
must be an even function of $\nu$ for $\nu \to \infty$ in order to give opposite signs for the 
amplitudes of $pp$ and $\bar{p}p$ scattering, see \eqref{6.14}, \eqref{6.15}, 
and it must again have the cut structure shown 
in figure \ref{model:fig7}. We make, as for the pomeron amplitude 
in \eqref{6.16}, a power law ansatz for $A^{\mathbbm O}$, 
\begin{equation}\label{6.29}
A^{\mathbbm O} (\nu,t) =
\big[ 3 \beta_{{\mathbbm O}pp} F_1(t) \big]^2
\eta_{\mathbbm O}
\left[ 4 \alpha_{\mathbbm O}'^2 (\nu_2 -\nu) (\nu-\nu_1) \right]^{\frac{1}{2} (\alpha_{\mathbbm O}(t) -1)},
\quad
\eta_{\mathbbm O} = \pm 1\,.
\end{equation}
Here the coupling constant $\beta_{{\mathbbm O} pp}$ of dimension (mass)$^{-1}$, 
the factor $\eta_{\mathbbm O}$, and the trajectory function 
$\alpha_{\mathbbm O}(t)$ are, of course, unknown. They are to be taken 
as free parameters, hopefully to be determined some day by experiment. 
For dimensional reasons we have inserted the factor 
$\alpha'^2_{{\mathbbm O}}$ in \eqref{6.29}. In \eqref{3.14} we have made 
an ansatz for $\alpha_{\mathbbm O}(t)$ as a linear function of $t$
\be\label{6.29a}
\alpha_{\mathbbm O}(t) = 1 + \epsilon_{\mathbbm O} +\alpha_{\mathbbm O}'\,t\,.
\ee
For a `decent' odderon we should have
\be\label{6.29b}
|\epsilon_{\mathbbm O}| \ll 1\,.
\ee
Of course, the pomeron amplitude must dominate over the odderon one 
for $s\to\infty$ in order to ensure positive total $pp$ and $\bar pp$ 
cross sections. Thus, we must require
\be\label{6.29c}
\epsilon_{\mathbbm O} \le \epsilon_{\mathbbm P}\,;
\ee
see \eqref{3.8} and \eqref{3.14}. The factor $\eta_{\mathbbm O}$ 
remains as a free parameter. For lack of other information we shall 
set $\alpha'_{\mathbbm O}= \alpha'_{\mathbbm P}=0.25 \,\mbox{GeV}^{-2}$ 
as already listed in \eqref{3.14}. We could also have used a scale factor 
different from $4\alpha'^2_{\mathbbm O}$ in \eqref{6.29}. 

From \eqref{6.8a} to \eqref{6.15} and \eqref{6.29} we obtain for 
high energies, $s\to\infty$,
\begin{align}\label{6.30}
\langle p(p_1'), p(p_2') | {\cal T} | p(p_1), p(p_2) \rangle\big|_{\mathbbm O} =&\,
[ 3 \beta_{{\mathbbm O}pp} F_1(t) ]^2 
\eta_{\mathbbm O}
(- i s \alpha_{\mathbbm O}' )^{\alpha_{\mathbbm O}(t) -1} \nn\\
& \times\bar{u}(p_1') \gamma^\mu u(p_1)
\bar{u}(p_2') \gamma_\mu u(p_2)\,,
\\
\nn
\label{6.31}
\langle \bar{p}(p_1'), p(p_2') | {\cal T} | \bar{p}(p_1), p(p_2) \rangle\big|_{\mathbbm O} =&\,
- [ 3 \beta_{{\mathbbm O}pp} F_1(t) ]^2 
\eta_{\mathbbm O}
(- i s \alpha_{\mathbbm O}' )^{\alpha_{\mathbbm O}(t) -1} \nn \\
& \times \bar{v}(p_1) \gamma^\mu v(p_1')
\bar{u}(p_2') \gamma_\mu u(p_2)\,.
\end{align}
The relative minus sign between the $pp$ and the $\bar pp$ amplitudes is, 
of course, what is required for a $C=-1$ exchange. Thus we see that the 
odderon may be considered as an effective vector exchange in the sense 
of QFT. The corresponding coupling Lagrangian can be taken as 
\be\label{6.32}
{\cal L}_{{\mathbbm O}pp}'(x) =
-3 M_0 \beta_{{\mathbbm O}pp}
{\mathbbm O}_\mu(x) \,
\bar{\psi}_p(x) \gamma^\mu \psi_p(x)
\ee
where, for dimensional reason, 
we have inserted a factor $M_0\equiv 1\,\mbox{GeV}$ to make the coupling 
constant $M_0\beta_{{\mathbbm O} pp}$ dimensionless. From \eqref{6.32} 
we get in the standard way of QFT using also isospin symmetry
the ${\mathbbm O} NN$ vertices \eqref{3.46}. 
Choosing the effective odderon propagator according 
to \eqref{3.13}, \eqref{3.14} gives then exactly the $pp$ and $\bar pp$ 
amplitudes \eqref{6.30} and \eqref{6.31}, respectively.

To summarise: the odderon may be considered as effective vector 
exchange in the sense of QFT; see also \cite{Ewerz:2003xi} and 
sections 6.2 and 6.3 of \cite{Nachtmann:1991ua}. 
This makes the odderon in some sense 
theoretically simpler than the pomeron. The odderon is supposed to 
be an isoscalar object. This fixes the ${\mathbbm O}nn$ and 
${\mathbbm O}pp$ vertices in \eqref{3.46} to be equal. 
While the general structure of the effective odderon propagator 
\eqref{3.13} is fixed by the odderon's vector property the details 
are completely open. A linear odderon trajectory \eqref{3.14} and 
a scale factor equal to $\alpha'_{\mathbbm O}$ in the power law 
in \eqref{3.13} are just guesses to be confirmed or refuted 
by experiment. 

Now we discuss the ${\mathbbm O}\gamma f_2$ coupling. For lack 
of other information we make for the effective coupling Lagrangian 
an ansatz in analogy to the one for $f_2 \gamma \gamma$ in \eqref{5.15}. 
We get then 
\be\label{6.37a}
\begin{split}
{\cal L}_{{\mathbbm O}\gamma f_2}' = &
\left[ a_{{\mathbbm O}\gamma f_2} (\del_\kappa F_{\mu\nu}) 
\del_\lambda (\del^\mu {\mathbbm O}^\nu - \del^\nu {\mathbbm O}^\mu )
+ b_{{\mathbbm O}\gamma f_2} F_{\mu\kappa} 
(\del^\mu {\mathbbm O}_\lambda -\del_\lambda {\mathbbm O}^\mu)
+ (\kappa \leftrightarrow \lambda) \right]
\\
&\times \left[ g^{\kappa \kappa'} g^{\lambda \lambda'} 
- \frac{1}{4} g^{\kappa \lambda} g^{\kappa' \lambda'}
\right] \phi_{\kappa' \lambda'} \,.
\end{split}
\ee
Since we have a coupling involving one photon we define also coupling 
parameters where one factor of $e$ is taken out: 
\be\label{6.37b}
\hat{a}_{{\mathbbm O}\gamma f_2} = \frac{a_{{\mathbbm O}\gamma f_2}}{e}
\qquad
\hat{b}_{{\mathbbm O}\gamma f_2} = \frac{b_{{\mathbbm O}\gamma f_2}}{e}
\,.
\ee
These $\hat{a}_{{\mathbbm O}\gamma f_2}$ and 
$\hat{b}_{{\mathbbm O}\gamma f_2}$ should then be of `normal' size 
on the hadronic scale. The ${\mathbbm O}\gamma f_2$ vertex is easily 
obtained from \eqref{6.37a} and is listed in \eqref{3.48} where appropriate 
form factors are included. 

\subsection{Reggeon Exchanges and Total Cross Sections}
\label{Reggeon exchanges and total cross sections}

It is well known that in high-energy nucleon-nucleon scattering 
the subdominant terms after pomeron exchange are due to the 
reggeon exchanges $f_{2R},a_{2R}, \omega_R$, and $\rho_R$; 
see for instance \cite{Donnachie:2002en}. The reggeons $f_{2R}$ 
and $a_{2R}$ have charge conjugation $C=+1$, the reggeons 
$\omega_R$ and $\rho_R$ have $C=-1$. We construct, therefore, 
effective propagators and vertices for these $C=+1$ and $C=-1$ 
reggeons in complete analogy to those of the pomeron and the 
odderon, respectively.

Let us start with the $C=+1$ exchanges $f_{2R}$ and $a_{2R}$. 
We consider them as effective spin 2 exchanges. 
In analogy to pomeron exchange we 
make for the effective ${\mathbbm R}_+$ propagators and 
${\mathbbm R}_{+}NN$ vertices (${\mathbbm R}_+=f_{2R},a_{2R}$)
the ans\"atze \eqref{3.9}, \eqref{3.33}, and \eqref{3.34}. Here we 
are taking into account that $f_{2R}$ is an isoscalar and $a_{2R}$ 
the third component of an isovector. We assume the so-called 
degeneracy to hold, that is, equal trajectories for $f_{2R}$ 
and $a_{2R}$. This is a good representation of the experimental 
findings; see \cite{Donnachie:2002en}. 
The numbers quoted for this ${\mathbbm R}_+$ trajectory 
in \eqref{3.10} are taken from figure 2.8 and section 3.1 of \cite{Donnachie:2002en}. 
The overall sign factor of the ${\mathbbm R}_+$ propagators can either 
be taken from experiment, see below, or from the signature-factor argument. 
For the latter see for instance (2.18) and (3.11) of \cite{Donnachie:2002en} 
and (6.8.15) of \cite{Collins:1977jy}. The values for the couplings given 
in \eqref{3.33A} and \eqref{3.34A} are discussed below. 

We treat the $C=-1$ reggeons ${\mathbb R}_-=\omega_R,\rho_R$ as 
effective vector exchanges. Taking their isospin quantum numbers into 
account we obtain the propagators \eqref{3.11} and the vertices 
\eqref{3.38} and \eqref{3.40}. The overall signs of the ${\mathbbm R}_-$ 
propagators can be determined from experiment or from the signature-factor 
argument. Again we assume degeneracy of the $\omega_R$ and $\rho_R$ 
trajectories and take the parameters 
of the ${\mathbbm R}_-$ trajectory from figure 2.8 and section 3.1 
of \cite{Donnachie:2002en}. 
As is shown there, the parameters of the ${\mathbbm R}_+$ 
and ${\mathbbm R}_-$ trajectories can be taken as equal. Of course, 
in our framework it would be easy to relax this assumption if so 
required by experiment. The values for the couplings in \eqref{3.38A} 
and \eqref{3.40A} are discussed below. 

Now we can write down the expressions for the elastic amplitudes 
and cross sections for nucleon-nucleon scattering at high energies. 
Here we shall take into account the hadronic exchanges and neglect 
$\gamma$ exchange which is only relevant for very small $|t|$. 
Also the data which we shall consider refer only to the strong-interaction 
parts of the amplitudes. 
We get the following, using \eqref{6.23}, for $pp$ and $pn$ scattering. 
For ease of writing we set for $p$ and $n$ also $N(I_3)$ with 
$I_3=+1/2$ and $-1/2$, respectively. 
\begin{align}\label{6.33}
\langle p(p_1', \lambda_1'), N(I_3,p_2',\lambda_2') | {\cal T} | p&(p_1,\lambda_1), N(I_3,p_2,\lambda_2) \rangle 
\nn\\
=i\,2 s\,\delta_{\lambda_1'\lambda_1}\delta_{\lambda_2'\lambda_2}
[F_1(t)]^2 
\bigg\{ 
&\,(3 \beta_{{\mathbbm P}NN})^2
(- i s \alpha_{\mathbbm P}' )^{\alpha_{\mathbbm P}(t) -1} 
\nn\\
&
+ \left[ (g_{f_{2R}pp})^2 + (-1)^{I_3-\frac{1}{2}}(g_{a_{2R}pp})^2 \right]
M_0^{-2} 
(- i s \alpha_{{\mathbbm R}_+}')^{\alpha_{{\mathbbm R}_+}(t) -1} 
\nn\\
&
+ \left[ (g_{\omega_{R}pp})^2 + (-1)^{I_3-\frac{1}{2}}(g_{\rho_{R}pp})^2 \right]
M_-^{-2} i\;
(- i s \alpha_{{\mathbbm R}_-}')^{\alpha_{{\mathbbm R}_-}(t) -1} 
\nn\\
&
+ (3 \beta_{{\mathbbm O}pp})^2
(-i \eta_{\mathbbm O})
(- i s \alpha_{\mathbbm O}' )^{\alpha_{\mathbbm O}(t) -1} 
\bigg\}.
\end{align}
For $\bar pp$ and $\bar pn$ scattering we obtain 
\begin{align}\label{6.34}
\langle \bar{p}(p_1', \lambda_1'), N(I_3,p_2',\lambda_2') | {\cal T} | \bar{p}&(p_1,\lambda_1), N(I_3,p_2,\lambda_2) \rangle 
\nn\\
= i\,2 s\, \delta_{\lambda_1'\lambda_1}\delta_{\lambda_2'\lambda_2}
[F_1(t)]^2 
\bigg\{
&\,(3 \beta_{{\mathbbm P}NN})^2
(- i s \alpha_{\mathbbm P}' )^{\alpha_{\mathbbm P}(t) -1} 
\nn\\
&
+ \left[ (g_{f_{2R}pp})^2 + (-1)^{I_3-\frac{1}{2}}(g_{a_{2R}pp})^2 \right]
M_0^{-2} 
(- i s \alpha_{{\mathbbm R}_+}')^{\alpha_{{\mathbbm R}_+}(t) -1} 
\nn\\
&
- \left[ (g_{\omega_{R}pp})^2 + (-1)^{I_3-\frac{1}{2}}(g_{\rho_{R}pp})^2 \right]
M_-^{-2} i
(- i s \alpha_{{\mathbbm R}_-}')^{\alpha_{{\mathbbm R}_-}(t) -1} 
\nn\\
&
- (3 \beta_{{\mathbbm O}pp})^2
(-i \eta_{\mathbbm O})
(- i s \alpha_{\mathbbm O}' )^{\alpha_{\mathbbm O}(t) -1} 
\bigg\}.
\end{align}
For the total cross sections of $pp,pn,\bar pp$, and $\bar pn$ scattering 
we obtain from the optical theorem for high energies
\begin{align}\label{6.35}
\sigma_{\rm tot}(p,&\,N(I_3))  =
\frac{1}{4s} 
\sum_{\lambda_1, \lambda_2}
\text{Im}
\langle p(p_1, \lambda_1), N(I_3,p_2,\lambda_2) | {\cal T} | p(p_1,\lambda_1), N(I_3,p_2,\lambda_2) \rangle
\nn\\
=2 \bigg\{&
(3\beta_{{\mathbbm P}NN})^2 \cos\left[\frac{\pi}{2}(\alpha_{\mathbbm P}(0) -1 )\right]
(s \alpha_{\mathbbm P}' )^{\alpha_{\mathbbm P}(0) -1}
\nn\\
& 
+ \left[ (g_{f_{2R}pp})^2 + (-1)^{I_3-\frac{1}{2}}(g_{a_{2R}pp})^2 \right]
M_0^{-2} \cos\left[\frac{\pi}{2}(\alpha_{{\mathbbm R}_+}(0) -1 )\right]
(s \alpha_{{\mathbbm R}_+}')^{\alpha_{{\mathbbm R}_+}(0) -1} 
\nn\\
&
- \left[ (g_{\omega_{R}pp})^2 + (-1)^{I_3-\frac{1}{2}}(g_{\rho_{R}pp})^2 \right]
M_-^{-2} \cos\left[\frac{\pi}{2}\alpha_{{\mathbbm R}_-}(0)\right]
(s \alpha_{{\mathbbm R}_-}')^{\alpha_{{\mathbbm R}_-}(0) -1} 
\nn\\
&
+ \eta_{\mathbbm O}
( 3 \beta_{{\mathbbm O}pp})^2
\cos\left[\frac{\pi}{2}\alpha_{{\mathbbm O}}(0)\right]
(s \alpha_{{\mathbbm O}}')^{\alpha_{{\mathbbm O}}(0) -1} 
\bigg\},
\end{align}
\begin{align}\label{6.36}
\sigma_{\rm tot}(\bar{p},&\,N(I_3)) 
\nn\\
= 2 \bigg\{&
(3\beta_{{\mathbbm P}NN})^2 \cos\left[\frac{\pi}{2}(\alpha_{\mathbbm P}(0) -1 )\right]
(s \alpha_{\mathbbm P}' )^{\alpha_{\mathbbm P}(0) -1}
\nn\\
& 
+ \left[ (g_{f_{2R}pp})^2 + (-1)^{I_3-\frac{1}{2}}(g_{a_{2R}pp})^2 \right]
M_0^{-2} \cos\left[\frac{\pi}{2}(\alpha_{{\mathbbm R}_+}(0) -1 )\right]
(s \alpha_{{\mathbbm R}_+}')^{\alpha_{{\mathbbm R}_+}(0) -1} 
\nn\\
&
+ \left[ (g_{\omega_{R}pp})^2 + (-1)^{I_3-\frac{1}{2}}(g_{\rho_{R}pp})^2 \right]
M_-^{-2} \cos\left[\frac{\pi}{2}\alpha_{{\mathbbm R}_-}(0)\right]
(s \alpha_{{\mathbbm R}_-}')^{\alpha_{{\mathbbm R}_-}(0) -1} 
\nn\\
&
- \eta_{\mathbbm O}
( 3 \beta_{{\mathbbm O}pp})^2
\cos\left[\frac{\pi}{2}\alpha_{{\mathbbm O}}(0)\right]
(s \alpha_{{\mathbbm O}}')^{\alpha_{{\mathbbm O}}(0) -1} 
\bigg\}.
\end{align}

In figures 3.1 and 3.2 of \cite{Donnachie:2002en} the following fits 
to the experimental total cross sections are presented
\ba\label{6.37}
\sigma_\mathrm{tot} (a,b) = X_{ab} \left( \frac{s}{M_0^2}\right)^{0.0808} 
+ Y_{ab} \left( \frac{s}{M_0^2}\right)^{-0.4525}  \,.
\ea
Here $(a,b)=(p,p),(p,n),(\bar p,p),(\bar p,n)$, and $M_0=1\, \mbox{GeV}$. 
The numbers given in \cite{Donnachie:2002en} are 
\be\label{6.38}
X_{ab} \equiv X = 21.70 \,\mbox{mb} \:\,\widehat{=} \:\,55.73 \,\mbox{GeV}^{-2}
\ee
and for $Y_{ab}$ as listed in table \ref{model:table2}. 
\begin{table}[ht]\centering
\renewcommand{\arraystretch}{1.2}
\begin{tabular}{cc|c|c}
$a$ & $b$ & \multicolumn{2}{c}{$Y_{ab}$}\\
\hline
$p$ & $p$ & 56.08 mb & 144.02 GeV$^{-2}$\\
$p$ & $n$ & 54.77 mb & 140.66 GeV$^{-2}$\\ 
$\bar p$ & $p$ & 98.39 mb & 252.68 GeV$^{-2}$\\ 
$\bar p$ &  $n$ & 92.71 mb & 238.10 GeV$^{-2}$
\end{tabular}
\caption{\label{model:table2}Values for $Y_{ab}$ in \eqref{6.37} from figures 3.1 and 3.2 
of \cite{Donnachie:2002en}.}
\end{table}

An odderon contribution is apparently not needed to fit these total 
cross section data. But we note that for $\alpha_{{\mathbbm O}}(0)\approx 1$ 
we have 
\be\label{6.39}
\cos \left[\frac{\pi}{2} \alpha_{\mathbbm{O}}(0) \right] \approx 0 \,.
\ee
Thus, as is well known, an odderon with $\alpha_{{\mathbbm O}}(0)\approx 1$ 
will hardly be visible in total cross sections; see \eqref{6.35} and \eqref{6.36}. 

Comparing now \eqref{6.35} and \eqref{6.36} with \eqref{6.37} we can first 
of all extract the intercepts of the pomeron, ${\mathbbm R}_+$, 
and ${\mathbbm R}_-$ trajectories, 
\begin{equation}\label{6.40}
\begin{split}
\alpha_{\mathbbm P}(0) &= 1 + \epsilon_{\mathbbm P} =
1.0808\,, 
\\
\alpha_{{\mathbbm R}_+}(0) = \alpha_{{\mathbbm R}_-}(0) &=1 - 0.4525 =
0.5475\,.
\end{split}
\end{equation}
These values are quoted in \eqref{3.8}, \eqref{3.10} and \eqref{3.12}. 
Furthermore, we get
\be\label{6.41}
(\beta_{{\mathbbm P}NN})^2 =
X\, 
\left[ 18 \cos \left(\frac{\pi}{2} \epsilon_{\mathbbm P}\right) 
(M_0^2 \alpha_{\mathbbm P}')^{\epsilon_{\mathbbm P}}
\right]^{-1},
\ee
\begin{align}\label{6.42}
 (M_0)^{-2} g_{f_{2R}pp}^2 =& \,[ Y_{pp} + Y_{pn} + Y_{\bar{p}p} + Y_{\bar{p}n} ] \nn \\
 & \times \left[ 8 \cos\left(\frac{\pi}{2} (\alpha_{{\mathbbm R}_+}(0) -1)\right) 
 (M_0^2 \alpha_{{\mathbbm R}_+}')^{\alpha_{{\mathbbm R}_+}(0)-1} \right]^{-1},
 \\
 \label{6.43}
 (M_0)^{-2} g_{a_{2R}pp}^2 =&\, [ Y_{pp} - Y_{pn} + Y_{\bar{p}p} - Y_{\bar{p}n} ]  \nn\\
 &\times \left[ 8 \cos\left(\frac{\pi}{2} (\alpha_{{\mathbbm R}_+}(0) -1)\right) 
 (M_0^2 \alpha_{{\mathbbm R}_+}')^{\alpha_{{\mathbbm R}_+}(0)-1} \right]^{-1},
\\
\label{6.44}
(M_{-})^{-2} g_{\omega_{R}pp}^2 = &\,[ -Y_{pp} - Y_{pn} + Y_{\bar{p}p} + Y_{\bar{p}n} ] \nn\\
&\times \left[ 8 \cos\left(\frac{\pi}{2} \alpha_{{\mathbbm R}_-}(0)\right) 
(M_0^2 \alpha_{{\mathbbm R}_-}')^{\alpha_{{\mathbbm R}_-}(0)-1} \right]^{-1},
\\
\label{6.45}
(M_{-})^{-2} g_{\rho_{R}pp}^2 =& \,[ -Y_{pp} + Y_{pn} + Y_{\bar{p}p} - Y_{\bar{p}n} ] \nn \\
&\times \left[ 8 \cos\left(\frac{\pi}{2} \alpha_{{\mathbbm R}_-}(0)\right) 
(M_0^2 \alpha_{{\mathbbm R}_-}')^{\alpha_{{\mathbbm R}_-}(0)-1} \right]^{-1}.
\end{align}
From \eqref{6.41} we get with $X$ from \eqref{6.38} and 
$\epsilon_{\mathbbm P}$, $\alpha'_{\mathbbm P}$ from \eqref{3.8}
\be\label{6.46}
(\beta_{{\mathbbm P}NN})^2 = 3.49~\text{GeV}^{-2} \,.
\ee
Assuming now $\beta_{{\mathbbm P}NN}>0$ gives 
\be\label{6.47}
\beta_{{\mathbbm P}NN} = 1.87 \,\mbox{GeV}^{-1} 
\ee
as listed in \eqref{3.30A}. This is the standard DL value for 
$\beta_{{\mathbbm P}NN}$ given in 
\cite{Donnachie:1983hf,Donnachie:1985iz,Donnachie:1987gu}. 
Inserting the numerical values for $Y_{ab}$ from table \ref{model:table2} in \eqref{6.42} 
to \eqref{6.45} we get the results for the squared couplings listed 
in table \ref{model:table3}. We can only give estimates for the 
errors on these numbers, 
maybe of order $5\%$ for the $f_{2R}$ and $\omega_R$ values and 
significantly larger for the $a_{2R}$ and $\rho_R$ values.
\begin{table}[ht]\centering
\renewcommand{\arraystretch}{1.2}
\begin{tabular}{l|r}
&[GeV$^{-2}$] \\ 
\hline
$(M_0)^{-2} g^2_{f_{2R}pp}$ &  121.95  \\
$(M_0)^{-2}g^2_{a_{2R}pp}$& 2.82  \\
$(M_-)^{-2}g^2_{\omega_Rpp}$& 37.65 \\
$(M_-)^{-2}g^2_{\rho_Rpp}$ & 2.05  
\end{tabular}
\caption{\label{model:table3}
Numerical values for the squared couplings from \eqref{6.42} 
to \eqref{6.45} and table \ref{model:table2}.}
\end{table}

We shall now assume
\be\label{6.47a}
g_{f_{2R}pp} > 0\,, \qquad
g_{a_{2R}pp} > 0\,.
\ee
From table \ref{model:table3} we get then
\be\label{6.47b}
g_{f_{2R}pp}  =11.04\,, \qquad
g_{a_{2R}pp}  = 1.68\,.
\ee
These are the values adopted in \eqref{3.33A} and \eqref{3.34A}, respectively.

\begin{figure}[t]
\begin{center}
\includegraphics[height=85pt]{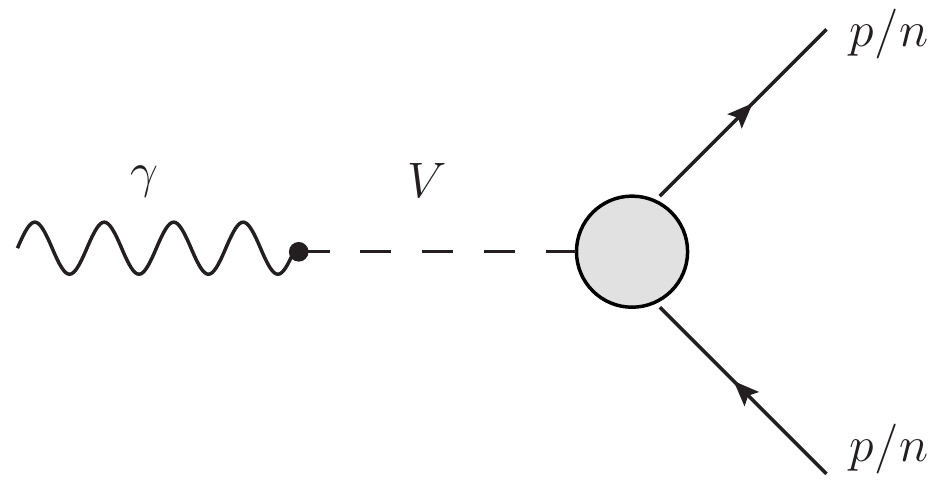}
\caption{VMD diagrams for the nucleon electromagnetic form factors ($V=\omega, \rho$).
\label{model:fig8}}
\end{center}
\end{figure}

To obtain numerical estimates for the couplings $g_{\omega_Rpp}$ 
and $g_{\rho_Rpp}$ themselves, as well as for the mass parameter 
$M_-$ we proceed as follows. We consider the electromagnetic form 
factors of proton and neutron at zero momentum transfer and assume 
the VMD model with $V=\omega,\rho$ to hold at this kinematic point. 
In the corresponding diagrams shown in figure \ref{model:fig8} we have the 
$\gamma V$ couplings which 
are known, see \eqref{3.20} and \eqref{3.21}, and the couplings 
$Vpp$ and $Vnn$. We shall now {\em assume} that these latter 
couplings are the same as the reggeon couplings 
$V_Rpp$, $V_Rnn~(V=\omega,\rho)$:
\begin{equation}
\label{6.54a}
g_{Vpp} \approx g_{V_Rpp}\,, \quad 
g_{Vnn} \approx g_{V_Rnn} \quad \mbox{for $V=\omega, \rho$} \,.
\end{equation}
We obtain then from the diagrams of figure \ref{model:fig8} and the known 
charges of $p$ and $n$ with the isospin relations $g_{\rho_R nn}=-g_{\rho_R pp}$ 
and $g_{\omega_R nn}=g_{\omega_R pp}$ 
\be\label{6.48}
\frac{1}{\gamma_\rho} g_{\rho_R pp} + \frac{1}{\gamma_\omega} g_{\omega_R pp} =1\,,\qquad
-\frac{1}{\gamma_\rho} g_{\rho_R pp} + \frac{1}{\gamma_\omega} g_{\omega_R pp} =0\,;
\ee
so that 
\be\label{6.49}
g_{\omega_R pp} = \frac{1}{2} \gamma_\omega = \sqrt{\pi}
\bigg( \frac{4\pi}{\gamma_\omega^2}\bigg)^{-1/2}\,, \qquad
g_{\rho_R pp} = \frac{1}{2} \gamma_\rho = \sqrt{\pi}
\bigg( \frac{4\pi}{\gamma_\rho^2}\bigg)^{-1/2}\,.
\ee
With the numbers for $4\pi/\gamma^2_V$ from \eqref{3.21} this gives 
\begin{align}\label{6.50a}
g_{\omega_R pp} &= 8.65 \pm 0.15\,,
\\
\label{6.50b}
g_{\rho_R pp} &= 2.52 \pm 0.06\,,
\\
\label{6.51}
\frac{g_{\omega_R pp}}{g_{\rho_R pp}} &= 3.43 \pm 0.10 \,.
\end{align}
This has to be compared to
\be\label{6.52}
\left|\frac{g_{\omega_R pp}}{g_{\rho_R pp}} \right| = 4.29
\ee
from table \ref{model:table3} where our error estimate is at least $10\%$. The central values 
for $|g_{\omega_Rpp}/ g_{\rho_Rpp}|$ from \eqref{6.51} and \eqref{6.52} 
differ only by $\approx 20\%$, a typical accuracy for VMD arguments. 
This gives credence to our assumption \eqref{6.54a}. We shall, therefore, 
fix the signs of $g_{\omega_Rpp}$ and $g_{\rho_R pp}$ from the 
VMD relations \eqref{6.49}, \eqref{6.50a}, \eqref{6.50b}. 
Since $(M_-)^2g^2_{\omega_Rpp}$ is much bigger and better determined 
than $(M_-)^2g^2_{\rho_Rpp}$ from table \ref{model:table3} we shall in the following 
take for $g_{\omega_Rpp}$ the VMD value from \eqref{6.50a} and 
calculate $M_-$ from table \ref{model:table3}. This gives 
\be\label{6.53}
M_- = \frac{g_{\omega_R pp}}{\sqrt{37.65}} \, \mbox{GeV} = 1.41 \, \mbox{GeV} 
\ee
which is very reasonable for a hadronic mass parameter. 
From table \ref{model:table3} we get then 
\be\label{6.54}
g_{\rho_R pp} = 2.02 \,.
\ee
The values for $M_-$ from \eqref{6.53}, for $g_{\omega_Rpp}$ 
from \eqref{6.50a}, and for $g_{\rho_Rpp}$ from \eqref{6.54} are 
taken as default values for the parameters of the propagators 
and vertices in \eqref{3.12}, \eqref{3.38A}, and \eqref{3.40A}, respectively. 

In this section we have obtained default values for our coupling parameters 
by considering just the imaginary parts of the forward-scattering amplitudes 
\eqref{6.33}, \eqref{6.34}. Of course, the next step should be to confront the 
model with the extensive knowledge about elastic scattering; for reviews see 
for instance \cite{Donnachie:2002en,ref8,Fiore:2008tp}. For recent work on 
exclusive central production where also elastic scattering is treated see 
\cite{Lebiedowicz:2009pj}. Also the TOTEM results \cite{Antchev:2013gaa} 
from LHC should be taken into account. They have been discussed in the 
context of Regge theory for example in \cite{Donnachie:2013xia}. 
We leave the discussion of these issues in view of our model for future work. 

\boldmath
\section{Pion-Proton and $\rho$-Proton Scattering}
\label{Pion-proton and rho-proton scattering}
\unboldmath

\subsection{Pion-Proton Scattering}
\label{Pion-proton scattering}

We consider here the reactions
\ba\label{7.1}
\pi^\pm (p_1) + p(p_2,\lambda_2) \longrightarrow \pi^\pm (p'_1) + p(p'_2,\lambda'_2) 
\ea
at high energies. Again we set 
\be\label{7.2}
s = (p_1 + p_2)^2 \,, \qquad
t = (p_1 - p'_1)^2 \,.
\ee
The diagrams for these reactions with ${\mathbbm P}, f_{2R}, \rho_R$ 
and $\gamma$ exchange are shown in figure \ref{model:fig2}. 
We shall only consider the hadronic exchanges in the following. 
Note that $G$-parity invariance of strong interactions forbids 
$a_{2R}\pi\pi$, $\omega_R\pi\pi$, and ${\mathbbm O}\pi\pi$ 
vertices, cf.\ table \ref{table1}. As in sections \ref{Pomeron exchange} and 
\ref{Reggeon exchanges and total cross sections} we shall treat 
the pomeron and $f_{2R}$ exchanges as effective rank-2 tensor 
exchanges. The ${\mathbbm P}$ and $f_{2R}$ propagators are 
already constructed. It remains to construct the ${\mathbbm P}\pi\pi$ 
and $f_{2R}\pi\pi$ vertices. In analogy to ${\cal L}'_{{\mathbbm P} pp}$ 
in \eqref{6.26} and ${\cal L}'_{f_2\pi\pi}$ in \eqref{5.6} we make 
the following ansatz for the ${\mathbbm P}\pi\pi$ coupling Lagrangian
\be\label{7.3}
\begin{split}
{\cal L}_{{\mathbbm P}\pi\pi}' (x)= &
2 \beta_{{\mathbbm P}\pi\pi} {\mathbbm P}_{\kappa \lambda} (x)
\left(g^{\kappa\mu} g^{\lambda\nu} - \frac{1}{4}
g^{\kappa\lambda} g^{\mu\nu}\right)
\\
&\times 
\sum_{a=1}^3
\left[
\frac{1}{2} \pi_a(x) \partial_\mu \partial_\nu \pi_a(x)
+\frac{1}{2} (\partial_\mu \partial_\nu \pi_a(x)) \pi_a(x)
- (\partial_\mu \pi_a(x)) (\partial_\nu \pi_a(x))
\right] \,.
\end{split}
\ee
This leads to the ${\mathbbm P}\pi\pi$ vertex \eqref{3.31}. 
The coupling constant $\beta_{{\mathbbm P}\pi\pi}$ of dimension 
$\mathrm{mass}^{-1}$ will be determined below. Our ansatz for the $f_{2R}\pi\pi$ 
vertex is completely analogous to the ${\mathbbm P}\pi\pi$ vertex 
and is given in \eqref{3.36}. For the $\rho_R\pi\pi$ vertex we 
can take the $\rho\pi\pi$ vertex \eqref{3.26a} as a model. 
Our coupling Lagrangian reads (compare \eqref{4.3a}) 
\be\label{7.4}
{\cal L}_{\rho_R\pi\pi}' =
\frac{1}{2} g_{\rho_R\pi\pi}\;
\rho_R^\mu(x)
[ i \pi^+(x) \overleftrightarrow{\partial_\mu} \pi^-(x) ]\,.
\ee
This leads to the vertex \eqref{3.42} with $g_{\rho_R\pi\pi}$ to be 
determined below. 

We are now ready to write down the amplitudes 
for the reactions \eqref{7.1}. For high energies we find the following, 
using \eqref{6.23}, 
\begin{equation}\label{7.4a}
\begin{split}
\langle \pi^\pm(p_1'), p(p_2',\lambda_2') | {\cal T} | \pi^\pm(p_1), p(p_2,&\,\lambda_2) \rangle 
\\
=i\; 2 s\; \delta_{\lambda_2'\lambda_2} F_1(t) F_M(t)
\bigg\{ &\,
6 \beta_{{\mathbbm P}\pi\pi} \beta_{{\mathbbm P}NN}
(- i s \alpha_{\mathbbm P}' )^{\alpha_{\mathbbm P}(t) -1} 
\\
&
+
\frac{1}{2} g_{f_{2R}\pi\pi}\; g_{f_{2R}pp}
M_0^{-2}
(- i s \alpha_{{\mathbbm R}_+}')^{\alpha_{{\mathbbm R}_+}(t) -1} 
\\
&
\pm
\frac{i}{2} g_{\rho_{R}\pi\pi}\; g_{\rho_{R}pp}
M_-^{-2} 
(- i s \alpha_{{\mathbbm R}_-}')^{\alpha_{{\mathbbm R}_-}(t) -1}
\bigg\}.
\end{split}
\end{equation}
From \eqref{7.4a} we find via the optical theorem for large $s$
\begin{equation}\label{7.5}
\begin{split}
\sigma_{\rm tot}(\pi^\pm, p) =&\,
\frac{1}{2s} 
\sum_{\lambda_2}
\text{Im}\,
\langle \pi^\pm(p_1), p(p_2,\lambda_2) | {\cal T} | \pi^\pm(p_1), p(p_2,\lambda_2) \rangle 
\\
= &\,
2 \bigg\{
6\beta_{{\mathbbm P}\pi\pi}\,\beta_{{\mathbbm P}NN} \cos\left[\frac{\pi}{2}(\alpha_{\mathbbm P}(0) -1 )\right]
(s \alpha_{\mathbbm P}' )^{\alpha_{\mathbbm P}(0) -1}
\\
& \hspace*{.5cm}
+ \frac{1}{2} g_{f_{2R}\pi\pi}\;g_{f_{2R}pp} 
M_0^{-2} \cos\left[\frac{\pi}{2}(\alpha_{{\mathbbm R}_+}(0) -1 )\right]
(s \alpha_{{\mathbbm R}_+}')^{\alpha_{{\mathbbm R}_+}(0) -1} 
\\
&\hspace*{.5cm}
\mp
\frac{1}{2} g_{\rho_{R}\pi\pi}\;g_{\rho_{R}pp} 
M_-^{-2} \cos\left[\frac{\pi}{2}\alpha_{{\mathbbm R}_-}(0)\right]
(s \alpha_{{\mathbbm R}_-}')^{\alpha_{{\mathbbm R}_-}(0) -1}
\bigg\}.
\end{split}
\end{equation}

In figure 3.1 of \cite{Donnachie:2002en} the following fit to the 
experimental total cross sections $\sigma_{\text {tot}}(\pi^\pm, p)$ 
is presented
\be\label{7.6}
\sigma_{\rm tot}(\pi^\pm, p) = X_{\pi p} 
\bigg(\frac{s}{M_0^2}\bigg)^{0.0808}
+
Y_{\pi^\pm p}
\bigg(\frac{s}{M_0^2}\bigg)^{-0.4525} \,.
\ee
Here
\be\label{7.7}
\begin{split}
X_{\pi p} &= 13.63~\text{mb} \;\widehat{=}\; 35.00~\text{GeV}^{-2}\,,\\
Y_{\pi^+ p} &= 27.56~\text{mb} \;\widehat{=}\; 70.78~\text{GeV}^{-2}\,,\\
Y_{\pi^- p} &= 36.02~\text{mb} \;\widehat{=}\; 92.51~\text{GeV}^{-2}\,.
\end{split}
\ee
From \eqref{7.5} and \eqref{7.6} and with \eqref{3.8}, \eqref{3.10}, \eqref{3.12} 
we get
\begin{align}\label{7.8}
\beta_{{\mathbbm P}\pi\pi}\,\beta_{{\mathbbm P}NN} &=
X_{\pi p}
\left[
12
\cos\left[\frac{\pi}{2}(\alpha_{\mathbbm P}(0) -1 )\right]
(M_0^2 \alpha_{\mathbbm P}' )^{\alpha_{\mathbbm P}(0) -1}
\right]^{-1}\,,
\\
\label{7.9}
(M_0)^{-2} g_{f_{2R}\pi\pi}\;g_{f_{2R}pp} &= (Y_{\pi^+ p} + Y_{\pi^- p})
\left[
2 \cos\left[\frac{\pi}{2}(\alpha_{{\mathbbm R}_+}(0) -1 )\right]
(M_0^2 \alpha_{{\mathbbm R}_+}')^{\alpha_{{\mathbbm R}_+}(0) -1} 
\right]^{-1}\,,
\\
\label{7.10}
(M_-)^{-2} g_{\rho_{R}\pi\pi}\;g_{\rho_{R}pp}  &= (Y_{\pi^- p} - Y_{\pi^+ p})
\left[
2 \cos\left[\frac{\pi}{2}\alpha_{{\mathbbm R}_-}(0)\right]
(M_0^2 \alpha_{{\mathbbm R}_-}')^{\alpha_{{\mathbbm R}_-}(0) -1}
\right]^{-1}.
\end{align}
With $X_{\pi p}$ and $Y_{\pi^\pm p}$ from \eqref{7.7} and the values 
for the pomeron and Regge trajectories from \eqref{3.8}, \eqref{3.10}, 
and \eqref{3.12} we get from \eqref{7.8} to \eqref{7.10}
\begin{align}\label{7.11}
\beta_{{\mathbbm P}\pi\pi}\,\beta_{{\mathbbm P}NN} &=
3.29~\text{GeV}^{-2}\,,
\\
\label{7.12}
(M_0)^{-2} g_{f_{2R}\pi\pi}\;g_{f_{2R}pp} &= 
102.72~\text{GeV}^{-2}\,,
\\
\label{7.13}
(M_-)^{-2} g_{\rho_{R}\pi\pi}\;g_{\rho_{R}pp}  &= 
15.88~\text{GeV}^{-2}\,.
\end{align}

We can use \eqref{7.11} to \eqref{7.13} and the results of 
section \ref{Reggeon exchanges and total cross sections} 
for the pomeron and reggeon couplings to the proton and 
extract the corresponding couplings to pions. From \eqref{7.11} 
we find with $\beta_{{\mathbbm P}NN}$ from \eqref{3.30A}
\be\label{7.14}
\beta_{{\mathbbm P}\pi\pi} = 1.76 \,\mbox{GeV}^{-1}\,.
\ee
With $M_-=1.41 \,\mbox{GeV}$ from \eqref{6.53}, $g_{f_{2R}pp}$  from 
\eqref{3.33A} and $g_{\rho_Rpp}$ from \eqref{3.40A} we find
\begin{align}\label{7.15}
g_{f_{2R}\pi\pi} &= 9.30\,,\\
\label{7.16}
g_{\rho_{R}\pi\pi} &= 15.63\,.
\end{align}
These are the default values listed in \eqref{3.31A}, \eqref{3.36A} 
and \eqref{3.42A}, respectively.

Now we recall that from $\Gamma(f_2\to\pi\pi)$ we determined 
$|g_{f_2\pi\pi}|=9.26\pm 0.15$ in \eqref{5.12}. This is surprisingly 
close to what we get from \eqref{7.15} identifying $g_{f_{2R}\pi\pi}$ 
and $g_{f_2\pi\pi}$. This motivates us to extract the sign of 
$g_{f_2\pi\pi}$ from this identification which leads to \eqref{5.13} 
and \eqref{3.28param}. 

For $g_{\rho\pi\pi}$ \eqref{3.27} and $g_{\rho_R\pi\pi}$ \eqref{7.16} 
we find agreement in sign and in magnitude to $\approx 25\%$. 
Given the error on $g_{\rho_R\pi\pi}$ which should certainly be 
taken to be at least $10\%$ we find this result rather satisfactory. 
It gives again support to our hypothesis of relating ordinary and 
reggeon vertices. 

\boldmath
\subsection{$\rho^0$-Proton Scattering}
\label{rho-proton scattering}
\unboldmath

In this section we consider elastic $\rho^0$-proton scattering:
\be\label{7.17}
\rho^0 (p_1,\epsilon_1) + p(p_2,\lambda_2) \longrightarrow 
\rho^0 (p'_1,\epsilon'_1) + p(p'_2,\lambda'_2) \,.
\ee
Here $\lambda_2,\lambda'_2,\in\{1/2,-1/2\}$ are the proton 
helicities and $\epsilon_1,\epsilon'_1$ the $\rho^0$ polarisation 
vectors. Of course, the reaction \eqref{7.17} is hard to observe 
directly in experiments. But \eqref{7.17} will be an important 
ingredient in our study of photoproduction reactions in 
\cite{searchfortheodderon}. The diagrams for elastic $\rho p$ scattering at 
high energies are shown in figure \ref{model:fig3}. We are here interested 
in $\rho^0 p$ scattering \eqref{7.17}  
where $C=-1$ exchange is forbidden. 
Also $a_{2R}$ exchange is forbidden by isospin invariance. 
We are left with ${\mathbbm P}$ and $f_{2R}$ exchange 
which -- in our approach -- are both treated as rank-two tensor exchanges. 
The general structure of the ${\mathbbm P}\rho\rho$ and 
the $f_{2R}\rho\rho$ vertices must, therefore, be as we found it 
for the $f_2\gamma\gamma$ vertex; see \eqref{3.29new}. This leads to 
our ans{\"a}tze \eqref{3.32} and \eqref{3.37}. 
The parameters listed in \eqref{3.32A} and \eqref{3.37AA} 
will be discussed below. The amplitude for \eqref{7.17} is now 
easily written down. 
For large $s$ we find with \eqref{6.23} 
\begin{equation}\label{7.18}
\begin{split}
\langle \rho^0(p_1',\epsilon_1'), p(p_2',\lambda_2') | {\cal T} | \rho^0(&\,p_1,\epsilon_1), p(p_2,\lambda_2) \rangle 
\\
= - i \delta_{\lambda_2'\lambda_2} \frac{1}{2s}
F_1(t)  F_M(t) &
 (\epsilon_1'^\mu)^* \epsilon_1^\nu
 (p_2'+p_2)^\kappa (p_2'+p_2)^\lambda
\\
\times \bigg\{
\Gamma_{\mu\nu\kappa\lambda}^{(0)}(p_1',-p_1)
&
\left[
6 \beta_{{\mathbbm P}NN} a_{{\mathbbm P}\rho\rho} 
(- i s \alpha_{\mathbbm P}' )^{\alpha_{\mathbbm P}(t) -1} 
\right.
\\
& \left.
+2 M_0^{-1}
g_{f_{2R}pp} \, a_{f_{2R}\rho\rho}
(- i s \alpha_{{\mathbbm R}_+}')^{\alpha_{{\mathbbm R}_+}(t) -1} 
\right]
\\
-\Gamma_{\mu\nu\kappa\lambda}^{(2)}(p_1',-p_1)
&\left[
3 \beta_{{\mathbbm P}NN} b_{{\mathbbm P}\rho\rho} 
(- i s \alpha_{\mathbbm P}' )^{\alpha_{\mathbbm P}(t) -1} 
\right.
\\
& \left.
+M_0^{-1} g_{f_{2R}pp} \, b_{f_{2R}\rho\rho}
(- i s \alpha_{{\mathbbm R}_+}')^{\alpha_{{\mathbbm R}_+}(t) -1} 
\right]
\bigg\} \,.
\end{split}
\end{equation}
Using the optical theorem we can calculate from \eqref{7.18} the 
total cross sections for $\rho^0$ mesons of definite helicities 
on unpolarised protons. Working in the c.\,m.\ system of the 
reaction \eqref{7.17} we define the polarisation vectors for $\rho^0$ 
of four-momentum $p_1=(p_1^0,\vec{p}_1)^T$ and helicity 
$m\in\{\pm 1,0\}$ as follows
\begin{equation}\label{7.19}
\epsilon^{(\pm 1)\mu} = \mp \frac{1}{\sqrt{2}}
\begin{pmatrix} 0 \\ \vec{e_1} \pm i\vec{e_2} \end{pmatrix}\,, \qquad
\epsilon^{(0)\mu} = \frac{1}{m_\rho} 
\begin{pmatrix} |\vec{p_1}| \\ \vec{e_3} p_1^0 \end{pmatrix} \,.
\end{equation}
Here $\vec{e}_1,\vec{e}_2,\vec{e}_3$ are right-handed Cartesian 
basis vectors with $\vec e_3=\vec p_1/|\vec p_1|$. From \eqref{7.18} we get
\begin{equation}\label{7.20}
\begin{split}
\sigma_{\rm tot}&(\rho^0 (\epsilon^{(m)}), p) =
\frac{1}{2s} \sum_{\lambda_2} \text{Im}
\langle \rho^0(p_1,\epsilon^{(m)}), p(p_2,\lambda_2) | {\cal T} | \rho^0(p_1,\epsilon^{(m)}), p(p_2,\lambda_2) \rangle 
\\
=&\, 3\beta_{{\mathbbm P}NN}
\left[
2 m_\rho^2\; a_{{\mathbbm P}\rho\rho} + (1 -\delta_{m,0}) b_{{\mathbbm P}\rho\rho}
\right]
\cos\left[\frac{\pi}{2}(\alpha_{\mathbbm P}(0) -1 )\right]
(s \alpha_{\mathbbm P}' )^{\alpha_{\mathbbm P}(0) -1}
\\
& 
+g_{f_{2R}pp} M_0^{-1}
\left[ 
2 m_\rho^2 a_{f_{2R}\rho\rho} + (1 -\delta_{m,0}) b_{f_{2R}\rho\rho}
\right] 
\cos\left[\frac{\pi}{2}(\alpha_{{\mathbbm R}_+}(0) -1 )\right]
(s \alpha_{{\mathbbm R}_+}')^{\alpha_{{\mathbbm R}_+}(0) -1} \,.
\end{split}
\end{equation}
Note that in our approach the total cross sections for $\rho^0$ mesons 
with longitudinal $(m=0)$ and transverse $(m=\pm 1)$ polarisation 
are different. 

\begin{figure}[htb]
\begin{center}
\includegraphics[width=0.55\textwidth]{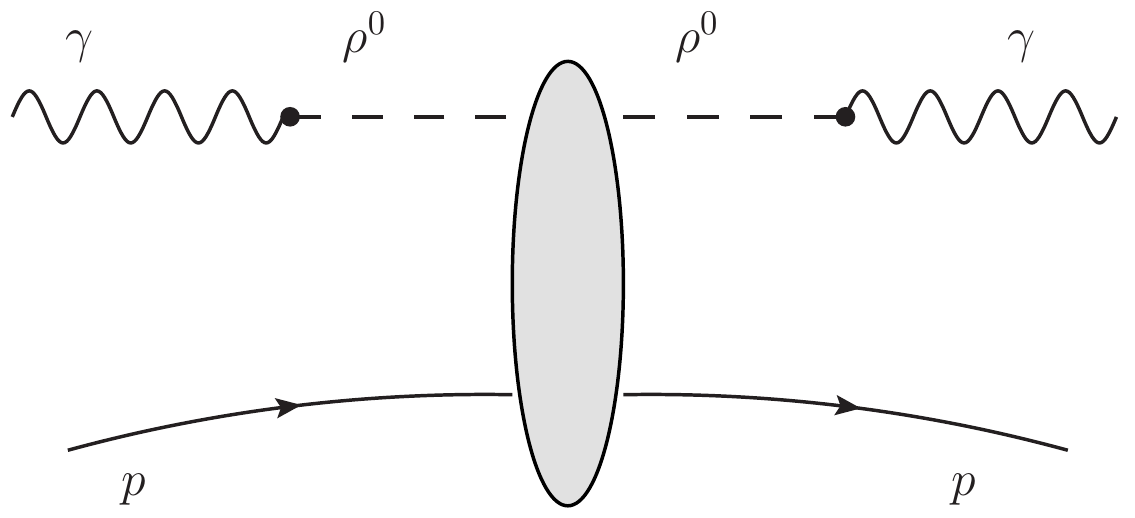}
\caption{
VMD diagram for $\gamma p$ forward scattering with intermediate 
$\rho^0$ meson. \label{fig1000}}
\end{center}
\end{figure}

\begin{figure}[ht]
\begin{center}
\includegraphics[width=0.9\textwidth]{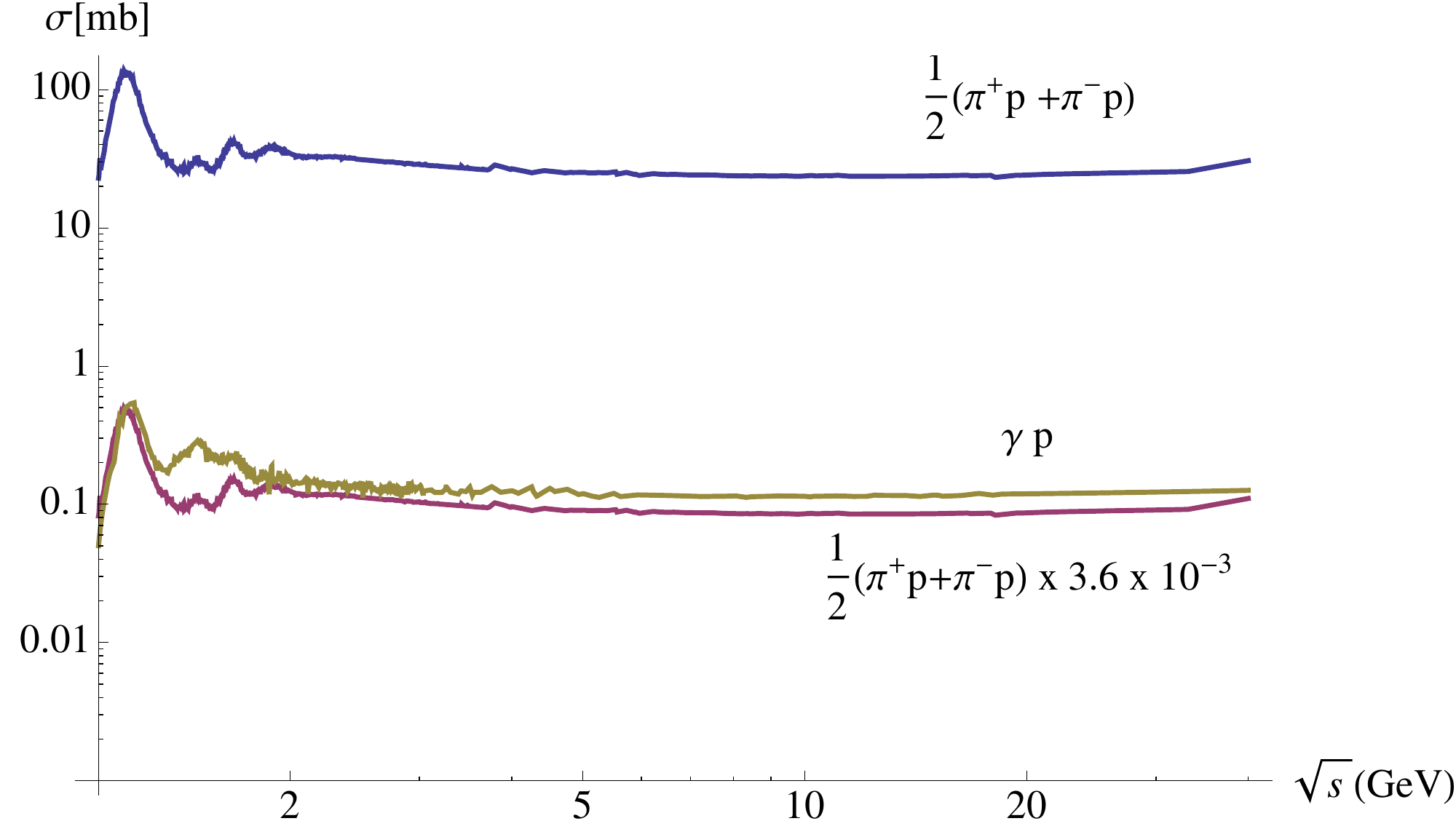}
\caption{
Data for $\frac{1}{2}\left( \sigma_{\rm tot} (\pi^+, p) 
+ \sigma_{\rm tot} (\pi^-, p) \right)$ and rescaled by a factor 
$3.6 \times 10^{-3}$ compared to $\sigma_{\rm tot} (\gamma, p)$ 
as function of the c.\,m.\ energy $\sqrt{s}$. Data are  
from \cite{Durham}, the $\pi^\pm p$ data have been interpolated 
for computing their average. 
\label{fig1001}}
\end{center}
\end{figure}

In order to get estimates for the ${\mathbbm P}\rho\rho$ 
and $f_{2R}\rho\rho$ coupling constants we make the assumption that at 
high energies the total cross section for {\em transversely} polarised 
$\rho^0$ mesons equals the average of the $\pi^+p$ and $\pi^-p$ 
cross sections. 
\newline
{\em Assumption:}
\be\label{7.21}
\sigma_{\rm tot}(\rho^0(\epsilon^{(m)}), p)  =
\frac{1}{2} \big[
\sigma_{\rm tot}(\pi^+, p) + \sigma_{\rm tot}(\pi^-, p) 
\big] 
\qquad \text{for } m=\pm 1.
\ee
We can motivate \eqref{7.21} by a VMD argument. Let us calculate the total 
$\gamma p$ cross section from the VMD diagram in figure \ref{fig1000} 
taking into account only the $\rho^0$ contribution since the $\gamma \rho^0$ 
coupling is much larger than the $\gamma \omega$ and $\gamma \phi$ 
couplings; see \eqref{3.20} to \eqref{3.21}. 
Since real photons have only transverse polarisation we get from figure 
\ref{fig1000} with \eqref{3.20} and \eqref{3.2a} 
\be\label{7.22a}
\sigma_{\rm tot} (\gamma, p) = \frac{e^2}{4 \pi} 
\left(\frac{4 \pi}{\gamma_\rho^2}\right) \frac{1}{2} 
\sum_{m=\pm 1} \sigma_{\rm tot} (\rho^0(\epsilon^{(m)}), p) \,. 
\ee
In the spirit of VMD the cross section for on-shell $\rho^0$ 
mesons is to be inserted here. With the assumption \eqref{7.21} 
we should have, therefore, 
\be\label{7.22b}
\begin{split}
\sigma_{\rm tot} (\gamma, p) &= \frac{e^2}{4 \pi} 
\left(\frac{4 \pi}{\gamma_\rho^2}\right) \frac{1}{2} 
\left[ \sigma_{\rm tot} (\pi^+, p) + \sigma_{\rm tot} (\pi^-, p) \right]
\\
&=3.6 \times 10^{-3} \, \frac{1}{2} 
\left[ \sigma_{\rm tot} (\pi^+, p) + \sigma_{\rm tot} (\pi^-, p) \right]  \,.
\end{split}
\ee
This relation is compared to data in figure \ref{fig1001}. 
The agreement is quite reasonable. Note that $\sigma_{\rm tot} (\gamma, p)$ 
should be a little higher than the r.\,h.\,s.\ of \eqref{7.22b} due 
to the contribution from $\omega$ and $\phi$ mesons in this energy range. 

With the $\pi p$ total cross sections from \eqref{7.5} we obtain 
now from \eqref{7.20} and \eqref{7.21} 
\begin{align}\label{7.22}
3 \beta_{{\mathbbm P}NN} 
( 2 m_\rho^2 a_{{\mathbbm P}\rho\rho} + b_{{\mathbbm P}\rho\rho} ) &=
12 \beta_{{\mathbbm P}NN}\; \beta_{{\mathbbm P}\pi\pi}\,,
\\
\label{7.23}
g_{f_{2R}pp} M_0^{-1}
( 2 m_\rho^2 a_{f_{2R}\rho\rho} + b_{f_{2R}\rho\rho} ) &=
g_{f_{2R}pp}\; g_{f_{2R}\pi\pi} M_0^{-2}\,.
\end{align}
From \eqref{7.22} and \eqref{7.14} we get 
\be\label{7.24}
2 m_\rho^2\; a_{{\mathbbm P}\rho\rho} + b_{{\mathbbm P}\rho\rho} =
4 \beta_{{\mathbbm P}\pi\pi} = 7.04~\text{GeV}^{-1}\,.
\ee
This is the relation listed in \eqref{3.32A}. Combining \eqref{7.23} 
and \eqref{7.15} gives
\be\label{7.25}
2 m_\rho^2\; a_{f_{2R}\rho\rho} + b_{f_{2R}\rho\rho} =
M_0^{-1} g_{f_{2R}\pi\pi} =
9.30~\text{GeV}^{-1}\,.
\ee

\begin{figure}[ht]
\begin{center}
\includegraphics[height=125pt]{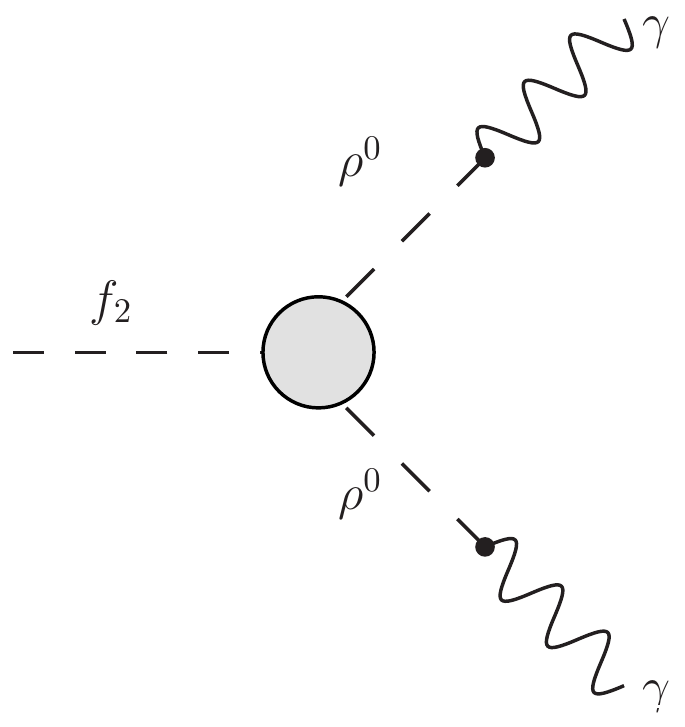}
\caption{VMD diagram for $f_2\to\gamma\gamma$ with 
intermediate $\rho^0$. 
\label{model:fig9}}
\end{center}
\end{figure}

For the couplings $a_{f_{2R}\rho\rho}$ and $b_{f_{2R}\rho\rho}$ we 
can go a step further and give estimates for them individually. 
We shall assume that the couplings of the meson $f_2$ to 
$\rho^0\rho^0$ and of the reggeon $f_{2R}$ to $\rho^0\rho^0$ are equal: 
\be\label{7.26}
a_{f_{2}\rho\rho} \approx a_{f_{2R}\rho\rho}\,,\qquad
b_{f_{2}\rho\rho} \approx b_{f_{2R}\rho\rho}\,.
\ee
Here $a_{f_{2}\rho\rho}$ and $b_{f_{2}\rho\rho}$ are defined as the 
$f_2\gamma\gamma$ couplings in \eqref{3.29new} but for $\rho^0$ 
in place of $\gamma$. 
The next step is to relate the $f_2\rho\rho$ couplings to the 
$f_2\gamma\gamma$ couplings using VMD. Taking into account 
only the $\rho^0$ meson, that is, neglecting contributions 
from the $\omega$ and $\phi$ mesons we get for $f_2 \to \gamma \gamma$ 
the VMD diagram in figure \ref{model:fig9}. With 
\eqref{3.20}, \eqref{3.21} and \eqref{3.29new} this gives 
\be\label{7.27}
a_{f_2 \gamma\gamma} =\frac{e^2}{\gamma_\rho^2} a_{f_2 \rho\rho}\,\qquad
b_{f_2 \gamma\gamma} =\frac{e^2}{\gamma_\rho^2} b_{f_2 \rho\rho}\,.
\ee
With \eqref{7.26} and the numbers from \eqref{3.29a} we get as estimates 
\begin{align}
\label{7.29}
a_{f_{2}\rho\rho} &=a_{f_{2R}\rho\rho}
= a_{f_2 \gamma\gamma} \left( \alpha \frac{4\pi}{\gamma_\rho^2}\right)^{-1} 
= 2.92~\text{GeV}^{-3}\,,\\
\label{7.30}
b_{f_{2}\rho\rho} &=b_{f_{2R}\rho\rho}
= b_{f_2 \gamma\gamma} \left( \alpha \frac{4\pi}{\gamma_\rho^2}\right)^{-1} 
= 5.02~\text{GeV}^{-1}\,.
\end{align}
Note that we have assumed here 
\be\label{7.30a}
 a_{f_2 \gamma\gamma}>0\,,\qquad b_{f_2 \gamma\gamma}>0\,.
\ee
From \eqref{7.29} and \eqref{7.30} we get 
\be\label{7.31}
2 m_\rho^2 a_{f_{2R} \rho\rho} + b_{f_{2R} \rho\rho} = 8.52~\text{GeV}^{-1}
\ee
which agrees with \eqref{7.25} to better than $10\%$. We take this as 
supporting the above sign assumptions \eqref{7.30a} for 
$a_{f_2\gamma\gamma}$ and $b_{f_2\gamma\gamma}$. 
The values for $a_{f_{2R}\rho\rho}$ from \eqref{7.29} and for 
$b_{f_{2R}\rho\rho}$ from \eqref{7.30} are taken as default values 
in \eqref{3.37AA}. 

For the coupling parameters of $\rho_R \rho^0 f_2$ in \eqref{3.45} we take again 
as estimates 
\begin{equation}
\label{7.33a}
a_{\rho_R \rho^0 f_2} \approx a_{f_{2R}\rho\rho}\,, \qquad 
b_{\rho_R \rho^0 f_2} \approx b_{f_{2R}\rho\rho}\,.
\end{equation}
With \eqref{7.29} and \eqref{7.30} this gives the default values quoted in 
\eqref{3.45A}. 

The $f_2$ can also couple to $\omega \omega$ and $\omega_R \omega$. 
We have neglected this in our VMD argument above, see figure \ref{model:fig9}, 
since the $\gamma$-$\omega$ coupling is much smaller than the 
$\gamma$-$\rho^0$ coupling; see \eqref{3.21}. A simple guess is that 
the couplings of $f_2$ to the $\omega$ and $\rho^0$ mesons are of the 
same size: 
\be
\label{7.35A}
\begin{split}
a_{\omega \omega f_2} &\approx a_{\omega_R \omega f_2} 
\approx a_{\rho_R\rho^0 f_2}
\,,
\\
b_{\omega \omega f_2} &\approx b_{\omega_R \omega f_2} 
\approx b_{\rho_R\rho^0 f_2}\,.
\end{split}
\ee
Here the $\omega_R \omega f_2$ vertex is assumed to be of the form 
\eqref{3.44} and analogously for the $\omega \omega f_2$ vertex.

\begin{figure}[htb]
\begin{center}
\includegraphics[height=125pt]{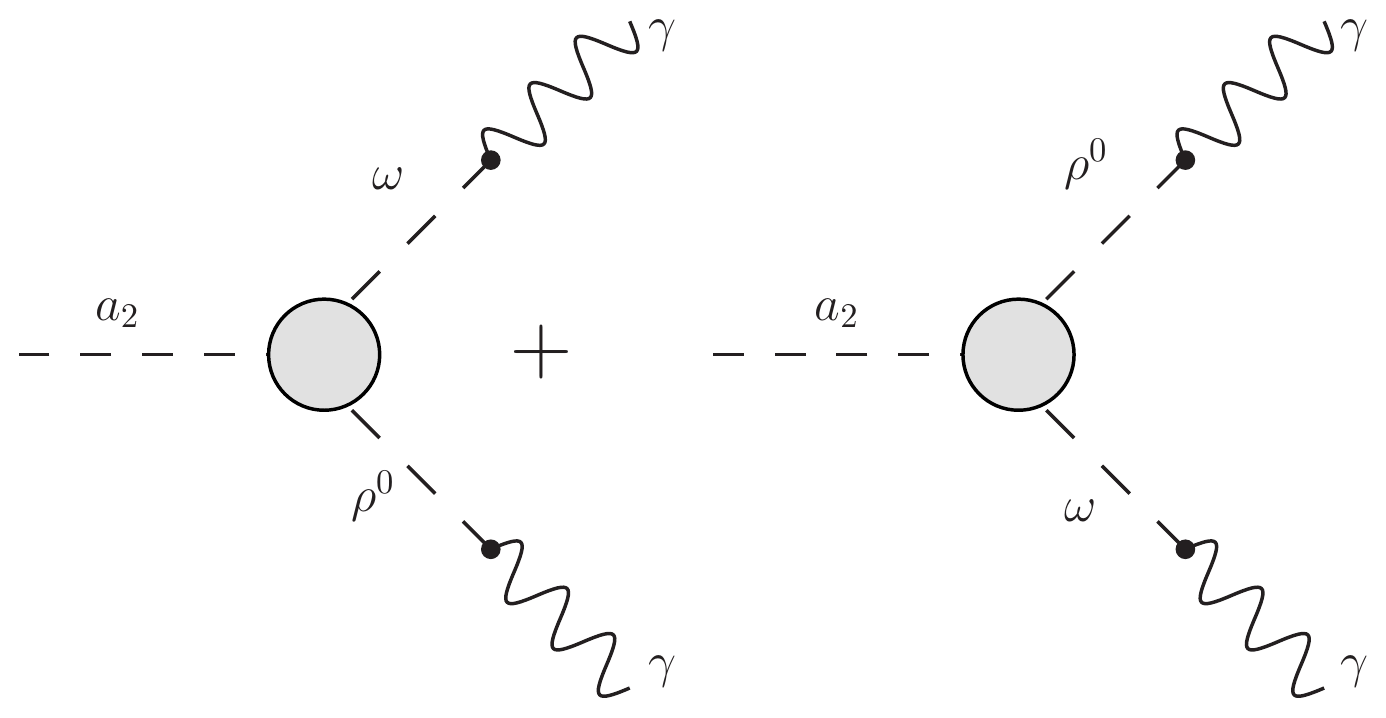} 
\caption{VMD diagrams for $a_2\to\gamma\gamma$. 
\label{model:fig10}}
\end{center}
\end{figure}

At the end of this section we make some remarks 
on the vertices for $a_2\gamma\gamma$ \eqref{3.29b} and 
$a_{2R}\omega\rho$ \eqref{3.37a}. We shall again use VMD to 
relate these vertices. Neglecting a possible contribution from the 
$\phi$ meson we have for $a_2\gamma\gamma$ the VMD diagrams 
shown in figure \ref{model:fig10}. 
This gives in analogy to the discussion above 
\be\label{7.32}
a_{a_2 \gamma\gamma} = 2\frac{e^2}{\gamma_\rho \gamma_\omega} a_{a_2 \omega\rho}\,,\qquad
b_{a_2 \gamma\gamma} = 2\frac{e^2}{\gamma_\rho \gamma_\omega} b_{a_2 \omega\rho}\,.
\ee
Here the coupling constants $a_{a_2 \omega \rho}$ and $b_{a_2 \omega \rho}$ 
are defined as in \eqref{3.29b} for $a_2\gamma\gamma$ but with $\gamma \gamma$ 
replaced by $\omega\rho$. With the assumptions 
\be\label{7.33}
a_{a_2 \omega\rho} \approx a_{a_{2R} \omega\rho}\,,\qquad
b_{a_2 \omega\rho} \approx b_{a_{2R} \omega\rho}
\ee
we get from \eqref{7.32}
\be\label{7.34}
\begin{split}
a_{a_{2R}\omega\rho} &= (2\alpha)^{-1} 
\bigg( \frac{4\pi}{\gamma_\rho^2} \frac{4\pi}{\gamma_\omega^2} \bigg)^{-1/2} a_{a_2 \gamma\gamma}\,,\\
b_{a_{2R}\omega\rho} &= (2\alpha)^{-1} 
\bigg( \frac{4\pi}{\gamma_\rho^2} \frac{4\pi}{\gamma_\omega^2} \bigg)^{-1/2} b_{a_2 \gamma\gamma}\,.
\end{split}
\ee

In \eqref{5.24} we have obtained a number for the sum of the 
squared $a_2\gamma\gamma$ couplings. In order to 
estimate the couplings individually we shall make the {\sl ad hoc} 
assumption that
\be\label{7.35}
m_{a_2}^3 a_{a_2 \gamma\gamma} = m_{f_2}^3 a_{f_2 \gamma\gamma} 
\cdot r\,,\qquad
m_{a_2}b_{a_2 \gamma\gamma} = m_{f_2} b_{f_2 \gamma\gamma} \cdot r
\ee
with a real scale factor $r$. Then $r^2$ is given by the ratio of the r.\,h.\ sides 
of \eqref{5.24} and \eqref{5.19a} and we get 
\be\label{7.36}
r^2 = \frac{3.60}{11.62} \,,\qquad
r = \pm 0.56 \,.
\ee
Here we must leave the sign of $r$ open. 
With the numbers for $a_{f_2\gamma\gamma}$ and 
$b_{f_2\gamma\gamma}$ from \eqref{3.29a} we have now 
\be\label{7.37}
a_{a_2 \gamma\gamma} = \pm \frac{e^2}{4\pi} 0.74~\text{GeV}^{-3}\,,\qquad
b_{a_2 \gamma\gamma} = \pm \frac{e^2}{4\pi} 1.35~\text{GeV}^{-1}\,.
\ee
From \eqref{7.34}, \eqref{7.37} and $\gamma^2_\rho/(4\pi)$, 
$\gamma^2_\omega/4\pi$ from \eqref{3.21} we get as estimates
\be\label{7.38}
a_{a_{2R} \omega\rho} = \pm 2.56~\text{GeV}^{-3}\,,\qquad
b_{a_{2R} \omega\rho} = \pm 4.68~\text{GeV}^{-1}\,.
\ee
The numbers from \eqref{7.37} and \eqref{7.38} are taken 
as default values in \eqref{3.29b} and \eqref{3.37ab}, respectively. 
We emphasize that these should be considered as very rough estimates. 

\section{Conclusions}
\label{Conclusions}

In this article we have proposed a model for soft high-energy reactions. 
The model is formulated in terms of effective propagators and vertices 
for the exchange objects: pomeron, odderon and reggeons. The vertices 
are constructed according to the standard rules of QFT and the effective 
propagators take into account the crossing properties of the amplitudes 
and the power laws, in the c.\,m.\ energy squared $s$, as suggested by 
the Regge ansatz. We have given the Feynman rules for our model in 
section \ref{Propagators and vertices} and someone only interested in 
applying the model for concrete calculations of amplitudes just has to read 
that section. The justification of the propagators and vertices of section 
\ref{Propagators and vertices} has been given in sections \ref{Vector mesons} 
to \ref{Pion-proton and rho-proton scattering}. In section \ref{Vector mesons} 
we have dealt with vector mesons and we have given an interpretation of 
the time-honoured vector-meson-dominance (VMD) model which we find 
quite satisfactory. We have shown how, starting from a perfectly 
gauge-invariant coupling of the photon to the vector mesons 
$V=\rho^0, \omega,\phi$, one arrives at the standard VMD couplings; 
see section \ref{sec:siffforms}. In section \ref{Pomeron exchange} 
we have dealt with the pomeron. We have formulated pomeron exchange 
as an effective rank-two tensor exchange, that is, as spin 2 exchange. 
For the amplitudes of $pp$ and $\bar{p} p$ elastic scattering this leads 
at high energies to the same expressions as obtained from the 
Donnachie-Landshoff pomeron. We emphasise that applying the 
standard QFT rules to the exchange of our spin 2 pomeron 
automatically gives equality of the pomeron part of the $pp$ and $\bar{p} p$ elastic 
amplitudes. Furthermore, we have found it easy to write down QFT vertices 
for the coupling of our tensor pomeron to two vector particles. 
This is important for applications of our model to photon-induced reactions 
where the question of gauge invariance arises; see \cite{searchfortheodderon}. 
In section \ref{Odderon exchange} we have discussed the odderon as an 
effective vector exchange. In section 
\ref{Reggeon exchanges and total cross sections} we have presented our 
ansatz for reggeon exchanges. We consider the $f_2$ and $a_2$ reggeons 
as effective rank-two tensor exchanges and the $\rho$ and $\omega$ 
reggeons as effective vector exchanges. 

We note that already many years ago attempts were made to understand 
the pomeron as a tensor; see \cite{Freund:1962,Freund:1971sh,Carlitz:1971ee}. 
In these papers typically it was tried to relate the properties of the pomeron 
to those of mesonic trajectories. But we believe now that the pomeron 
is predominantly a gluonic object. Thus, its properties are not -- and should 
not be -- directly related to those of the reggeons. Our present model 
for the pomeron takes this into account and is, therefore, completely 
different from the models of \cite{Freund:1962,Freund:1971sh,Carlitz:1971ee}. 

Our attempts to determine the parameters of our model as far as possible 
from data have been presented in sections 
\ref{Reggeon exchanges and total cross sections}, \ref{Pion-proton scattering}, 
and \ref{rho-proton scattering}. We have, in particular, considered fits to 
total cross-section data as presented in \cite{Donnachie:2002en}. In this way 
we have obtained default values for most of the parameters of our model. 
Clearly, it would be desirable to test the model in detail with a global fit to 
the available data. This is beyond the scope of the present work and must 
be left for future investigations. In this article we have restricted ourselves 
to the soft high-energy reactions listed in section \ref{The reactions}. 
It is straightforward to extend the model to other reactions of this type, 
for instance, reactions involving kaons, $\phi$ mesons, and hyperons. 
Also central production of mesons in proton-proton collisions can be 
discussed, 
\begin{equation}
p + p \longrightarrow p + \mathrm{meson} + p \,.
\end{equation}
Such reactions are of particular interest since they have been studied 
experimentally and can be studied further in future experiments. 
A paper on central production based on our model is in preparation 
\cite{LNStoappear}. The model may also be applied to reactions suitable 
for odderon searches and to reactions involving polarised protons. 

To summarise: the model for soft high-energy reactions which we have 
developed in this article has as main purpose a practical one. The model 
should allow easy calculations, using standard rules of QFT, for a variety 
of soft reactions. The hope is that it will provide a sort of standard against 
which experimentalists may compare their data in detail, that is, including 
all angular and spin distributions. 

\section*{Acknowledgments}

The authors would like to thank M.\ Albrow, S.-U.\ Chung, M.\ Diehl, 
L.\ Jenkovszky, P. Lebiedowicz, A.\ Martin, 
M.\ Sauter, W.\ Sch\"afer, R.\ Schicker, A.\ Sch\"oning, A.\ Szczurek, 
and O.\ Teryaev for useful discussions. 
The work of C.\,E.\ was supported by the Alliance Program of the
Helmholtz Association (HA216/EMMI). 

\appendix

\boldmath
\section{The $f_2$ Propagator in Detail}
\unboldmath
\label{appA}

Here we construct in detail our model for the $f_2$ propagator. 
Throughout this section we work in pure strong interaction 
dynamics neglecting electromagnetic effects. We start by constructing 
for given four-momentum $k$ with $k^0>0$ and $k^2>0$ the symmetric 
rank 2 polarisation tensors corresponding to spin 2, spin 1 and spin 0. 
For this we assume $\vec{k}\neq 0$, use the coordinate system \eqref{5.2} 
and define as in \eqref{5.3} but with $m_{f_2}$ replaced by $\sqrt{k^2}$ 
\be\label{A.1}
\epsilon ^{(\pm) \mu}(k)=\mp\frac{1}{\sqrt{2}}
\begin{pmatrix}0\\\vec{e}_1\pm i\vec{e}_2\end{pmatrix}\,,
\qquad
\epsilon^{(0)\mu}(k)=\frac{1}{\sqrt{k^2}}
\begin{pmatrix}|\vec{k}|\\ \vec{e}_3k^0\end{pmatrix} \,.
\ee
A basis for the polarisation tensors of spin 2, $\epsilon^{(m)\mu\nu}(k)$ 
with $m\in \{-2,\dots 2\}$, is defined as in \eqref{5.1}, but with 
$\epsilon^{(m_i)\mu}(k)$ from \eqref{A.1}. A basis for spin 1 is given by
\be\label{A.2}
\eta^{(m)\mu\nu}(k)
=\frac{1}{\sqrt{2k^2}}(k^\mu\epsilon^{(m)\nu}(k)+\epsilon^{(m)\mu}(k)k^\nu)\,,
\quad m\in\{1,0,-1\}\,.
\ee
We have two possibilities for spin $0$ polarisation tensors:
\be\label{A.3}
\begin{split}
\xi^{(1)\mu\nu}(k)&=\frac{k^\mu k^\nu}{k^2}\,,
\\
\xi^{(2)\mu\nu}(k)&=\frac{1}{\sqrt{3}}\left(-g^{\mu\nu}+\frac{k^\mu k^\nu}{k^2}\right) \,.
\end{split}
\ee
The following relations are easily checked:
\begin{align}
\left(\epsilon^{(m)\mu}(k)\right)^*
\epsilon^{(n)}_\mu(k)&=-\delta_{mn}\,,\label{A.4}\\
\left(\epsilon^{(m)\mu\nu}(k)\right)^*\epsilon^{(n)}_{\mu\nu}(k)&=\delta_{mn}\,,\label{A.5}\\
\left(\epsilon^{(m)\mu\nu}(k)\right)^*\eta^{(n)}_{\mu\nu}(k)&=0\,,\label{A.6}\\
\left(\epsilon^{(m)\mu\nu}(k)\right)^*\xi^{(a)}_{\mu\nu}(k)&=0\,,\label{A.7}\\
\left(\eta^{(m)\mu\nu}(k)\right)^*\eta^{(n)}_{\mu\nu}(k)&=-\delta_{mn} \,,\label{A.8}\\
\left(\eta^{(m)\mu\nu}(k)\right)^*\xi^{(a)}_{\mu\nu}(k)&=0\,,\label{A.9}\\
\left(\xi^{(a)\mu\nu}(k)\right)^*\xi^{(b)}_{\mu\nu}(k)&=\delta_{ab} \,.
\label{A.10}
\end{align}
Here it is understood that $m,n,a,b$ run over the appropriate index ranges.

Now we construct the tensors $P^{(i)}$ \eqref{5.102} by defining
\be\label{A.11}
\begin{split}
P^{(2)}_{\mu\nu,\kappa\lambda}(k)=& 
\sum^2_{m=-2}\epsilon^{(m)}_{\mu\nu}(k)\left(\epsilon^{(m)}_{\kappa\lambda}(k)\right)^*
\\
=&\,\frac{1}{2}\left(-g_{\mu \kappa}+\frac{k_\mu k_\kappa}{k^2}\right)
\left(-g_{\nu\lambda}+\frac{k_\nu k_\lambda}{k^2}\right) 
+\frac{1}{2}\left(-g_{\mu\lambda}+\frac{k_\mu k_\lambda}{k^2}\right)
\left(-g_{\nu \kappa}+\frac{k_\nu k_\kappa}{k^2}\right) 
\\
&-\frac{1}{3}\left(-g_{\mu\nu} +\frac{k_\mu k_\nu}{k^2}\right)
\left(-g_{\kappa\lambda}+\frac{k_{\kappa} k_\lambda}{k^2}\right)\,,
\end{split}
\ee
\be\label{A.12}
\begin{split}
P^{(1)}_{\mu\nu,\kappa\lambda}(k)=&\sum^1_{m=-1}\eta^{(m)}_{\mu\nu}(k)
\left(\eta^{(m)}_{\kappa\lambda}(k)\right)^* 
\\
=&\,\frac{1}{2k^2}\Bigg\{k_\mu k_{\kappa}
\left(-g_{\nu\lambda}+\frac{k_\nu k_\lambda}{k^2}\right) 
+k_\mu k_\lambda
\left(-g_{\nu \kappa}+\frac{k_\nu k_{\kappa}}{k^2}\right) 
\\
& \hspace*{30pt}
+k_\nu k_{\kappa}
\left(-g_{\mu\lambda}+\frac{k_\mu k_\lambda}{k^2}\right)
+k_\nu k_\lambda
\left(-g_{\mu \kappa}+\frac{k_\mu k_{\kappa}}{k^2}\right)\Bigg\}\,,
\end{split}
\ee
\be\label{A.13}
P^{(0,ab)}_{\mu\nu,\kappa\lambda}(k)=\xi^{(a)}_{\mu\nu}(k)
\left(\xi^{(b)}_{\kappa\lambda}(k)\right)^*\,,
\ee
\be\label{A.14}
P^{(0,11)}_{\mu\nu,\kappa\lambda}(k)=\frac{k_\mu k_\nu}{k^2}\frac{k_{\kappa}k_\lambda}{k^2}\,,
\ee
\be\label{A.15}
P^{(0,22)}_{\mu\nu,\kappa\lambda}(k)=\frac13
\left(-g_{\mu\nu}+\frac{k_\mu k_\nu}{k^2}\right)
\left(-g_{\kappa\lambda}+\frac{k_{\kappa}k_\lambda}{k^2}\right)\,,
\ee
\be\label{A.16}
P^{(0,12)}_{\mu\nu,\kappa\lambda}(k)=\frac{k_\mu k_\nu}{k^2}\frac{1}{\sqrt{3}}
\left(-g_{\kappa\lambda}+\frac{k_{\kappa}k_\lambda}{k^2}\right)\,,
\ee
\be\label{A.17}
P^{(0,21)}_{\mu\nu,\kappa\lambda}(k)=\frac{1}{\sqrt{3}}
\left(-g_{\mu\nu}+\frac{k_\mu k_\nu}{k^2}\right)
\frac{k_{\kappa}k_\lambda}{k^2}\,.
\ee
We have the following relations
\be\label{A.18}
\begin{split}
P^{(2)}(k)P^{(2)}(k)&=P^{(2)}(k)\,,\\
P^{(2)}(k)P^{(1)}(k)&=0\,,\\
P^{(2)}(k)P^{(0,ab)}(k)&=0\,,\\
P^{(1)}(k)P^{(1)}(k)&=-P^{(1)}(k)\,,\\
P^{(1)}(k)P^{(0,ab)}(k)&=0\,,\\
P^{(0,ab)}(k)P^{(0,cd)}(k)&=\delta_{bc}P^{(0,ad)}(k)\,.
\end{split}
\ee
Here we use matrix notation. That is, the first eq.\ 
of \eqref{A.18} reads in detail 
\be\label{A.19}
P^{(2)}_{\mu\nu,\rho\sigma}(k)P^{(2)\rho\sigma}{}_{\kappa\lambda}(k) = 
P^{(2)}_{\mu\nu,\kappa\lambda}(k)\,,
\ee
etc. Furthermore, we get the trace relations 
\be\label{A.20}
\begin{split}
P^{(2)}_{\mu\nu} {}^{\mu\nu}(k)&=5\,,\\
P^{(1)}_{\mu\nu} {}^{\mu\nu}(k)&=-3\,,\\
P^{(0,ab)}_{\mu\nu} {}^{\mu\nu}(k)&=\delta_{ab}\,,
\end{split}
\ee
and the completeness relation
\be\label{A.21}
\left(P^{(2)}(k)-P^{(1)}(k)+P^{(0,11)}(k)+P^{(0,22)}(k)\right)_{\mu\nu,\kappa\lambda}
=\frac12(g_{\mu \kappa}g_{\nu\lambda}+g_{\mu\lambda} g_{\nu \kappa}) \,.
\ee

Now we come to the general tensor-field propagator \eqref{5.100}. 
With the help of the $P^{(i)}$, \eqref{A.11} to \eqref{A.17}, we can write 
down the expansion \eqref{5.102} where $\Delta^{(2)}(k^2)$, $\Delta^{(1)}(k^2)$, 
and $\Delta^{(0)}_{ab}(k^2)$ are invariant functions. 
For the free propagator \eqref{5.101} these read 
\be\label{A.22}
\begin{split}
\left. \Delta^{(2)}(k^2) \right|_{\rm free}&=\frac{1}{k^2-m^2_{f_2}+i\epsilon}\,,\\
\left. \Delta^{(1)}(k^2) \right|_{\rm free}&=\frac{1}{m^2_{f_2}}\,,\\
\left. \Delta^{(0)}_{11}(k^2) \right|_{\rm free}&=\frac23\frac{k^2-m^2_{f_2}}{(m_{f_2})^4}\,,\\
\left. \Delta^{(0)}_{12}(k^2) \right|_{\rm free}= \left. \Delta^{(0)}_{21}(k^2) \right|_{\rm free}&=
-\frac{1}{\sqrt{3}}\frac{1}{m^2_{f_2}}\,,\\
\left. \Delta^{(0)}_{22}(k^2) \right|_{\rm free}&=0 \,.
\end{split}
\ee
We analyze now the full tensor-field propagator \eqref{5.102} in the same 
spirit as done for the $\gamma$-$\rho$-$\omega$ system in \cite{Melikhov:2003hs}. 
From \eqref{5.100} we get 
\be\label{A.23}
\Delta^{(T)}_{\mu\nu,\kappa\lambda}(k)-\big(\Delta^{(T)}_{\kappa\lambda,\mu\nu}(k)\big)^*
=\frac{1}{i} \int d^4x \,e^{ikx}\langle 0|\big \{\phi_{\mu\nu}(x)\phi_{\kappa\lambda}(0)
+\phi_{\kappa\lambda}(0)\phi_{\mu\nu}(x)\big\}|0\rangle \,.
\ee
Inserting a complete set of intermediate states we get for $k^0>0$ 
\be\label{A.24}
\Delta^{(T)}_{\mu\nu,\kappa\lambda}(k)-\big(\Delta^{(T)}_{\kappa\lambda,\mu\nu}(k)\big)^*
=\frac{1}{i}\sum_X(2\pi)^4\delta^{(4)}(k-p_X) \langle 0|\phi_{\mu\nu}(0)|X(p_X)\rangle
\langle X(p_X)|\phi_{\kappa\lambda}(0)|0\rangle .
\ee
For the following we need the inverse propagator which 
in matrix notation is found to be 
\be\label{A.25}
\big(\Delta^{(T)}\big)^{-1} (k)=P^{(2)}(k)\big(\Delta^{(2)}(k^2)\big)^{-1}
+P^{(1)}(k)\big(\Delta^{(1)}(k^2)\big)^{-1}
+P^{(0,ab)}(k)\big(\big(\Delta^{(0)}\big)^{-1} (k^2)\big)_{ab} \,.
\ee
As done in \cite{Melikhov:2003hs} for the vector propagators we define now reduced 
matrix elements by
\be\label{A.25a}
\begin{split}
&\langle X(p_X)\|\phi_{\mu\nu}\|0\rangle= \langle X(p_X)|\phi^{\kappa\lambda}(0)|0\rangle
\big(\big(\Delta^{(T)}\big)^{-1}(p_X)\big)_{\kappa\lambda,\mu\nu}\,,\\
&\langle 0\|\phi_{\mu\nu}\|X(p_X)\rangle=\langle X(p_X)\|\phi_{\mu\nu}\|0\rangle^* \,.
\end{split}
\ee
We get from \eqref{A.24} and \eqref{A.25}, always for $k^0>0$, 
\be\label{A.26}
\begin{split}
\frac{1}{2i}\big[\big(\Delta^{(T)}\big)^{-1}(k)-
\big(\big(\Delta^{(T)}\big)^{-1}(k)\big)^\dagger\big]_{\mu\nu,\kappa\lambda}
=&\frac12\sum_X(2\pi)^4\delta^{(4)}(k-p_X)
\\
&\times \langle 0\|\phi_{\mu\nu}\|X(p_X)\rangle
\langle X(p_X)\|\phi_{\kappa\lambda}\|0\rangle \,.
\end{split}
\ee
Contracting with $P^{(2)}(k)$ gives
\be\label{A.27}
\begin{split}
&\frac{1}{2i}\left[(\Delta^{(2)}(k^2))^{-1}- 
((\Delta^{(2)}(k^2))^{-1})^*\right] \\ 
&=\frac{1}{2}\sum_X(2\pi)^4\delta^{(4)}(k-p_X)
\,\frac{1}{5}\! \sum^2_{m=-2}(\epsilon^{(m)\mu\nu}(k))^*
\langle 0\|\phi_{\mu\nu}\|X(p_X)\rangle 
\langle X(p_X)\|\phi_{\kappa\lambda}\|0\rangle \epsilon^{(m)\kappa\lambda}(k)\,.
\end{split}
\ee
As already discussed in section \ref{Tensor mesons}, the function $\Delta^{(2)}(k^2)$ is 
the one where the $f_2$ resonance appears. We shall {\em define} the 
squared $f_2$ mass as zero point of the real part of the inverse of $\Delta^{(2)}$:
\be\label{A.28}
\text{Re}\,\big(\Delta^{(2)}(m^2_{f_2})\big)^{-1}=0\,.
\ee
Then we will normalise the tensor field $\phi_{\mu\nu}$ such that 
\be\label{A.29}
\big(\Delta^{(2)}(0)\big)^{-1}=-m^2_{f_2}
\ee
and define $B_{f_2}(k^2)$ by 
\be\label{A.30}
\big(\Delta^{(2)}(k^2)\big)^{-1}=-m^2_{f_2}+k^2+B_{f_2}(k^2)\,.
\ee
With \eqref{A.28} and \eqref{A.29} we must require
\be\label{A.31}
B_{f_2}(0)=0\,, \quad\quad\text{Re}\,B_{f_2}(m^2_{f_2})=0\,.
\ee
From \eqref{A.27} we get, setting $s=k^2$, 
\be\label{A.32}
\begin{split}
\mathrm{Im}\,B_{f_2}(s)=&\,\frac{1}{2}\sum_X(2\pi)^4\delta^{(4)}(k-p_X) \,
\frac15\sum^2_{m=-2}\big(\epsilon^{(m)\mu\nu}(k)\big)^*
\langle 0\|\phi_{\mu\nu}\|X(p_X)\rangle\\
&\times \langle X(p_X)\|\phi_{\kappa\lambda}\|0\rangle\, \epsilon^{(m)\kappa\lambda}(k) \,.
\end{split}
\ee
For an `on-shell' $f_2$ we have
\be\label{A.33}
\langle X(k)|{\cal T}|f_2(k,\epsilon^{(m)}\big)\rangle= 
-\langle X(k)\|\phi_{\mu\nu}\|0\rangle \,\epsilon^{(m)\mu\nu}(k) \,,
\ee
as is  easily seen from the reduction formula, and
\be\label{A.33a}
\mathrm{Im}\, B_{f_2}(m^2_{f_2})=m_{f_2}\Gamma_{f_2}\,.
\ee
We assume now that $B_{f_2}(s)$ satisfies a once subtracted dispersion relation: 
\be\label{A.34}
B_{f_2}(s)=B_{f_2}(0)+sB_{f_2}'(0) 
+\frac{s^2}{\pi}\int^\infty_{4m^2_\pi}ds'
\frac{ \mathrm{Im} \,B_{f_2}(s')}{(s')^2(s'-s-i\epsilon)} \,.
\ee
In pure strong interactions the lowest intermediate states $|X\rangle$ 
in \eqref{A.32} are the two-pion states and, therefore, the dispersion 
integral in \eqref{A.34} starts at $s'=4m^2_\pi$. We set now
\be\label{A.36}
R_{f_2}(s)=\frac s\pi \,\mbox{V.\,P.} \int^\infty_{4m^2_\pi}ds'
\frac{\mathrm{Im}\,B_{f_2}(s')}{(s')^2(s'-s)}
\ee
where V.\,P.\ means the principal value prescription. The expressions 
for $B_{f_2}(s)$ and $(\Delta^{(2)}(s))^{-1}$ satisfying all the constraints 
\eqref{A.28}, \eqref{A.29} and \eqref{A.31} read then 
\be\label{A.37}
B_{f_2}(s)= s\big[R_{f_2}(s)-R_{f_2}(m^2_{f_2})\big] 
+i \,\mathrm{Im} \,B_{f_2}(s) \,,
\ee
\be\label{A.39}
\big(\Delta^{(2)}(s)\big)^{-1}=-m^2_{f_2}+s+s\big[R_{f_2}(s)-R_{f_2}(m^2_{f_2})\big] 
+i\, \mathrm{Im} \,B_{f_2}(s) \,.
\ee

It remains to construct $\mathrm{Im}\,B_{f_2}(s)$ explicitly. For the $\pi\pi$ intermediate 
states in \eqref{A.32} we get from \eqref{3.28} 
\be\label{A.40}
\left. \mathrm{Im} \,B_{f_2}(s)\right|_{\pi\pi}=\frac{1}{320\pi}\left|\frac{g_{f_2\pi\pi}F^{(f_2\pi\pi)}(s)}{M_0}\right|^2
s^2\left[1-\frac{4m^2_\pi}{s}\right]^{5/2}\theta(s-4m^2_\pi) \,,
\ee
\be\label{A.40a}
\begin{split}
\mathrm{Im} \,B_{f_2}(s)|_{\pi^+\pi^-}&=\frac{2}{3}\, \mathrm{Im}\, B_{f_2}(s)|_{\pi\pi} \,,\\
\mathrm{Im} \,B_{f_2}(s)|_{\pi^0\pi^0}&=\frac{1}{3}\, \mathrm{Im}\,B_{f_2}(s)|_{\pi\pi} \,.
\end{split}
\ee
Since the branching ratio 
$\Gamma(f_2\to\pi\pi)/\Gamma_{f_2}\cong 0.85$, see \eqref{5.11}, 
we set as an {\em approximation}
\be\label{A.41}
\mathrm{Im}\,B_{f_2}(s)=\frac{\Gamma_{f_2}}{\Gamma(f_2\to\pi\pi)} \,\mathrm{Im} \,B_{f_2}(s)|_{\pi\pi} \,.
\ee
Inserting this in \eqref{A.36} to \eqref{A.39} leads to the $f_2$ propagator 
given in \eqref{5.104} to \eqref{5.107}. 

\begin{figure}[ht]
\begin{center}
\includegraphics[height=100pt]{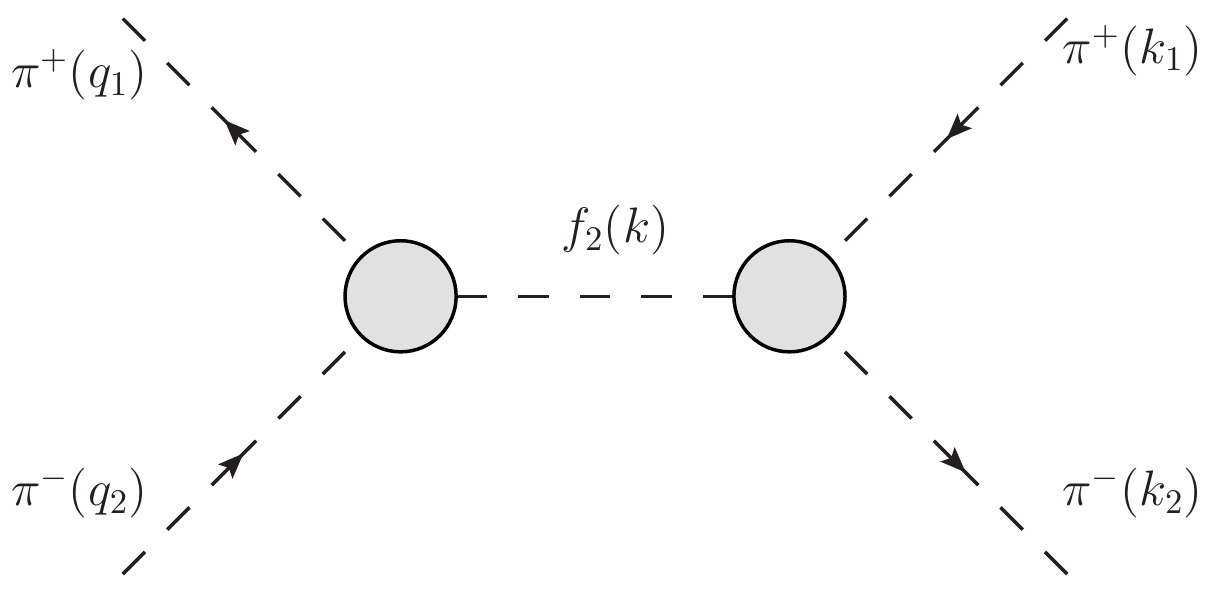}
\caption{Scattering $\pi^+\pi^-\to\pi^+\pi^-$ with exchange of $f_2$ in the $s$ channel.
\label{f2_prop_fig300}}
\end{center}
\end{figure}

Finally, we consider $\pi^+\pi^-$ elastic scattering with exchange of 
an $f_2$ meson in the $s$ channel, see figure \ref{f2_prop_fig300},
\be\label{A.42}
\pi^+(q_1)+\pi^-(q_2)\longrightarrow \pi^+(k_1)+\pi^-(k_2)\,,\qquad
k=q_1+q_2=k_1+k_2\,.
\ee
With the vertex \eqref{3.28} and the invariant function $\Delta^{(2)}(k^2)$ 
from \eqref{5.104} we obtain for the T-matrix element in the c.\,m.\ 
system of the reaction \eqref{A.42} 
\be\label{A.43}
\langle \pi^+(\vec{k}_1),\pi^-(-\vec{k}_1)|\mathcal{T}|\pi^+(\vec{q}_1),\pi^-(-\vec{q}_1)\rangle
=16\pi\left(1-\frac{4m^2_\pi}{s}\right)^{-1/2} 5a_2(s)P_2(\cos \vartheta) \,.
\ee
Here $\vartheta$ is the scattering angle,
\be\label{A.44}
\cos \vartheta=\frac{{\vec{k}}_1\cdot {\vec{q}}_1}{|{\vec{k}}_1||{\vec{q}}_1|}\,,
\ee
$P_2$ is the Legendre polynomial and 
\ba\label{A.45}
a_2(s)=-\mathrm{Im} \,B_{f_2}(s)|_{\pi^+\pi^-}\,\Delta^{(2)}(s)\,.
\ea
As it should be for $f_2$ exchange, scattering is only for orbital angular 
momentum $l=2$, $a_2(s)$ is the $l=2$ partial-wave amplitude, 
and the partial-wave unitary relation is respected. 
Indeed, with $\delta_2(s)$ the phase shift and $\eta_2(s)$ the inelasticity  
parameter we find from \eqref{A.45}
\be\label{A.46}
\begin{split}
\eta_2(s) & e^{2i\delta_2(s)}= \, 2ia_2(s)+1 \\
=&\,\Big\{s-m^2_{f_2}+s\big[R_{f_2}(s)-R_{f_2}(m^2_{f_2})\big]
-i\big[ \mathrm{Im}\,B_{f_2}(s)|_{\pi^+\pi^-}-\mathrm{Im}\,\tilde B_{f_2}(s)\big]\Big\}\\
&\times 
\Big\{s-m^2_{f_2}+s\big[R_{f_2}(s)-R_{f_2}(m^2_{f_2})\big] 
+i\big[\mathrm{Im}\,B_{f_2}(s)|_{\pi^+\pi^-}+\mathrm{Im}\,\tilde B_{f_2}(s)\big]\Big\}^{-1}\,.
\end{split}
\ee
Here we set 
\be\label{A.47}
\mathrm{Im}\,\tilde B_{f_2}(s)=\mathrm{Im}\,B_{f_2}(s)-\mathrm{Im}\,B_{f_2}(s)|_{\pi^+\pi^-}\,,
\ee
that is, $\mathrm{Im}\,\tilde B_{f_2}(s)$ is defined as in \eqref{A.32} but with the sum over 
intermediate states $|X\rangle$ excluding the $\pi^+\pi^-$ states. 
Clearly we still have
\be\label{A.48}
\mathrm{Im}\,\tilde B_{f_2}(s)\geq 0\,.
\ee
For instance, with our approximation \eqref{A.41} we get
\be\label{A.48a}
\mathrm{Im}\,\tilde B_{f_2}(s)=
\left(\frac{\Gamma_{f_2}}{\Gamma (f_2\to\pi\pi)}-\frac{2}{3}\right)
\mathrm{Im}\,B_{f_2}(s)|_{\pi\pi}>0\,.
\ee
From \eqref{A.46} and \eqref{A.48} we find
\be\label{A.49}
\left|\eta_2(s)\right|\leq 1
\ee
as it should be.

To conclude this section we note that our model for the $f_2$ propagator 
could easily be improved by also modelling the amplitudes for the 
states $|X\rangle\neq |\pi\pi\rangle$ in \eqref{A.32}. These would 
be in first instance the four-pion and the $K\bar K$ states. We estimate 
that our $f_2$ propagator \eqref{5.104} to \eqref{5.107} should 
be a reasonable model for, say, $0<s<10\,\mbox{GeV}^2$. 
For larger $s$ we expect that further multi-particle states $|X\rangle$ in 
the sum \eqref{A.32} will become important. It may then be more reasonable 
to consider quark-antiquark states as final states $|X\rangle$ 
in \eqref{A.32}. But this is beyond the scope of the present article. 

\section{The Pomeron as Coherent Sum of Exchanges with Spin 2, 4, 6, etc}
\label{appB}

Here we show how to write the pomeron exchange amplitude for $pp$ 
scattering \eqref{6.27} as coherent sum of exchanges with spin $2$, $4$, $6$, etc. 
The technique is similar to the one used in section 6.2 of \cite{Nachtmann:1991ua}. 
We start from \eqref{6.8a} and \eqref{6.17} which give for the pomeron part of 
the $pp$ scattering amplitude for large $\nu$  
\be\label{B.1}
\begin{split}
\langle p(p'_1),p(p'_2)|{\cal T}|p(p_1),p(p_2)\rangle|_{\mathbbm P} \equiv& 
\,{\cal T}^{\mathbbm P}_{fi} 
=T^{(pp){\mathbbm P}}(\nu,t)\bar u(p'_1)\gamma^\mu u(p_1)\bar u(p'_2)\gamma_\mu u(p_2)\\
=&\,f(t)\exp \left[-i\pi\frac12\big(\alpha_{\mathbbm P}(t)-2\big)\right] 
(4\alpha'^{2}_{\mathbbm P}\nu^2)^{\frac12(\alpha_{\mathbbm P}(t)-2)}2\alpha'_{\mathbbm P}\nu \\
&\times \bar u(p'_1)\gamma^\mu u(p_1)\bar u(p'_2)\gamma_\mu u(p_2)\,.
\end{split}
\ee
In the following we represent $\nu$ as 
\be\label{B.1a}
\nu=\frac14(p_1+p'_1)\cdot (p_2+p'_2) \,.
\ee
Now we use the identity
\be\label{B.2}
(4\alpha'^{2}_{\mathbbm P}\nu^2)^{\frac12(\alpha_{\mathbbm P}(t)-2)}=
\frac{1}{\Gamma\left(1-\frac12\alpha_{\mathbbm P}(t)\right)}\int^\infty_0 d\tau \,
\tau^{-\frac12\alpha_{\mathbbm P}(t)}
\exp (-4\alpha'^{2}_{\mathbbm P}\nu^2 \tau) \,.
\ee
Inserting \eqref{B.2} in \eqref{B.1} we get
\be\label{B.3}
{\cal T}^{\mathbbm P}_{fi}=\tilde f(t)\int^\infty_0 d\tau\,\tau^{-\frac12\alpha_{\mathbbm P}(t)}
\exp (-4\alpha'^{2}_{\mathbbm P}\nu^2 \tau)\,
2\alpha'_{\mathbbm P}\nu\, \bar u(p'_1)\gamma^\mu u(p_1)\bar u(p'_2)\gamma_\mu u(p_2)
\ee
where 
\be\label{B4}
\tilde f(t)=f(t)\exp \left[-i\pi\frac12(\alpha_{\mathbbm P}(t)-2)\right]
\Gamma^{-1}\left(1-\frac12\alpha_{\mathbbm P}(t)\right).
\ee
Expanding the exponential function in \eqref{B.3} we find
\be\label{B.5}
{\cal T}^{\mathbbm P}_{f_i}=\tilde f(t)\int^\infty_0 d\tau \,\tau^{-\frac12\alpha_{\mathbbm P}(t)}
\sum^\infty_{n=1}\frac{(-\tau)^{(n-1)}}{(n-1)!}(2\alpha'_{\mathbbm P}\nu)^{2n-1} \,
\bar u(p'_1)\gamma^\mu u(p_1)\bar u (p'_2)\gamma_\mu u(p_2)\,.
\ee
For large $\nu$ we have, using \eqref{6.23} and \eqref{B.1a} 
\be\label{B.6}
\begin{split}
(2\alpha'_{\mathbbm P}\nu)^{2n-1}\,
\bar u(p'_1)\gamma^\mu u(p_1)\bar u(p'_2)\gamma_\mu u(p_2)
=&\left(\frac12\alpha'_{\mathbbm P}\right)^{2n-1}\bar u(p'_1)R^{\mu_1\dots \mu_{2n}}
(p'_1,p_1)u(p_1)\\
&\times \bar u(p'_2)R_{\mu_1\dots \mu_{2n}}(p'_2,p_2) u(p_2)
\end{split}
\ee
where
\be\label{B.7}
\begin{split}
R^{\mu_1\dots \mu_{2n}}(p',p)=&\,
\frac{1}{2n}\big\{\gamma^{\mu_1}(p'+p)^{\mu_2}\dots(p'+p)^{\mu_{2n}}
+(p'+p)^{\mu_1}\gamma^{\mu_2}\dots(p'+p)^{\mu_{2n}}\\
&\,+ \dots +(p'+p)^{\mu_1}\dots (p'+p)^{\mu_{2n-1}}\gamma^{\mu_{2n}}\big\}\,.
\end{split}
\ee
Inserting \eqref{B.7} in \eqref{B.6} we can write ${\cal T}_{fi}^{\mathbbm P}$ 
as coherent sum of exchanges with spin $2$, $4$, $6$, \dots, 
\be\label{B.8}
\begin{split}
{\cal T}^{\mathbbm P}_{fi}=&\,\tilde f(t)\int^\infty_0 d\tau \,\tau^{-\frac12\alpha_{\mathbbm P}(t)}
\sum^\infty_{n=1}\frac{(-\tau)^{n-1}}{(n-1)!}\left(\frac12\alpha'_{\mathbbm P}\right)^{2n-1}
\\
&\times\bar u(p'_1)R^{\mu_1\dots \mu_{2n}}(p'_1,p_1)u(p_1)
\bar u(p'_2)R_{\mu_1\dots \mu_{2n}}(p'_2,p_2)u(p_2)\,.
\end{split}
\ee
Note that the infinite sum in \eqref{B.8} starts with spin $2$ 
and {\em not} with spin $0$ exchange. For spin $0$ exchange we 
would have to have a structure
\be\label{B.9}
{\cal T}^{\mathbbm P}_{fi} \propto \bar u(p'_1)u(p_1)\bar u(p'_2) u(p_2) \,.
\ee
For high-energy small-angle scattering this would give 
$s$-channel-helicity-conserving, single- and double-helicity-flip amplitudes 
of the same order of magnitude. This is not what is seen by experiment which 
finds predominantly $s$-channel-helicity conservation; for recent $pp$ 
experimental results see for instance \cite{Adamczyk:2012kn,sandacz-talk}. 
As is well known, one can understand this from QCD where gluons couple in a chirality 
conserving way to quarks. For high energy quark-quark scattering this 
leads to quark-helicity conservation, as can be seen, for instance, 
from the general analysis of this process in \cite{Nachtmann:1991ua}. 
There and in \cite{Nachtmann:1996kt} it is shown -- in a model -- how this leads also to 
helicity conservation for hadron-hadron scattering. 

Finally we remark that the odderon-exchange amplitude \eqref{6.30}, \eqref{6.31} 
can be understood as arising from the coherent sum of spin 1, 3, 5, \dots exchanges. 
This can easily be shown with the help of the above techniques.

\end{document}